\documentclass[a4paper]{article}

\usepackage[super]{natbib}

\usepackage[english]{babel}
\usepackage[utf8]{inputenc}
\usepackage[T1]{fontenc}

\usepackage{tcolorbox}

\usepackage{caption}
\usepackage{listings}
\usepackage{graphicx}

\usepackage[autostyle]{csquotes}

\usepackage[a4paper,top=3cm,bottom=2cm,left=3cm,right=3cm,marginparwidth=1.75cm]{geometry}

\usepackage{authblk}
\usepackage{rotating}
\usepackage{csvsimple}
\usepackage{float}

\begin{document}


\title{\huge Bugs as Features (Part I): Concepts and Foundations for the Compositional Data Analysis of the Microbiome-Gut-Brain Axis}
\author[1, 2, +]{Thomaz F. S. Bastiaanssen}
\author[3*]{Thomas P. Quinn}
\author[4*]{Amy Loughman}

\affil[1]{\footnotesize APC Microbiome Ireland, University College Cork, Ireland.
}
\affil[2]{\footnotesize Department of Anatomy and Neuroscience, University College Cork, Ireland.
}
\affil[3]{\footnotesize Independent Scientist, Geelong, Australia
}
\affil[4]{\footnotesize IMPACT (the Institute for Mental and Physical Health and Clinical Translation), Food and Mood Centre, Deakin University, Australia.
}
\affil[+]{\footnotesize Corresponding author: thomazbastiaanssen@gmail.com}
\affil[*]{\footnotesize Joint senior authors}

\date{}

\Affilfont{\fontsize{4}{4}}

\maketitle

\begin{abstract}
    
There has been a growing acknowledgement of the involvement of the gut microbiome - the collection of microbes that reside in our gut - in regulating our mood and behaviour. This phenomenon is referred to as the microbiome-gut-brain axis. While our techniques to measure the presence and abundance of these microbes have been steadily improving, the analysis of microbiome data is non-trivial. 

Here, we present a perspective on the concepts and foundations of data analysis and interpretation of microbiome experiments with a focus on the microbiome-gut-brain axis domain. We give an overview of foundational considerations prior to commencing analysis alongside the core microbiome analysis approaches of alpha diversity, beta diversity, differential feature abundance and functional inference. We emphasize the compositional data analysis (CoDA) paradigm. 

Further, this perspective features an extensive and heavily annotated microbiome analysis in R in the supplementary materials, as a resource for new and experienced bioinformaticians alike. 

\end{abstract}

\newpage

\section{Introduction}
Microorganisms can be found in large numbers in almost all environments, including in and on the human body. The largest collection of microbes on humans can be found in the gut and is referred to as the gut microbiome. According to recent estimates, the human gut microbiome typically consists of around $3*10^{13}$ microbes, weighing approximately 200 grams \citep{RN1}. In terms of genetic diversity, the microbiome outmatches its human host more than three orders of magnitude, and has co-evolved with their eukaryotic hosts \citep{tierney2019landscape}. Whilst the gut microbiome typically refers to microbes housed in the large intestine, there are microbial niches throughout the gastrointestinal tract that have relevance for health states, in particular in the oral cavity (e.g., tongue, plaque, gingival surfaces) and the small intestine. For completeness, we will use 'gut microbiome' or 'gut microbiota' to encompass any and all microbial communities along the gastrointestinal tract.

\subsection{The Microbiome-Gut-Brain Axis}
With the advent of high-throughput sequencing, the gut microbiome has become a popular subject of investigation, as evidenced by large scientific endeavours designed to map the human microbiome to health and disease \citep{RN3, RN7, RN4, RN5, RN6}. As part of these efforts, it has become increasingly clear that the microbiome is in constant bidirectional communication with the host, and that both systems influence each other on multiple levels. 
The bidirectional communication between the microbiome and the host brain is referred to as the microbiome-gut-brain axis. There are several ways in which this communication occurs, for example in the production of neuroactive compounds and metabolites like short-chain fatty acids, modulating the immune system and direct stimulation of the vagus nerve 
\citep{RN8}. Besides playing an important role in gut-brain communication during health and homeostasis, the microbiome has also been found to be affected by psychotropic medication \citep{RN10, RN11}. In some cases, the microbiome can even metabolize psychotropic medication such as L-DOPA, which is commonly prescribed for Parkinson’s disease
\citep{RN12}.

The oral microbiome, commonly sampled via saliva, has also been reported to associate with depressive states, risk of dementia, metabolic health and cardiovascular disease \citep{intro1,intro2, intro3, intro4}. Similarly to the large intestinal microbiome-gut-brain axis, there are numerous routes of communication between oral microbes and the central nervous system, including direct translocation of microbes via facial nerves, the olfactory system and the blood stream, as well as indirect neuroinflammatory effects via systemic inflammation caused by periodontal infection. The small intestine is more difficult to study, however it has established roles in carbohydrate metabolism, bile acid deconjugation and micronutrient storage \citep{martinez2018small}. It is implicated in gastrointestinal  pathophysiologies such as environmental enteric dysfunction, pouchitis, IBS,  of relevance to the pathophysiology \citep{kastl2020structure}. 

\subsection{A perspective on microbiome bioinformatics analysis}
 Microbiome analysis is complex, and the discoveries about methods and biology alike are evolving constantly. This perspective aims to make microbiome data analysis less daunting by presenting a concise description of the key steps involved. Although there are many reasonable approaches to analysing the microbiome, we set out to provide the reader with at least one such approach. In this article, we present an overview of the various methods used to analyse, interpret, and visualise microbiome studies. While the text below focuses on high-level concepts, we also include a fully reproducible analysis in the Supplement, written in RMarkdown, that takes our readers through a complete start-to-finish analysis of microbiome data. 

This two-part perspective series and the accompanying supplementary materials were written with an audience specialised in biological psychiatry in mind and many of the examples in this article reflect this. However, we argue the points discussed here can be applied to most, if not all, host-microbiome experiments. An overview of a typical microbiome analysis can be found in \textbf{Fig. 1}. 

\begin{figure}[hbtp]
\includegraphics[width=\textwidth]{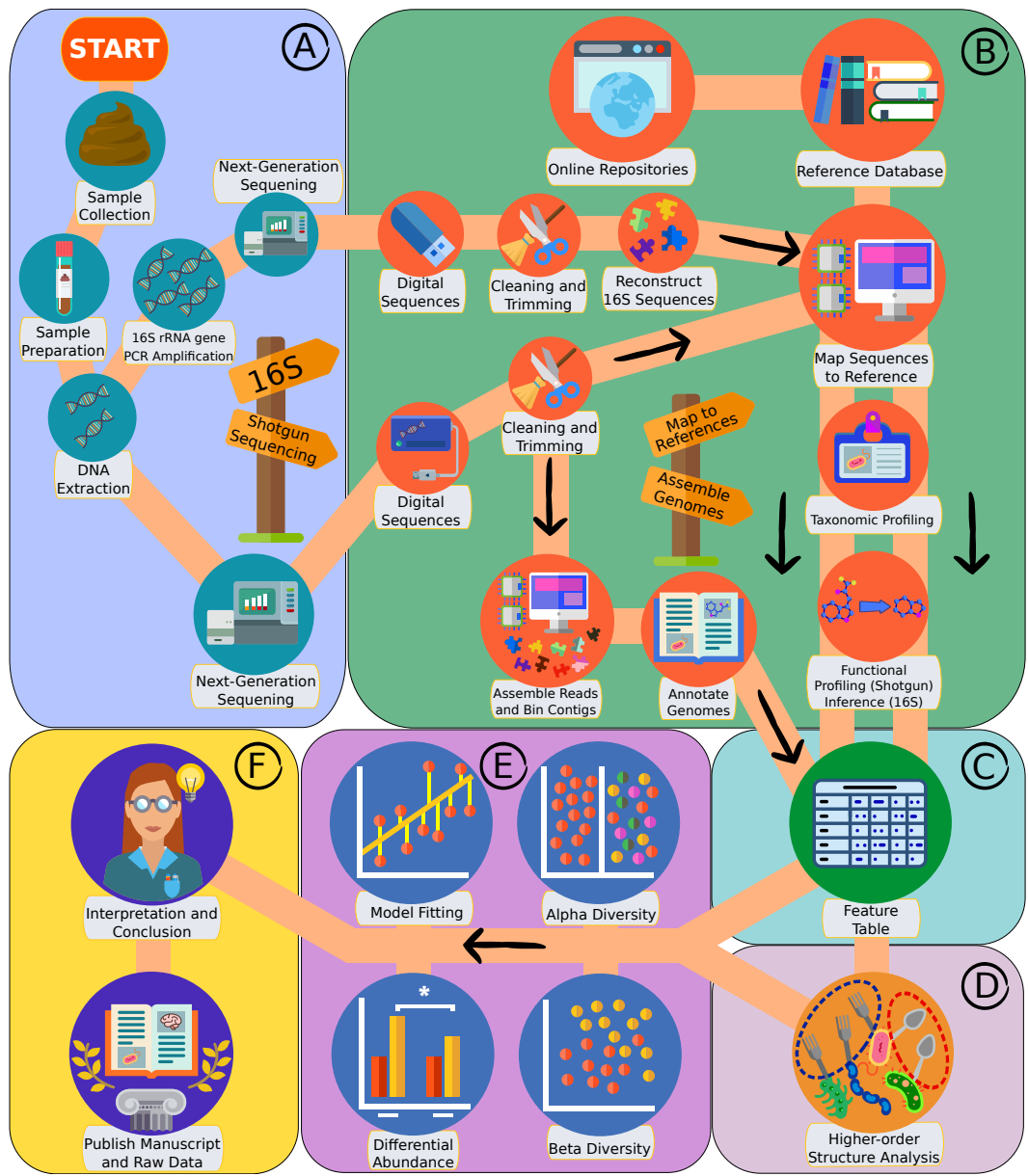}
\centering
\caption{\textbf{From Stool to Story. Overview of what a typical gut microbiome analysis may look like.} \textbf{A)} Shows the pre-digital part of the pipeline. Genetic material is isolated and digitized, either using 16S rRNA gene or metagenomic shotgun sequencing. In \textbf{B)} the digitized reads are annotated based on taxonomy and/or function. This process is distinct between data from 16S and metagenomic shotgun sequencing. In the case of shotgun sequencing, reads can be mapped directly to reference genomes, or reads can be assembled to metagenome-assembled genomes (MAGs) which can in turn be annotated for taxonomy and functional content. Often, both approaches are employed within the same study. In \textbf{C)}, the features are tallied up into feature tables. In \textbf{D)}, higher-order structures are identified and derived from the feature table. Examples include mesoscale structures such as interaction networks, trophic layers, ecological guilds and functional modules. In \textbf{E)} the features of the microbiome are assessed statistically. Special attention should be given here to controlling the false discovery rate (FDR). Finally, in \textbf{F)}, the findings are interpreted and presented for peer-review. In tandem with publication, raw data should be made available to other researchers by uploading to a repository like the SRA or ENA. For a more in-depth comparison of sequencing protocols, we refer the reader to specialized reviews \citep{seq_protocols, szostak2022standardisation}. }
\end{figure}
\newpage

\section{Getting ready for the analysis}

\subsection{Pre-registration}
Pre-registration is a mainstay of reproducible science, and is becoming an increasingly common practice \citep{RN15, RN13, RN14}. We stress here that we are not advocating for the pre-registration of any and all microbiome studies. Indeed, exploratory studies play a crucial role in mapping ``microbial dark matter'' and allow for subsequent hypothesis generation. Pre-registration involves documenting hypotheses and an analysis plan for a study prior to examining data and running the analysis. It is a practical commitment to avoid ‘fishing’ and the selective presentation of results on the basis of significance, and to mitigate against the known cognitive biases of human reasoning (e.g., confirmation bias) \citep{RN16}. In microbiome science, where the control of the Type 1 error rate is critical, and the reproducibility of findings is particularly challenging \citep{RN17}, pre-registration is especially important. Early indications suggest that the practice has reduced publication bias for positive results \citep{RN18}, and can therefore improve the integrity of published research. 
Pre-registration tools prompt researchers to describe their study and research questions, and then generate a date-stamped document that can be published with a digital object identifier (DOI) either immediately or after a user-defined period of embargo (e.g., following publication). The pre-registration document thus serves as a public record of the planned analyses and analytic strategy which can be referenced in resulting publications to affirm that the findings reflect a hypothesis-driven analysis. Pre-processing steps should also be specified a priori where practicable, since these will affect downstream results. There are a number of free tools and guidelines for pre-registration, including the Open Science Framework (osf.io) and As Predicted (aspredicted.org). These are akin to clinical trial registration sites, and are suitable for observational and experimental studies alike. Guidance as to relevant details to include in pre-registration and study design more broadly can be sought from emerging consensus checklists such as STORMS \citep{RN19}.
It is important to note that within a pre-registration framework, exploratory and post-hoc analyses are still entirely valid. Indeed, within a relatively young and rapidly evolving field such as microbiome science, it is appropriate to continue hypothesis generation and exploration. However, exploratory analyses should be presented as such, and should be clearly distinguished from confirmatory hypothesis testing \citep{RN16}. Pre-registration may include a discussion about the power calculations used to select the sample size, which we discuss in the companion piece to this article \citep{part2}. The special case of experimental designs involving Faecal Microbiota Transplantation (FMT) is also discussed there. Also see \cite{ferdous2022rise}.

\subsubsection{Considering potential confounding factors}
A key challenge of microbiome research, and in particular observational studies of the human microbiome, is delineating the variable-of-interest from other factors that influence the ecosystem. Genetics, ethnicity, early life factors such as modes of birth and feeding and stress, habitual diet, environmental exposures and medication use are just some of the important contributors to the human microbiome, and are also often related to the outcome of interest \citep{RN51, RN10, RN52, RN53}. These are always worth considering in microbiome research, especially in the context of causal inference modelling. The most appropriate way to account for these confounding factors is highly context-dependent and can be difficult to determine. Clearly communicated and reproducible methods are key. We will discuss three important and related approaches to statistically deal with these factors, a process referred to as \textit{deconfounding}: 
\begin{itemize}
    \item \textbf{Linear modeling}: One straightforward and common strategy to deal with these factors is to include them as covariates in a statistical model, most commonly a generalised linear model (GLM). These flexible models can be thought of as more general versions of commonly used ``named'' statistical tests such as the t-test, Pearson's correlation coefficient and the ANOVA. Please also see the demonstration found in the supplementary files, where we demonstrate and discuss including covariates in a linear model. Relatedly, one could compare two models, an $H_0$ (null) model just including the covariates used as confounders to an $H_1$ model that includes the microbiome feature of interest in addition to those covariates, using a log-likelihood test (LRT) or similar.
\end{itemize}
Often however, since a wide variety of factors affect both the microbiome and the brain side of the microbiome-gut-brain axis, it can be unclear how many and which factors to include as covariates.
\begin{itemize}
    \item \textbf{Stepwise variable selection}:  Also referred to as stepwise regression, is often used as a way to statistically determine which factors to include as covariates in a data-driven manner. In brief, this method involves sequentially adding or substracting covariates from a model and recalculating the test statistic, retaining those covariates that display statistical significance. 
\end{itemize}
We note here that statistical significance does not determine whether a variable is or isn't confounder. Ideally, biological knowledge should inform the inclusion or exclusion of a covariate in a model. The biostatistician should always consider the biological interpretation of including or excluding a covariate from a statistical model. Practically however, this may not always be feasible, especially in highly complex biological data sets such as from microbiome science. 
\begin{itemize}
\item \textbf{Causal inference}: A third approach that does require biological interpretation. Though not always the stated aim, causal inference is commonly the underlying motivation for studies of the microbiome. Consider psychiatry, for example. We may aim to estimate the effects that the gut microbiome has on some parameter of brain function, whether it be mood, behavior, cognition, or a neurodevelopmental indicator. Other valid study aims might include description (e.g., of the microbiome in people experiencing depression) and prediction (e.g., can we predict who will develop depression based on their microbial features); however, causal interpretations are often attributed even to these kinds of studies \citep{RN54}. In the companion piece to this article, we discuss the 5 phases of causal inference analysis adapted from Ponsonby \cite{RN55,part2}. These phases make use of a directed acyclic graph (DAG), as shown in \textbf{Fig. 2}, modelled on the example from the Zhu et al (2020) dataset expanded on in the accompanying demonstration \citep{zhu2020metagenome}.
\end{itemize}
Even within descriptive or predictive studies, it can be useful to examine if causal features such as dose response or temporality exist. Causal questions are commonly implied even in cross-sectional and associative human studies for example, in which the microbiome is not being manipulated, and its effect is therefore not being explicitly measured. For this reason, causal inference principles have broad relevance. Importantly, causal inference is not the same as assigning causality based on an observational study; rather, causal inference seeks to determine whether the data supports a causal hypothesis by performing statistical analyses within a causal framework. 

\begin{figure}[hbtp]
\includegraphics[width=\textwidth]{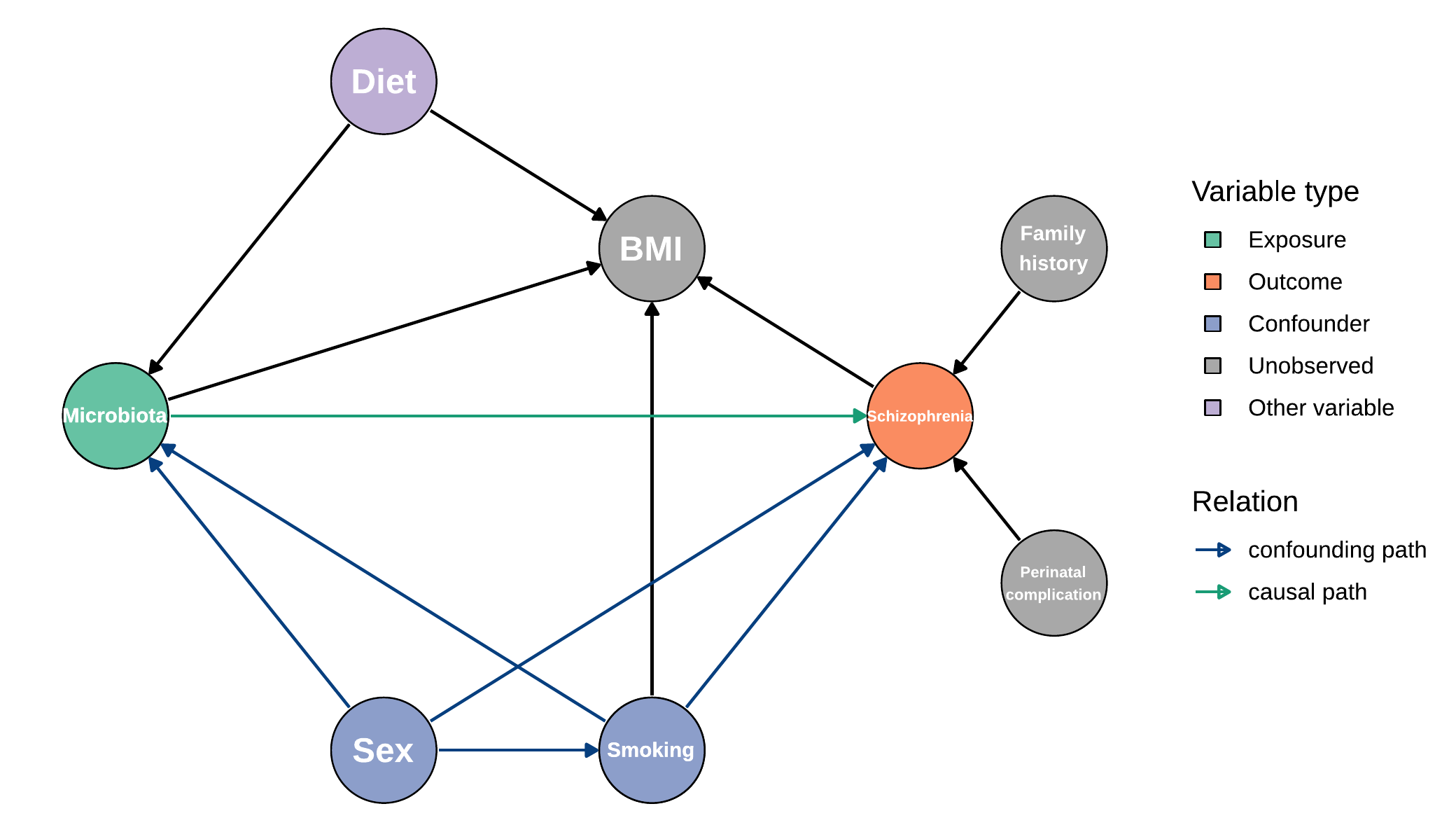}    
\centering
\caption{\textbf{An example DAG.} A DAG describes a hypothetical causal pathway, where an arrow from A → B suggests a causal relationship between A and B. Bidirectional relationships aren’t readily captured within this framework; the dominant direction needs to be selected. This DAG was created using the dagitty R library and reflects variables relevant to the Zhu et al. (2020) schizophrenia dataset used in the accompanying Rmarkdown script. One may wish to include measured variables only, but ideally the framework of constructs and variables is considered prior to (thereby informing) data collection. Some examples of relevant unobserved variables are shown here in grey. For readability, aim to have variables presented in temporal order from left to right. Under the assumptions of this DAG and with only three available covariates (sex, smoking status and body mass index), sex and smoking status comprise the minimal adjustment set required to estimate the total effect of gut microbiota on schizophrenia. }

\end{figure}

\subsubsection{The feature table}
Although 16S/amplicon and shotgun sequencing differ widely in execution, the type of data that is obtained tends to converge downstream in the analysis. After pre-processing, both 16S and shotgun sequencing methodologies yield a feature abundance table. A feature abundance table shows how many observations (i.e., counts), there were for each feature (e.g., microbe, function, gene, etc.) per sample. Some programs and frameworks, such as mOTUs and many shotgun sequencing tools, do not produce count tables but rather (relative) abundance tables (TSS-transformed). Fortunately, the compositional methods discussed in \textbf{Box 1} are still appropriate in these cases.
By convention, a feature abundance table should have features as columns and samples as rows. Although many software tools assume this organisation, there are notable exceptions, so it is always worth checking the software before proceeding with an analysis. It is tempting to directly correspond a count or abundance to a biological instance or abundance of a feature in a sample, but due to biases inherent to metagenomic sequencing \citep{RN34, McLarenBias, nearing2021identifying}, raw abundances should be pre-processed first, for example via normalization or log-ratio transformation. Sometimes counts and abundances are instead expressed as compositional data, which we discuss in \textbf{Box 1}. 
See supplementary sections \textbf{S1.2} and \textbf{S1.3} for tools to generate feature tables from 16S and shotgun sequencing data, respectively.

%
%
\begin{tcolorbox}
\textbf{BOX 1: 16S rRNA gene/amplicon vs. shotgun sequencing.} Generally speaking, two methods of microbiome sequencing are widely used. The first is 16S rRNA gene sequencing, also called amplicon sequencing or simply "16S", includes methods where an evolutionarily preserved genomic sequence is targeted and sequenced. The second is metagenomic shotgun sequencing, where all genetic material in a sample is targeted and sequenced.

\paragraph{}

While downstream bioinformatics analyses of both types of microbiome sequencing techniques converge, the actual techniques are distinct \citep{RN20}. When deciding whether to perform 16S or shotgun sequencing in host-microbiome experiments, broadly speaking, it is often preferable to perform shotgun sequencing, as the approach allows for higher resolution analysis and provides the researcher with direct information on the genetic content of a microbiome sample. Furthermore, shotgun sequencing allows for the reconstruction of uncharacterized genomes, enabling the researcher to investigate novel microbes. Shotgun sequencing also enables the user to track a specific microbe through several samples to perform transmission analysis (See \cite{valles2023person} and the section on FMT in the companion article \citep{part2}). We also note here that 16S sequencing results can potentially be biased due to 16Sr RNA gene copy number variation, though compositional techniques are robust to this (See \textbf{Box 2}).

\paragraph{}

There are however some scenarios where 16S is preferable. First of all, the price of 16S is lower than that of shotgun - though this difference is decreasing and shallow shotgun sequencing has been used as a cost-friendly alternative for 16S \citep{hillmann2018evaluating}. Furthermore, because 16S only targets a gene unique to microbes and has a PCR amplification step, the technique is preferable in samples with a low microbial biomass or with a large proportion of non-microbial (host) genetic material such as the tumour microbiome.
Notably, both techniques are biased towards detecting specific genetic sequences and thus by extent specific microbial taxa \citep{RN21}. Such biases are known to occur between metagenomic sequencing experiments of the same type, even between runs in the same laboratory \citep{McLarenBias, nearing2021identifying}.

\paragraph{}

While wet-lab protocols, including sequencing protocols, are outside of the scope of this article, they greatly impact the quality and content of the resulting microbiome sequencing data. For instance, the microbiome sampling kit itself can contribute a microbiome signature to a measurement \citep{de2021batch}. This can be especially impactful in environments with a low microbial biomass. There is an ongoing effort to standardize microbiome sequencing protocols, for which we refer the reader to specialized reviews \citep{seq_protocols, szostak2022standardisation}.

\paragraph{}

See supplementary sections \textbf{S1.1} \textbf{S1.2} and \textbf{S1.3} for recommendations on quality control (QC), pre-processing 16S sequencing data and shotgun sequencing data to generate feature tables, respectively.
\end{tcolorbox}
%
%

\subsubsection{Rare features and Rarefaction}
Before the microbiome analysis starts, it is common to filter out features by removing them entirely from the feature table. Testing fewer features reduces the magnitude of the FDR adjustment penalty, which in turn helps to increase the statistical power for the remaining tests. Most often, the filtered features represent rare taxa or rare genes. Commonly, features that are only detected in a certain percentage of samples are removed. This is referred to as prevalence filtering. Similarly, features that are only detected in low levels can be dropped. This is referred to as abundance filtering. 
It is worth considering the underlying reasons why a feature may have low prevalence or abundance in your data. Due to the count-based nature of sequencing data, low-abundance features are less likely to reach the limit of detection and come up as zeroes. It is easy to see how low count-abundance, perhaps due to low sequencing depth, can artificially increase feature variance \citep{erb_shrinkage}. In other words, the absence of evidence is not the evidence of absence.

Sometimes, one might wish to filter out features based on other metrics, such as variance \citep{RN45, RN44}. However, features should not be filtered based on their association with an outcome, as this could bias the test statistics and resulting p-value estimates in downstream statistical tests.
The total number of observations recorded for each sample in a feature table depends on the sequencing depth of the assay. Rarefaction is the practice of randomly removing observations from a sample until all samples have the same amount of observations. However, it has been described as an unnecessary and potentially counterproductive measure \citep{RN46}. It is more conventional now to address inter-sample differences in sequencing depth through effective library size normalization or log-ratio transformation \citep{RN47, RN36}. One notable exception is diversity analysis as discussed below \citep{RN49, RN48, RN50}. We argue that, while rarefaction is sometimes justifiable and even recommended, rarefaction should not be seen as the default approach.

\section{Linking the microbiome to host features}

\subsection{Diversity indices}
The microbiome is a complex ecosystem. The analysis and visualisation of the microbiome can be qualitatively distinct from other high-throughput sequencing data. Although the data arise from a molecular biology assay, several of the statistical approaches used in microbiome analysis originate from other fields, such as ecology. This makes microbiome science a clear beneficiary of interdisciplinary research.
Diversity, as popularized in ecology, is a way to quantify and understand variation in microbiome samples. Classically, diversity is separated into three related types: Alpha, Beta, and Gamma diversity \citep{RN63}. Alpha diversity refers to the degree of variation within a sample. Beta diversity refers to the degree of variation between samples. Gamma diversity refers to the total diversity in all samples (which can be  thought of as the Alpha diversity of all samples combined). In practice, Gamma diversity is rarely used. See \textbf{Fig. 3} for an overview of the different alpha and beta diversity indices. 

We expand on statistical considerations with diversity metrics, including applying statistical models to estimate the effects of variables on diversity, in supplementary section \textbf{S2.1} and \textbf{S2.2} respectively.

\begin{figure}[hbtp]
\includegraphics[width=\textwidth]{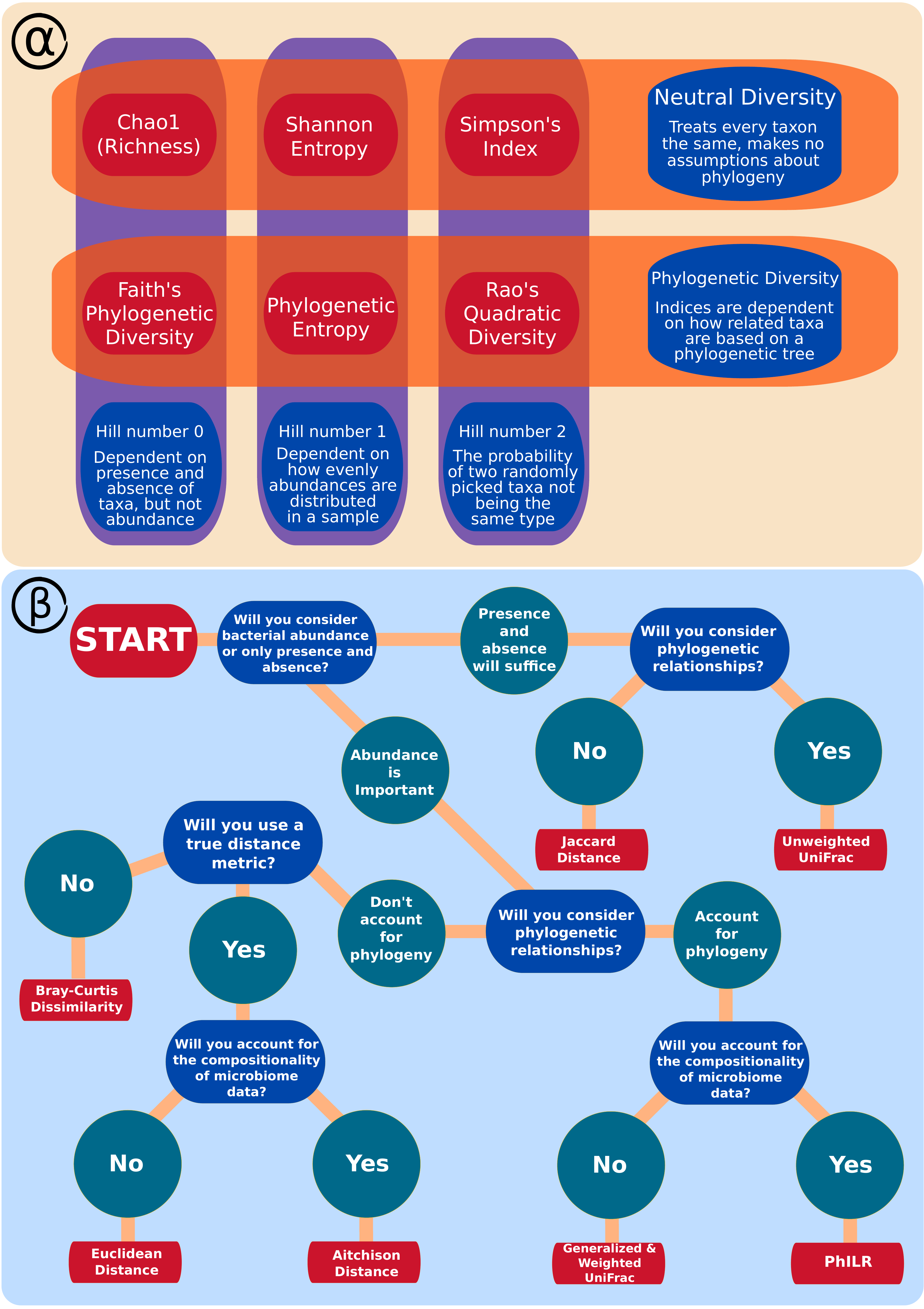}    
\centering
\caption{\textbf{Understanding Alpha and Beta diversity.}  {\boldmath $\alpha)$} Alpha Diversity metrics are related to each other. Commonly used alpha diversity metrics in the microbiome field can be classified along two axes. Here, we show the Hill number on the x-axis, and whether the index considers phylogeny on the y-axis.
 {\boldmath $\beta)$} Decision tree featuring common Beta diversity indices. Some Beta diversity indices are more suitable depending on the needs of the researcher. This decision tree recommends an index based on three common criteria: whether one wants to consider abundance, phylogeny, and/or true distance.
}

\end{figure}

\subsubsection{Alpha diversity – the diversity within samples}
Alpha diversity refers to the degree of complexity within a single microbiome sample. 
Many different but related alpha diversity indices exist, but their relation is unclear from their names. This can make the underlying principles confusing to understand. It is helpful to classify alpha diversity measures along two axes: the Hill number (0, 1, or 2) and whether it is phylogenetic (yes or no). 
Regarding the first axis, alpha diversity measures can be understood as being the result of a unifying equation in which a single parameter - called the Hill number - acts to vary the meaning of the equation, and thus define the alpha diversity measure. Every number gives a different alpha diversity metric. In practice, three Hill numbers are most often used: 0, 1 and 2. The number 0 defines Richness, or how many different features a sample has. The number 1 defines Evenness, or how equally the features in a sample are represented (equivalent to Shannon entropy). The number 2 defines Simpson’s Index, or the probability that two features picked at random do not have the same name (as a probability, it is bounded by 0 and 1). 
Regarding the second axis, other phylogenetic diversity (PD) measures, like Faith’s PD, extend alpha diversity by taking into account the coverage of all features (e.g., bacteria) on a phylogenetic tree. Typically, the more of the tree that is represented in a sample, the higher the diversity. \textbf{Fig. 3A} illustrates a classification of several popular alpha diversity measures.

\subsubsection{Beta diversity – the diversity between samples}
Beta diversity refers to the degree of difference between two microbiomes. It is worth appreciating the assumptions and limitations that come with describing the total difference between two complex ecosystems as a single number. There are many ways to measure the “difference” between two samples, and each one imparts a unique perspective on the data. In principle, one could use any dissimilarity or distance measure. Three common difference measures are:
\begin{itemize}
    \item \textbf{Jaccard’s Index}: This is a similarity measure that simply describes the proportion of unique taxa that are shared between two samples, without taking abundance into account. As such, one could interpret Jaccard’s Index as the fraction of unique taxa (not abundances) shared by two samples. If two samples have exactly the same microbe taxa, the Jaccard index will be 1. In the case that two samples share no microbe taxa, the Jaccard index will be 0. Subtracting Jaccard’s index from 1 makes it the Jaccard Distance measure. 
    \item \textbf{Euclidean Distance}: This is the geometric distance derived by applying the Pythagorean theorem, using every microbe as a separate dimension. It is computed by taking (the square root of) the sum of the squared differences in bacteria abundance. As in geometry, the minimum Euclidean distance is 0 while the maximum is unbounded. Euclidean Distance satisfies the triangle inequality, making it useful for certain geometric analyses, such as volatility analysis as discussed below. A related measure called \textbf{Aitchison Distance} is the Euclidean distance between log-ratio transformed data. This distance has a favorable property known as sub-compositional dominance (i.e., the removal of a taxa feature will never make two samples appear further apart) and is also equivalent to taking the Euclidean distance between all pairwise log-ratios \citep{RN47}.
    \item \textbf{Bray-Curtis Dissimilarity}: This dissimilarity measure is similar to Jaccard’s Index in that it ranges from 0-to-1, while also being similar to Euclidean distance in that it is computed from the differences between abundances. Bray-Curtis is calculated by summing the difference in abundance between each microbial taxon, and dividing it by the total microbial abundance of the two samples. Thus, one could interpret Bray-Curtis as the fraction of abundances unshared by two samples (compare with Jaccard Distance, which is the fraction of unique taxa unshared by two samples).
\end{itemize}
The three common difference measures listed above make use of bacteria presence or abundance without considering the phylogenetic relationship between the bacteria. Just as we can make alpha diversity phylogenetic, we can do the same with beta diversity. 
\begin{itemize}
    \item \textbf{UniFrac}: This distance makes use of phylogenetic information to measure the difference between samples. There are (at least) two types. The Unweighted UniFrac Distance considers the branch lengths of the phylogenetic tree along with microbial presence, and is defined as the sum of branch lengths unshared between the samples divided by the sum of branch lengths present in either sample. This measure has some analogy to Jaccard Distance in that an unweighted UniFrac distance of 1 means the two samples share no bacteria taxa in common. The Weighted UniFrac Distance further considers microbial abundance, and weighs each branch length in the Unweighted UniFrac formula by per-sample proportional abundances.
    \item \textbf{PhILR}: This method uses a log-ratio transformation called the isometric log-ratio (ILR) transformation which uses a phylogenetic tree to recast the microbiome variables as a series of log-contrasts called “balances” \citep{RN68}. PhILR offers two weighting options called taxon weighting and branch weighting. When both are disabled, the PhILR beta diversity is equivalent to Aitchison distance, although its use of phylogeny-based coordinates may yield a more interpretable ordination of the data. The taxon weighting provides a compositionally robust alternative to weighted Jaccard or Bray-Curtis measures, while the branch weighting provides a compositionally robust alternative to UniFrac measures.

\end{itemize}
\textbf{Fig. 3B} illustrates a decision tree that we as the authors use when selecting a beta diversity measure. As with alpha diversity, it is sometimes helpful to compare and contrast the results from multiple measures of beta diversity. 

%
%
\begin{tcolorbox}
\textbf{BOX 2: Microbiome data is compositional.} Compositional data refers to a type of data that can be described as a set of proportions, percentages, probabilities, or with a constant or arbitrary sum. Rather than the relative abundance or \textit{sizes} of the components, the \textit{ratios} between components hold information in compositional data \citep{aitchison1982statistical}. In the past few years, awareness is growing that microbiome datasets are compositional, which, if ignored, can lead to spurious results. The field of study of how to deal with compositional data is called Compositional Data Analysis (CoDA). There are excellent reviews on Compositional Data Analysis in general and how it relates to the microbiome in particular that we encourage our audience to read \citep{RN35, RN36, RN37}.

\paragraph{}

Compositional data analysis theory and practice is continually evolving, and a full review of the field is beyond the scope of this article. In this perspective, we will recommend, at a minimum, performing a centered log-ratio (CLR) or similar transformation (e.g., PhILR) on the count data before performing statistical analysis or visualizing the data. Notably, these log-ratio transformations do not require count data as input but could even be performed on expressly relative abundance data such as TSS-transformed feature tables. In contrast, transformations we expressly recommend against here to account for the effects of compositionality include log-transformations and Total-Sum Scaling (TSS) on their own.

\paragraph{}

However, there are three notable exceptions: 
\begin{itemize}
    \item First, alpha diversity should not be done on log-ratio transformed data. This is due to the nature of the formulas that are used to compute alpha diversity metrics, which all take feature abundance as input (\cite{hill1973diversity}, also see \textbf{Fig. 3A}). 
    \item Second, stacked bar plots should be generated using counts normalized to 1 or to percentages. This transformation is often referred to as Total-Sum Scaling (TSS).  Here, it's worth noting that stacked bar plots depict the proportion of observed reads, which is distinct from the actual sample (relative) abundance \citep{McLarenBias, nearing2021identifying}.   
    \item  Third, correlating taxa to each other, for example as part of a network analysis, warrants special attention. This is because one of the properties of compositional data is that features are inherently negatively correlated. Indeed, Karl Pearson warned against applying his namesake Pearson’s correlation coefficient on compositional data \citep{RN38}. Alternatives are available from the propr library \citep{RN39}, the SparCC library \citep{RN40}, and the SPIEC-EASI library \citep{RN41}.
\end{itemize}

Note that because microbiome count data typically has many zeros and the logarithm of a zero is undefined, the zeroes in microbiome count data must be addressed. Several reasonable solutions have been proposed, but it remains an open question as to which among these solutions performs best (though c.f., \cite{RN42} for one benchmark). In the demonstration found in the supplementary files, we employ an approach derived from the \cite{RN42}, where zeroes are replaced by $\frac{2}{3}$ of the lowest non-zero value. After the CLR-transformation, the values of features can take on any value (unlike count data, which cannot be negative). After transformation, classical statistical approaches can be applied as usual. 
\end{tcolorbox}
%
%

\subsection{Differential feature abundance}
In a nutshell, differential abundance (DA) analysis, refers to the practice of sequentially testing whether each individual feature (gene, taxon, function, etc) of a feature table is different based on our phenotypes, groups or other metadata.
DA is perhaps one of the most popular microbiome analyses. Like alpha and beta diversities, there are many approaches to measuring DA. In some cases, the features are deconfounded first by \textit{regressing out} the selected covariates (i.e. taking the residuals of a model fitted on the selected covariates, also see section \textbf{2.1.1}). Most methods follow the same general pattern: 
\begin{itemize}
    \item Apply a transformation to correct for variation in sequencing depth, compositionality and/or other biases (See \textbf{Box 1}).
    \item Perform a univariate statistical test for each taxon as a dependent variable with the sample meta-data as predictors, for instance by fitting a linear model. This is also a common time to account for confounders by including them as covariates.  
    \item Adjust the p-values for multiple testing, for example using Bonferroni, Storey’s q-value or Benjamini-Hochberg. We expand on multiple testing corrections in the companion piece to this article \citep{part2}. 
\end{itemize}
While this pattern is common for DA analysis, many packages and tool exist to assess differential abundance. Recently, there has been an effort to compare and contrast these tools \citep{nearing2022microbiome}. There is a striking heterogeneity in the performances of DA tools. We recommend compositional methods combined with linear modelling as a safe, consistent and well-understood default approach (see \textbf{Box 1}). Note that while most DA software do treat taxa as the dependent variables, one could just as well treat taxa as the predictors, as routinely done in machine learning applications.

\subsubsection{Functions}
While DA is most commonly used on taxonomic count data, it often makes more biological sense to investigate whether any particular functions rather than taxa in the microbiome can explain a phenotype. 
The way in which one gets to the functional feature table depends on the type of sequencing. In the case of 16S, Piphillin and PICRUSt2 are two options. Both of these tools infer what the metagenome of a sample might look like by mapping 16S sequences to a functional database (e.g., KEGG or MetaCyc) of marker genes from both fully sequenced microbial genomes and metagenome-assemled genomes (MAGs), then inferring a functional feature table based on the functions present in the reference genomes. In the case of shotgun sequencing, there are two main strategies:
\begin{itemize}
    \item If we mapped metagenomic shotgun sequencing reads to a database of reference genomes, the full microbial sequences are already available and we only need to identify genes and collate them, for example with the same KEGG or MetaCyc database. Tools like Woltka or HUMAnN3 in the biobakery suite are typically used to generate a functional feature table for shotgun data.
    \item If we rather generated metagenome-assemled genomes (MAGs), genes need to be detected using a program like Prodigal \citep{prodigal} or Bakta \citep{bakta} and annotated using a database like eggNOG, KEGG or MetaCyc. For a user friendly end-to-end pipeline to generate and annotate MAGs from metagenomic shotgun sequencing reads, we refer the reader to Metagenome-Atlas \citep{atlas}.  
 
\end{itemize}
Regardless of methodology, functional tables tend to contain a large amount of features. Frameworks such as the functional gut-brain modules and gut-metabolic modules (GBMs, GMMs) encompass microbial pathways that specifically cover specific functional aspects of the microbiome such as the potential to produce neuroactive compounds and to metabolise specific substrates \citep{neuroactive, GMMs}. These frameworks enable the interrogation of specific aspects of the microbiome-gut-brain axis. 

\subsubsection{Functional inference vs annotation}
Typically, functions are said to be inferred for 16S and assigned for shotgun sequencing. Strictly speaking, there is inference in both cases. However, in comparing 16S with shotgun sequencing, 16S functional inference can be thought of as a much bigger inferential leap than shotgun. With 16S, we have to first guess the entire genomic content based on a single sequence before inferring function, rather than inferring function directly from the reconstructed or mapped-to genome (as done in shotgun sequencing). This difference is so large that oftentimes inference is said to only happen with 16S. In both cases, a functional analysis is constrained by the validity and completeness of the functional database used to assign functional importance to the taxa or genes. A multitude of functional databases is currently available, with KEGG, UniRef90 and MetaCyc being among the most common general ones. Specialized databases, such as those covering antibiotic resistance genes (ARGs; ARDB, CARD) or carbohydrate metabolism (CAZy) are also available. These databases all have their own focus and frequently have different and sometimes unclear definitions of what a function entails and how a sequence is assigned to a function. While many databases are largely compatible, converting functional IDs between databases often requires some degree of manual curation, to the detriment of the field. Like taxonomic databases, functional databases are updated frequently and results may be affected as a consequence. Biologically speaking, the functional microbiome is known to be more consistent between hosts than the taxonomic microbiome, meaning that the results of functional analyses might generalize better \citep{RN79}. As databases expand over time, it is important to report the version number as part of the Methods.

\section{Beyond the foundations}
The design of host-microbiome experiments and the analysis and interpretation of the resulting data can be a daunting task. In this perspective, we set out to highlight and explain the foundational concepts to enable the reader to navigate and avoid the most common pitfalls. We have provided and referenced the tools to for the reader to customise  their own analysis. We do not claim this approach is the only reasonable way to perform microbiome analysis, only that it is a reasonable one. 
Generally, host-microbiome studies would benefit from reporting a characterization of the microbiome data in terms of alpha diversity, beta diversity and the general microbial composition using stacked bar plots or similar. During the statistical analysis of microbiome data, including modelling, correlating and differential abundance testing, it is important to consider the compositional nature of microbiome data, for example by first performing a CLR-transformation, and to account for the large number of tests performed, for example by performing the Benjamini-Hochberg procedure. In many cases, studies would also benefit from considering the functional potential of the microbiome rather than limiting analysis to the level of taxonomy. 

Microbiome analysis involves techniques and theory from a wide array of fields, including molecular biology, genetics, ecology and even mathematical geology. In the case of fields assessing host-microbe interactions, such as the microbiome-gut-brain axis field, expertise from additional fields such as immunology, psychology, psychiatry, pharmacology, neuroscience and nutrition becomes an additional requirement. In Part II, we present multidisciplinary techniques from across these and other fields to enrich and extend microbiome-gut-brain-axis \citep{part2}. Microbiome research is resource intensive, and there is a strong imperative to 'make the most' of collected data via innovative and robust analytic methods. We acknowledge the extensive body of knowledge and expertise required for comprehensive analysis of the microbiome -  including but not limited to the sheer number of specialist software tools needed for pre-processing and analysis. In this, and the following perspective piece, we strive to assist by showcasing some of these so that the reader and microbiome analyst from any scientific background can navigate the landscape with greater confidence. 

This perspective piece is not without limitations, we acknowledge that no single perspective can comprehensively cover the entire field of microbiome bioinformatics analysis. Indeed, we chose not to cover in silico metabolic modelling of the microbiome. We also focus on metagenomic analysis of the microbiome, rather than other methods to interrogate the microbiome such as metabolomic, metaproteomics or metatranscriptomics. 

The microbiome field is currently undergoing a phase of rapid growth and development. We anticipate that new tools, databases and approaches will slowly replace the current suite. In particular, we anticipate a movement towards more longitudinal experimental designs, as well as the integration of multiple omics approaches on the same microbiome in order to more clearly capture the metabolic functional capacity of the microbiome. For some of these extension topics, please refer to Part II \citep{part2}. In terms of statistical analysis, we encourage the ongoing adoption and further development of CoDA-oriented methodology. Open, freely available and well-documented resources like Bioconductor and CRAN are, and will continue to be, essential for the development and adoption of new bioinformatics tools and pipelines in the microbiome field as well as the broader scientific community. Of particular note, we would like to draw attention to the well-curated and maintained \texttt{curatedMetagenomicData} package on Bioconductor, which allows researchers to access large human microbiome datasets in feature table form, ready for analysis \citep{pasolli2017accessible}. 

Lastly, we would like to highlight that one of the major strengths of science is to build upon the previous findings of others. Often in the microbiome field, a large amount of data is gathered and a broad analysis is performed, perhaps linking the microbiome to a host condition. Whenever possible, raw sequencing data paired with anatomized metadata required to reproduce analyses should be made publicly available as it is essential for reliable and robust meta-analyses. Indeed, many journals require researchers to deposit their raw sequences to nucleotide repositories such as ENA, SRA and DRA. While these types of large-scale studies with broad hypotheses remain valuable to map out the interplay between the microbiome and host, we argue the field is ready to move towards hypothesis-driven experiments with the intent to uncover specific mechanisms and even tractable aspects of the microbiome to help improve both our understanding of the microbiome and our general health and wellbeing. These types of specific hypotheses should be formed based on observations from the large-scale exploratory studies and would also benefit from having bioinformaticians and biostatisticians present during the design stage. 

\section{Acknowledgements}
We would like to thank prof. Anne-Louise Ponsonby for her expert comments on DAGs, Dr. Darren L. Dahly for his insights on statistical analysis and prof. John F. Cryan for his continued encouragement and excellent advice. We are grateful for their help and support. We would also like to thank our partners. A substantial part of this manuscript was written during several periods of lockdown. We are grateful for your support and we love you. 

\section{Declarations}
The authors declare no competing interests.

APC Microbiome Ireland is a research centre funded by Science Foundation Ireland (SFI), through the Irish Governments’ national development plan (grant no. 12/RC/2273\_P2).


\newpage

\huge Supplementary Materials of\\ Bugs as Features (Part I): Concepts and Foundations for the Compositional Data Analysis of the Microbiome-Gut-Brain Axis

\normalsize	

\newpage

\section*{S1     16S rRNA gene/amplicon vs. shotgun sequencing}

\subsection*{S1.1 pre-processing and quality control (QC)}
Before 16s and shotgun sequening reads can be fed into specialized programs to analyse and generate feature abundance tables, the reads need to be checked for quality (QC). Typically, this entails two separate things:
\begin{itemize}
    \item Reads are filtered based on quality based on Phred scores. Phred scores give information on the probability a nucleotide is incorrect due to base-calling error. The relation between Phred score (Q) and the probability of error (P) is as follows: $Q = -10 _{10}log(P)$. As a rule of thumb, a Phred score of 30 (i.e. $ P = \frac{1}{1000}$ ) is often used as a cutoff. 
    \item Adapter sequences need to be trimmed. Due to the way in which microbiomes are sequenced, raw reads will contain short artificial sequences used to facilitate sequencing and amplification. These adapter sequences are straightforward to remove because we typically know what they are. The Cutadapt program was written specifically for this purpose \citep{martin2011cutadapt}. 
\end{itemize}

Preprocessing and QC are essential for microbiome sequencing experiments, as leaving in low quality reads and non-biological sequences like adapter sequences can and will lead to unreliable and possibly misleading annotations and downstream conclusions. Specialized free and open source QC programs like FASTQC \citep{andrews2010fastqc} can be used to assess fastq quality and trimmomatic \citep{bolger2014trimmomatic} to trim sequences. Over the years, numerous improvements and integrated end-to-end QC programs have been published \citep{RN30, 10.1093/bioinformatics/bty560}.

\subsection*{S1.2 Pre-processing 16S sequencing data}
We recommend the excellent, well-documented DADA2 framework to process 16S sequencing data to the feature count table level \citep{RN23, RN24}. The analysis of 16S sequencing typically begins by trimming reads, filtering them for quality based on a threshold, and removing chimera sequences. Then, a table of either operational taxonomic units (OTUs) or amplicon sequence variants (ASVs) is generated. The philosophy and process behind these two units differ meaningfully, and this has been comprehensively written about elsewhere \citep{RN23, RN24}. 

Importantly, due to a multitude of biases including PCR bias and copy number variation, the number of counts in a count table does not correspond to the absolute nor relative sample abundance of the corresponding taxa \citep{McLarenBias, silvermanPCR}. Compositional methods (See Box 2) tend to be robust to these biases, though alpha diversity metrics will be affected if not accounted for \citep{McLarenBias, silvermanPCR}. 
For the purpose of this review, both OTUs and ASVs can be seen as the highest taxonomic resolution that a specific method can distinguish. Practically, their resolution is close to the species or genus level, though taxonomic resolution is determined in a different manner. Although OTUs and ASVs are technically distinct, the two are interchangeable concepts when it comes to downstream statistical analysis. For recommendations on pre-processing 16S data we refer to this excellent workflow \citep{RN25}.

After generating a table of OTUs or ASVs, the next step is to assign taxonomy. In most cases, this is done by use of a reference database. Several such databases exist and some are better curated and maintained than others. 
The most commonly used databases are SILVA, RDP and GreenGenes. At time of writing, the SILVA database is widely regarded as the most accurate and extensive \citep{RN27, RN26}. Though the GTDB database has recently been adopted by many groups. Although the Greengenes database is still often used, it has not been updated since 2013 \citep{RN28}. Notably, GreenGenes2 was recently made available \citep{GreenGenes2}. 
While not covered in this review, the free and open source QIIME2 platform is an excellent and well-maintained Python-based resource for microbiome analysis \citep{RN29}. 

\subsection*{S1.3 Pre-processing shotgun sequencing data}
In the case of shotgun sequencing, it is also common to filter and trim reads in a fashion analogous to 16S data. Apart from this, non-microbial genetic material needs to be filtered out. This is often done by removing all reads that map to a reference genome of the host organism, as well as any other genomes that may be contaminants (e.g., plant genetic material from diet) using a tool like bowtie 2 \citep{bowtie2} or KneadData from the bioBakery suite \citep{RN30}. 

Unlike 16S, there are two major approaches to generate a feature abundance table in the metagenomic shotgun sequencing field. 
\begin{itemize}
\item First, there are reference-based approaches, where partially assembled reads are compared to a database in a manner reminiscent of 16S. Common tools used to process ‘raw’ data from shotgun sequencing include the aforementioned bioBakery suite \citep{RN30}, Kraken2 \citep{RN31} and Bracken \citep{RN32}. Not all tools that generate taxomomy-based abundance data also provide functional data, though HUMAnN 3 from the bioBakery suite \citep{RN30} and Woltka \citep{Woltka} do. A main advantage of the reference-based approaches is that they create common ground between studies that use the same database. Further, these approaches tend to be computationally less demanding, relatively. An important disadvantage of reference-based approaches is that genomes and features not documented in a database are not taken into account. 

\item Second, there are assembly-based approaches. Here genomes are reconstructed into Metagenome Assembled Genomes (MAGs). A major advantage of this approach is that the resulting MAGs reflect the actual observed genetic content, rather than mappping reads to a database. Because of this, the method is well-suited to discover new genomes and features, or to investigate structures like prophage and plasmids as well as genomes at a high resolution. A major disadvantage is that assembly-based methods tend to be particularly computationally expensive. Many tools exist to generate MAGs. Metagenome-Atlas is a sensible and well-documented end-to-end pipeline built for this purpose \citep{atlas}. 

\end{itemize} 
It is not uncommon to do both a reference and assembly based approach in the same study as the two methods can be used to answer different questions about the same data. Typically, taxonomic and functional classification of shotgun metagenomics data is more computationally expensive than its 16S sequencing analysis counterpart and thus often requires a server or computer cluster \citep{RN33}.

\section*{S2 Statistical considerations with diversity metrics}
\subsection*{S2.1 Statistical considerations with Alpha diversity}
Alpha diversity is used to summarize the entire microbiome composition as a single number. It is common to model alpha diversity as a dependent variable, using sample meta-data as the predictors (for example using linear models). When this is done, the literature has shown that a lower Alpha diversity is often associated with worse host health outcomes in humans \citep{RN64}. However, it should by no means be taken as principle that a higher Alpha diversity is strictly “better”, as there are many examples where elevated Alpha diversity indicates an abnormal or even unhealthy host state. For instance, in infants there is a high selection pressure on certain microbes, such as numerous species in the genera \textit{Bifidobacterium} and \textit{Lactobacillus} \citep{RN65}. Here, an increased Alpha diversity could indicate a lowered selection pressure, which could be indicative of health issues \citep{RN66}. 
There are at least three issues to consider when using alpha diversity for microbiome data analysis. 
\begin{itemize} 
\item First, all alpha diversity measures are sensitive to transcript-level measurement biases such as PCR bias \citep{McLarenBias}. This is recognized as a critical limitation of alpha diversity that cannot be resolved unless the PCR bias factors are already known a priori \citep{McLarenBias}. 
\item Second, some alpha diversity measures will change depending on the total number of observations (i.e., sequencing depth) recorded for a sample. It is often appropriate to “normalize” away differences in sequencing depth before comparing alpha diversity between samples. This can be done by dividing out total abundances to get proportions (e.g., in the case of Shannon entropy), or by performing rarefaction. Both procedures will equalise the number of observations between samples, so that they can be compared more fairly. In fact, many alpha diversity software tools will perform this “normalization” step automatically, though no normalization step is perfect. 
\item Third, all alpha diversity measures are sensitive to the number of rare taxa that get observed in samples, and thus are sensitive to sequencing depth. Failure to record the presence of a rare taxa, when it is in fact present, can make a sample appear less diverse than it is \citep{RN50}.
\end{itemize}
It is important to keep these three issues in mind when interpreting the results from an alpha diversity analysis. For example, in the case of very low microbial load due to, say, an antibiotics course, alpha diversity may appear higher than expected \citep{RN67}. This seemingly paradoxical phenomenon can be better understood when considering that there is a limited amount of sequencing material during the sequencing process, regardless of method used. In the case of an abundance of microbes, the most prevalent ones will use up most of the sequencing reagents, leaving little for the rarer taxa to be sequenced. In the case of a low bacterial load, there are no prevalent microbes to take up most of the material and thus the rarer taxa that happen to be in the sample will have a much higher likelihood to be sequenced, thus inflating the calculated diversity.

\subsection*{S2.2 Statistical considerations with Beta diversity}
There are two general strategies used to assess beta diversity:
\begin{itemize}
    \item Unconstrained (Qualitative) – visualization of samples plotted across an ordination of the data, such as a principal components analysis (PCA) or principal coordinates analysis (PCoA)
    \item Constrained (Quantitative) – explicit modelling of PCA/PcoA axes as a dependent variable, using the sample meta-data as the predictors, or some other formal comparison between the group centroids. Typical tests to assess significance here include PERMANOVA \citep{PERMANOVA} and $W_{d}^{*}$ \citep{wdstar}. 
\end{itemize}

When discussing Beta diversity, it is important to consider that microbiome data are compositional. This is because some common difference measures can have an irregular behavior when applied to compositional data (most notably, Euclidean distances). Fortunately, the study of compositional data has allowed for the development of tools and transformations that enable us to work with compositional data in virtually the same manner as regular data. In the case of Beta diversity, the preferred alternative to Euclidean distance is Aitchison distance \citep{RN47}. One clear advantage of Aitchison distance, which applies to the unweighted PhILR distance too, is that - unlike alpha diversity and other beta diversities - it is unaffected by PCR bias \citep{McLarenBias}.
We note here that there is an interesting parallel between compositional data and the probability vectors routinely studied in information theory. Similar to how Shannon entropy can be used to measure alpha diversity, other informatic metrics like Kullback-Leibler divergence could feasibly be used to measure beta diversity \citep{RN69}. Although these metrics are not commonplace in microbiome analysis, they are often used in machine learning, and may be more robust for the analysis of amalgamated data, for example genus-level or family-level data \citep{RN70}.

\end{document}


\maketitle

\hypertarget{introduction}{%
\section{0. Introduction}\label{introduction}}

Here, we will demonstrate how a microbiome analysis may look in
practice. For this demonstration, we have adapted some shotgun
metagenomic data from the \texttt{curatedMetagenomicData} library in R.
We're looking at a human cohort starring in the \emph{Metagenome-wide
association of gut microbiome features for schizophrenia} study (DOI:
10.1038/s41467-020-15457-9). After downloading the data it was
simplified by summing together all strains by genus. This will make it
easier to analyse without access to a server. For the set of operations
used to pre-process, please see section
\href{https://github.com/thomazbastiaanssen/Tjazi/blob/master/guidebook_sup/part2/README_part2.md\#gathering-and-preparing-our-data-1}{Download
and pre-process microbiome data} in the supplementary materials of part
II of this perspective piece. Briefly, in this data set, we have WGS
data from faecal samples from both patients with schizophrenia and
healthy volunteers, which will be referred to as ``healthy'' in the
Legends. This data has been included in the \texttt{Tjazi} library on
github for easy access purposes. Notably, we'll be using smoking status
and sex to demonstrate including a covariate in the analysis. All R code
used to transform, wrangle (reorganise) and plot the data is also shown
below as to hopefully provide a toolkit for aspiring and veteran
bioinformaticians alike. It should be noted that the analysis performed
here may not perfectly correspond to the one performed in the original
2020 manuscript, nor does the outcome (though they do generally agree in
that there is an effect of schizophrenia in the microbiome). This is
expected and indeed very common for microbiome studies. It is the result
of using a different statistical paradigm and should in no way discredit
the original analysis.

\hypertarget{code-chunk-load-our-libraries}{%
\subsubsection{Code chunk: Load our
libraries}\label{code-chunk-load-our-libraries}}

\begin{Shaded}
\begin{Highlighting}[]
\CommentTok{\#Statistical tools        Primarily PERMANOVA, alpha diversity and the CLR transformation.}
\FunctionTok{library}\NormalTok{(vegan)            }\CommentTok{\#install.packages("vegan")}
\FunctionTok{library}\NormalTok{(iNEXT)            }\CommentTok{\#install.packages("iNEXT")}
\FunctionTok{library}\NormalTok{(Tjazi)            }\CommentTok{\#devtools::install\_github("thomazbastiaanssen/Tjazi")}

\CommentTok{\#Data Wrangling}
\FunctionTok{library}\NormalTok{(tidyverse)        }\CommentTok{\#install.packages("tidyverse")}
\FunctionTok{library}\NormalTok{(knitr)            }\CommentTok{\#install.packages("knitr")}
\FunctionTok{library}\NormalTok{(waldo)            }\CommentTok{\#install.packages("waldo")}

\CommentTok{\#Plotting}
\FunctionTok{library}\NormalTok{(ggplot2)          }\CommentTok{\#install.packages("ggplot2")}
\FunctionTok{library}\NormalTok{(ggforce)          }\CommentTok{\#install.packages("ggforce")}
\FunctionTok{library}\NormalTok{(patchwork)        }\CommentTok{\#install.packages("patchwork")}
\FunctionTok{library}\NormalTok{(ggbeeswarm)       }\CommentTok{\#install.packages("ggbeeswarm")}
\FunctionTok{library}\NormalTok{(metafolio)        }\CommentTok{\#install.packages("metafolio")}

\CommentTok{\#Load prepared data from the schizophrenia study stored in the Tjazi library}
\FunctionTok{data}\NormalTok{(guidebook\_data)}
\end{Highlighting}
\end{Shaded}

\newpage

\hypertarget{code-chunk-load-our-count-table-and-perform-the-clr-transformation}{%
\subsubsection{Code chunk: Load our count table and perform the
CLR-transformation}\label{code-chunk-load-our-count-table-and-perform-the-clr-transformation}}

\begin{Shaded}
\begin{Highlighting}[]
\CommentTok{\#Disable strings automatically being read in as factors to avoid unintuitive behaviour.}
\FunctionTok{options}\NormalTok{(}\AttributeTok{stringsAsFactors =}\NormalTok{ F)}

\CommentTok{\#Set a seed for the purposes of reproducibility in this document.}
\FunctionTok{set.seed}\NormalTok{(}\DecValTok{1}\NormalTok{)}

\CommentTok{\#Load in the genus level count table and the metadata file. }
\CommentTok{\#Since we\textquotesingle{}re using prepared data, we already loaded it using \textasciigrave{}data(guidebook\_data)\textasciigrave{}, }
\CommentTok{\#but typically we\textquotesingle{}d do something like this:}
\CommentTok{\#}
\CommentTok{\#counts \textless{}{-} read.delim("genus\_level\_counts.csv", sep = ",", row.names = 1, header = T)}
\NormalTok{counts   }\OtherTok{\textless{}{-}}\NormalTok{ counts ; metadata }\OtherTok{\textless{}{-}}\NormalTok{ metadata}

\CommentTok{\#To be safe, let\textquotesingle{}s check whether our metadata and our count table have the same names.}
\FunctionTok{print}\NormalTok{(waldo}\SpecialCharTok{::}\FunctionTok{compare}\NormalTok{(}\FunctionTok{sort}\NormalTok{(metadata}\SpecialCharTok{$}\NormalTok{master\_ID), }\FunctionTok{sort}\NormalTok{(}\FunctionTok{colnames}\NormalTok{(counts)), }\AttributeTok{max\_diffs =} \DecValTok{5}\NormalTok{))}
\end{Highlighting}
\end{Shaded}

\begin{verbatim}
##     old               | new                               
## [1] "wHAXPI032581-18" - "wHAXPI032581.18" [1]             
## [2] "wHAXPI032582-19" - "wHAXPI032582.19" [2]             
## [3] "wHAXPI032583-21" - "wHAXPI032583.21" [3]             
## [4] "wHAXPI032584-22" - "wHAXPI032584.22" [4]             
## [5] "wHAXPI032585-23" - "wHAXPI032585.23" [5]             
## ... ...                 ...               and 166 more ...
\end{verbatim}

\begin{Shaded}
\begin{Highlighting}[]
\CommentTok{\#Looks like the metadata names contain dashes whereas the count table contains points. }
\CommentTok{\#We\textquotesingle{}ll change the dashes into dots in the metadata file. }
\NormalTok{metadata}\SpecialCharTok{$}\NormalTok{master\_ID }\OtherTok{\textless{}{-}} \FunctionTok{gsub}\NormalTok{(metadata}\SpecialCharTok{$}\NormalTok{master\_ID, }\AttributeTok{pattern =} \StringTok{"{-}"}\NormalTok{, }\AttributeTok{replacement =} \StringTok{"."}\NormalTok{)}

\CommentTok{\#Reorder the columns based on the metadata.}
\NormalTok{counts  }\OtherTok{\textless{}{-}}\NormalTok{ counts[,metadata}\SpecialCharTok{$}\NormalTok{master\_ID]}

\CommentTok{\#Fork off your count data so that you always have an untouched version handy.}
\NormalTok{genus   }\OtherTok{\textless{}{-}}\NormalTok{ counts}

\CommentTok{\#make sure our count data is all numbers}
\NormalTok{genus   }\OtherTok{\textless{}{-}} \FunctionTok{apply}\NormalTok{(genus,}\FunctionTok{c}\NormalTok{(}\DecValTok{1}\NormalTok{,}\DecValTok{2}\NormalTok{),}\ControlFlowTok{function}\NormalTok{(x) }\FunctionTok{as.numeric}\NormalTok{(}\FunctionTok{as.character}\NormalTok{(x)))}

\CommentTok{\#Remove features with prevalence \textless{} 10\% in two steps:}
\CommentTok{\#First, determine how often every feature is absent in a sample}
\NormalTok{n\_zeroes }\OtherTok{\textless{}{-}} \FunctionTok{rowSums}\NormalTok{(genus }\SpecialCharTok{==} \DecValTok{0}\NormalTok{)}

\CommentTok{\#Then, remove features that are absent in more than your threshold (90\% in this case).}
\NormalTok{genus    }\OtherTok{\textless{}{-}}\NormalTok{ genus[n\_zeroes }\SpecialCharTok{\textless{}=} \FunctionTok{round}\NormalTok{(}\FunctionTok{ncol}\NormalTok{(genus) }\SpecialCharTok{*} \FloatTok{0.90}\NormalTok{),]}
 
\CommentTok{\#Perform a CLR transformation {-} \#We are imputing zeroes using the \textquotesingle{}const\textquotesingle{} method }
\CommentTok{\#Essentially, we replace zeroes with 65\% of the next lowest value {-} see Lubbe et al 2021. }
\NormalTok{genus.exp }\OtherTok{\textless{}{-}} \FunctionTok{clr\_c}\NormalTok{(genus)}
\end{Highlighting}
\end{Shaded}

\begin{center}\rule{0.5\linewidth}{0.5pt}\end{center}

\newpage

\hypertarget{intermezzo-interpreting-centered-log-ratio-transformed-data}{%
\section{Intermezzo: Interpreting Centered Log-Ratio Transformed
Data}\label{intermezzo-interpreting-centered-log-ratio-transformed-data}}

The centered log-ratio (CLR) transformation may be the most common
approach to deal with compositional data, such as microbiome sequencing
data. We will not go into the reasons why this transformation is used
here - see the main text - but we will provide some material to help
form intuition on what the CLR transformation does and how to interpret
it. Let's start by taking a look at the mathematical notation.

Let's say we have microbiome sample which we will treat as a vector
called \(\bf{x}\) with size \(D\). We'll refer to the taxa - or more
generally the elements - of this vector \(\bf{x}\) as \({x}_1\) -
\({x}_D\). Then, CLR-transforming that vector \(\bf{x}\) would look like
this:

\[clr({\mathbf{x}}) = \left \lbrace \ln \left (\frac{{x}_{1}}{G({\mathbf{x}})} \right), \dots, \ln \left (\frac{{x}_{D}}{G({\mathbf{x}})} \right) \right \rbrace\]

Where \({G({\bf x})}\) is the geometric mean of \(\bf{x}\). Let's go
through it step by step.

You can calculate the geometric mean of a set of \emph{n} numbers by
multiplying them together and then taking the
\emph{n}\textsuperscript{th} root. Just like the `regular' mean, the
geometric mean says something about the center of your data.

Essentially what this says is that in order to get the CLR-transformed
values of a vector, you take every element of that vector, divide it by
the geometric mean of the entire vector and then take the natural
logarithm of the result and you're done.

We can deduce a few things about this transformation.

\begin{itemize}
\tightlist
\item
  First, since we're taking a natural logarithm,
  \(\frac{x_{n}}{G({\bf x})}\) can never be zero as the logarithm of
  zero is undefined. This means that we need to either replace or remove
  every zero in our data before we use this transformation. We expand on
  strategies for this in the main text.
\item
  Second, the possible range of our data has changed. Regular counts can
  go from 0 to infinity and relative abundance data can go from 0 to 1,
  but CLR-transformed data can go from negative infinity to positive
  infinity. The logarithm of a very small number divided by a very large
  number will be very negative.
\item
  Third, if \(x_{n}\) is exactly the same as the geometric mean
  \(G({\bf x})\), \(\frac{x_{n}}{G({\bf x})}\) will be 1 and thus
  \(clr(x_{n})\) will be 0 as the logarithm of 1 is equal to 0. This
  gives us some intuition about the size of CLR-transformed values.
  Going further on this, it means that an increase of 1 on a
  CLR-transformed scale corresponds to multiplying with \emph{e},
  Euler's number, which is approximately equal to 2.718282. Conversely,
  a decrease of 1 on a CLR-transformed scale corresponds to dividing by
  \emph{e}.
\end{itemize}

Furthermore there are a few points to keep in mind when interpreting
CLR-transformed values.

\begin{itemize}
\tightlist
\item
  First, the CLR-transformation is especially useful in the scenario
  where most features do not change, so that the geometric mean remains
  reasonably stable between your samples. If the geometric mean is very
  different between your samples, you're dividing by very different
  values between your samples.
\item
  Second, especially for microbiome sequencing experiments, we are
  usually dealing with how many reads we found for any given organism.
  Typically, we cannot relate this back to the absolute or even the
  relative abundances of those organisms, as all microbes have their own
  \emph{microbe-to-reads} conversion rate (again see the main text).
  Even so, the ratios between the reads are still highly informative.
\end{itemize}

The CLR-transformation is not a \emph{perfect solution} for
compositionality - in fact the idea of a solution to a type of data
seems a little odd - but in practice the CLR-transformation tends to be
a handy tool on the belt of a bioinformatician. Understanding what
exactly it does will greatly improve its utility and reduce the chance
of misinterpreting an analyisis.

\begin{center}\rule{0.5\linewidth}{0.5pt}\end{center}

\newpage

\hypertarget{stacked-barplots}{%
\section{1. Stacked Barplots}\label{stacked-barplots}}

Stacked barplots provide a visually appealing overview of the
composition of each sample. Normally, no tests are performed here, but
they can be helpful to give the data a visual check over and to show
obvious shifts in composition between groups. They could be seen as a
mix of alpha and beta diversity, as you can look at both the composition
of a single sample (alpha) and how much the samples differ from each
other (beta).

\hypertarget{code-chunk-generating-a-stacked-barplot-form-a-count-table}{%
\subsubsection{Code chunk: Generating a stacked barplot form a count
table}\label{code-chunk-generating-a-stacked-barplot-form-a-count-table}}

\begin{Shaded}
\begin{Highlighting}[]
\CommentTok{\#Fork off form the untransformed counts table}
\NormalTok{bargenus   }\OtherTok{\textless{}{-}}\NormalTok{ counts}

\CommentTok{\#Make into relative abundance}
\NormalTok{bargenus   }\OtherTok{\textless{}{-}} \FunctionTok{apply}\NormalTok{(bargenus, }\DecValTok{2}\NormalTok{, }\ControlFlowTok{function}\NormalTok{(i) i}\SpecialCharTok{/}\FunctionTok{sum}\NormalTok{(i)) }

\CommentTok{\#Define a cutoff for rare taxa in several steps:}
\CommentTok{\#first, determine the max \% abundance every feature ever shows up at }
\NormalTok{maxabundances }\OtherTok{\textless{}{-}} \FunctionTok{apply}\NormalTok{(bargenus, }\DecValTok{1}\NormalTok{, max)}

\CommentTok{\#Meanwhile, transpose the count table for future wrangling.}
\NormalTok{bargenus      }\OtherTok{\textless{}{-}} \FunctionTok{data.frame}\NormalTok{(}\FunctionTok{t}\NormalTok{(bargenus))}

\CommentTok{\#For every sample, sum up all rare taxa ( \textless{} 1\% at their highest in this case)}
\NormalTok{bargenus}\SpecialCharTok{$}\StringTok{\textasciigrave{}}\AttributeTok{Rare Taxa}\StringTok{\textasciigrave{}} \OtherTok{\textless{}{-}} \FunctionTok{rowSums}\NormalTok{(bargenus[,maxabundances }\SpecialCharTok{\textless{}} \FloatTok{0.01}\NormalTok{], }\AttributeTok{na.rm =} \ConstantTok{TRUE}\NormalTok{)}

\CommentTok{\#Remove the individual rare taxa now that they\textquotesingle{}re summed up}
\NormalTok{bargenus }\OtherTok{=}\NormalTok{ bargenus[,}\FunctionTok{c}\NormalTok{(maxabundances }\SpecialCharTok{\textgreater{}} \FloatTok{0.01}\NormalTok{, T) ] }\CommentTok{\#\textasciigrave{}T\textasciigrave{} to include the \textasciigrave{}Rare Taxa\textasciigrave{}.  }

\CommentTok{\#Prepare the data for ggplot by adding in metadata here}
\NormalTok{bargenus}\SpecialCharTok{$}\NormalTok{Group       }\OtherTok{=}\NormalTok{ metadata}\SpecialCharTok{$}\NormalTok{Group}
\NormalTok{bargenus}\SpecialCharTok{$}\NormalTok{ID          }\OtherTok{=}\NormalTok{ metadata}\SpecialCharTok{$}\NormalTok{master\_ID}

\CommentTok{\#Wrangle the data to long format for easy plotting}
\NormalTok{barlong }\OtherTok{=}\NormalTok{ bargenus }\SpecialCharTok{\%\textgreater{}\%} 
  \FunctionTok{pivot\_longer}\NormalTok{(}\SpecialCharTok{!}\FunctionTok{c}\NormalTok{(ID, Group), }\AttributeTok{names\_to =} \FunctionTok{c}\NormalTok{(}\StringTok{"Microbe"}\NormalTok{), }\AttributeTok{values\_to =} \StringTok{"value"}\NormalTok{) }\SpecialCharTok{\%\textgreater{}\%}
  \FunctionTok{mutate}\NormalTok{(}\AttributeTok{Microbe =} \FunctionTok{str\_replace}\NormalTok{(Microbe, }\StringTok{".*\_or\_"}\NormalTok{, }\StringTok{""}\NormalTok{))}

\CommentTok{\#Change the colour for the rare taxa to gray to make them stand out}
\NormalTok{cols }\OtherTok{=}\NormalTok{ metafolio}\SpecialCharTok{::}\FunctionTok{gg\_color\_hue}\NormalTok{(}\FunctionTok{length}\NormalTok{(}\FunctionTok{unique}\NormalTok{(barlong}\SpecialCharTok{$}\NormalTok{Microbe)))}
\NormalTok{cols[}\FunctionTok{unique}\NormalTok{(barlong}\SpecialCharTok{$}\NormalTok{Microbe)}\SpecialCharTok{==}\StringTok{"Rare Taxa"}\NormalTok{]}\OtherTok{=}\StringTok{"dark gray"}

\CommentTok{\#Create the stacked barplots using ggplot2}
\NormalTok{barlong }\SpecialCharTok{\%\textgreater{}\%}
  \FunctionTok{ggplot}\NormalTok{(}\FunctionTok{aes}\NormalTok{(}\AttributeTok{x =}\NormalTok{ ID, }\AttributeTok{y =}\NormalTok{ value, }\AttributeTok{fill =}\NormalTok{ Microbe)) }\SpecialCharTok{+} 
  \FunctionTok{geom\_bar}\NormalTok{(}\AttributeTok{stat =} \StringTok{"identity"}\NormalTok{, }\AttributeTok{col =} \StringTok{"black"}\NormalTok{, }\AttributeTok{linewidth =}\NormalTok{ .}\DecValTok{2}\NormalTok{, }\AttributeTok{width =} \DecValTok{1}\NormalTok{) }\SpecialCharTok{+} 
  \FunctionTok{facet\_row}\NormalTok{(}\SpecialCharTok{\textasciitilde{}}\NormalTok{Group, }\AttributeTok{scales =} \StringTok{"free\_x"}\NormalTok{) }\SpecialCharTok{+}
  \CommentTok{\#Adjust layout and appearance}
  \FunctionTok{scale\_fill\_manual}\NormalTok{(}\AttributeTok{values =}\NormalTok{ cols, }\AttributeTok{labels =} \FunctionTok{unique}\NormalTok{(}\FunctionTok{sub}\NormalTok{(}\StringTok{".*ales\_"}\NormalTok{, }\StringTok{""}\NormalTok{, barlong}\SpecialCharTok{$}\NormalTok{Microbe))) }\SpecialCharTok{+} 
  \FunctionTok{scale\_y\_continuous}\NormalTok{(}\AttributeTok{expand =} \FunctionTok{c}\NormalTok{(}\DecValTok{0}\NormalTok{, }\DecValTok{0}\NormalTok{)) }\SpecialCharTok{+}
  \FunctionTok{guides}\NormalTok{(}\AttributeTok{fill =} \FunctionTok{guide\_legend}\NormalTok{(}\AttributeTok{ncol =} \DecValTok{1}\NormalTok{, }\AttributeTok{keyheight =} \DecValTok{1}\NormalTok{, }\AttributeTok{title =} \StringTok{"Legend"}\NormalTok{)) }\SpecialCharTok{+} 
  \FunctionTok{theme\_bw}\NormalTok{() }\SpecialCharTok{+} \FunctionTok{xlab}\NormalTok{(}\StringTok{""}\NormalTok{) }\SpecialCharTok{+}  \FunctionTok{ylab}\NormalTok{(}\StringTok{"Proportion"}\NormalTok{) }\SpecialCharTok{+} 
  \FunctionTok{theme}\NormalTok{(}\AttributeTok{text =} \FunctionTok{element\_text}\NormalTok{(}\AttributeTok{size =} \DecValTok{14}\NormalTok{), }\AttributeTok{axis.text.x =} \FunctionTok{element\_blank}\NormalTok{())}
\end{Highlighting}
\end{Shaded}

\includegraphics{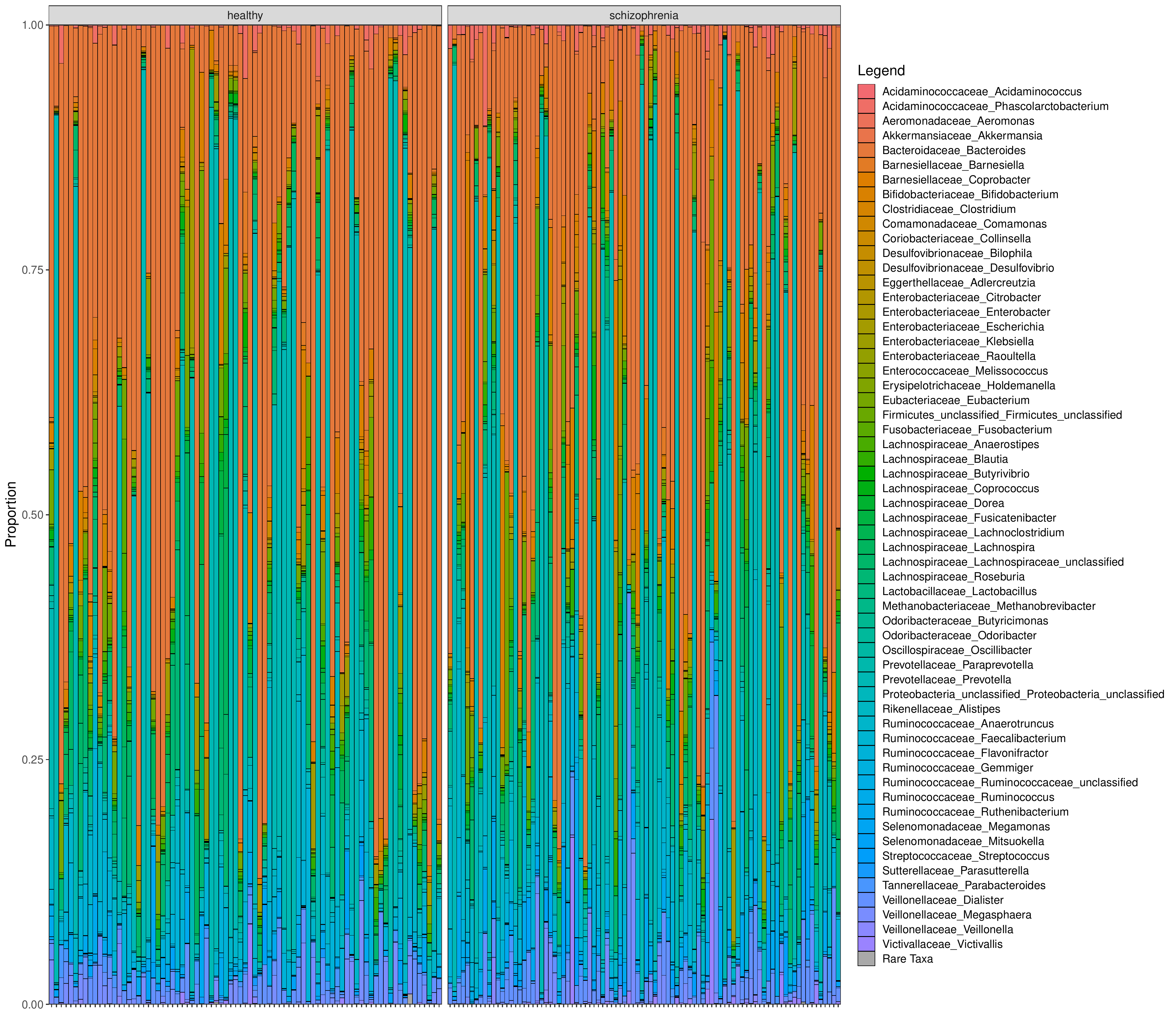}
Stacked barplots are helpful because they allow us to \emph{eyeball} the
data, giving us an idea of the experimental setup in the case of more
complex designs. They also allow us to get a general sense of the
effects we may expect and about the general levels of variance within
and between groups. In this case, nothing in particular stands out.
These samples look like they could have come from human microbiome
sequencing, which is exactly what we want!

\begin{center}\rule{0.5\linewidth}{0.5pt}\end{center}

\newpage

\hypertarget{alpha-diversity}{%
\section{2. Alpha Diversity}\label{alpha-diversity}}

Another staple in microbiome research is alpha diversity. In a nutshell,
alpha diversty is a set of measures that comment on how diverse,
complicated and/or rich a single sample is. The three most common
metrics for alpha diversity in microbiome research are Chao1, the
Simpson Index and Shannon Entropy.

\hypertarget{code-chunk-computing-and-plotting-alpha-diversity-from-a-count-table}{%
\subsubsection{Code chunk: Computing and plotting Alpha diversity from a
count
table}\label{code-chunk-computing-and-plotting-alpha-diversity-from-a-count-table}}

\begin{Shaded}
\begin{Highlighting}[]
\CommentTok{\#It is important to use the untouched count table here as we\textquotesingle{}re interested in rare taxa.}

\CommentTok{\#Compute alpha diversity using a wrapper around the iNEXT library, }
\CommentTok{\#which implements automatic rarefaction curves.}
\CommentTok{\#This step can take some time. }

\NormalTok{alpha\_diversity }\OtherTok{=} \FunctionTok{get\_asymptotic\_alpha}\NormalTok{(}\AttributeTok{species =}\NormalTok{ counts, }\AttributeTok{verbose =} \ConstantTok{FALSE}\NormalTok{) }

\CommentTok{\#Add metadata for plotting and stats. Make sure the count table and metadata match up!}
\NormalTok{alpha\_diversity}\SpecialCharTok{$}\NormalTok{Legend }\OtherTok{=}\NormalTok{ metadata}\SpecialCharTok{$}\NormalTok{Group}
\NormalTok{alpha\_diversity}\SpecialCharTok{$}\NormalTok{Sex    }\OtherTok{=}\NormalTok{ metadata}\SpecialCharTok{$}\NormalTok{Sex}
\NormalTok{alpha\_diversity}\SpecialCharTok{$}\NormalTok{Smoker }\OtherTok{=}\NormalTok{ metadata}\SpecialCharTok{$}\NormalTok{Smoker}


\CommentTok{\#Plot alpha diversity all at once using pipes}
\NormalTok{alpha\_diversity }\SpecialCharTok{\%\textgreater{}\%}

  \CommentTok{\#Wrangle the data to long format for easy plotting}
  \FunctionTok{pivot\_longer}\NormalTok{(}\SpecialCharTok{!}\FunctionTok{c}\NormalTok{(Legend, Sex, Smoker)) }\SpecialCharTok{\%\textgreater{}\%}

  \CommentTok{\#Pipe it all directly into ggplot2}
  \FunctionTok{ggplot}\NormalTok{(}\FunctionTok{aes}\NormalTok{(}\AttributeTok{x      =}\NormalTok{ Legend,}
             \AttributeTok{y      =}\NormalTok{ value, }
             \AttributeTok{fill   =}\NormalTok{ Legend, }
             \AttributeTok{shape  =}\NormalTok{ Sex, }
             \AttributeTok{group  =} \FunctionTok{interaction}\NormalTok{(Legend, Sex))) }\SpecialCharTok{+} 
  
  \CommentTok{\#Let\textquotesingle{}s use position\_dodge() to visually separate males and females by group}
  \FunctionTok{geom\_boxplot}\NormalTok{(}\AttributeTok{alpha =} \DecValTok{1}\SpecialCharTok{/}\DecValTok{2}\NormalTok{, }\AttributeTok{coef =} \DecValTok{100}\NormalTok{, }\AttributeTok{position =} \FunctionTok{position\_dodge}\NormalTok{(}\DecValTok{1}\NormalTok{)) }\SpecialCharTok{+} 
  \FunctionTok{geom\_beeswarm}\NormalTok{(}\AttributeTok{size =} \DecValTok{2}\NormalTok{, }\AttributeTok{cex =} \DecValTok{3}\NormalTok{, }\AttributeTok{dodge.width =} \DecValTok{1}\NormalTok{) }\SpecialCharTok{+} 
  \FunctionTok{facet\_wrap}\NormalTok{(}\SpecialCharTok{\textasciitilde{}}\NormalTok{name, }\AttributeTok{scales =} \StringTok{"free"}\NormalTok{) }\SpecialCharTok{+} \FunctionTok{theme\_bw}\NormalTok{()  }\SpecialCharTok{+}
  \FunctionTok{scale\_fill\_manual}\NormalTok{(}\AttributeTok{values =} \FunctionTok{c}\NormalTok{(}\StringTok{"healthy"}        \OtherTok{=} \StringTok{"\#fe9929"}\NormalTok{, }
                               \StringTok{"schizophrenia"}  \OtherTok{=} \StringTok{"\#8c6bb1"}\NormalTok{)) }\SpecialCharTok{+} 
  \FunctionTok{scale\_shape\_manual}\NormalTok{(}\AttributeTok{values =} \FunctionTok{c}\NormalTok{(}\StringTok{"female"} \OtherTok{=} \DecValTok{21}\NormalTok{, }
                                \StringTok{"male"}   \OtherTok{=} \DecValTok{22}\NormalTok{)) }\SpecialCharTok{+}  
  \FunctionTok{ylab}\NormalTok{(}\StringTok{""}\NormalTok{) }\SpecialCharTok{+} \FunctionTok{xlab}\NormalTok{(}\StringTok{""}\NormalTok{) }
\end{Highlighting}
\end{Shaded}

\includegraphics{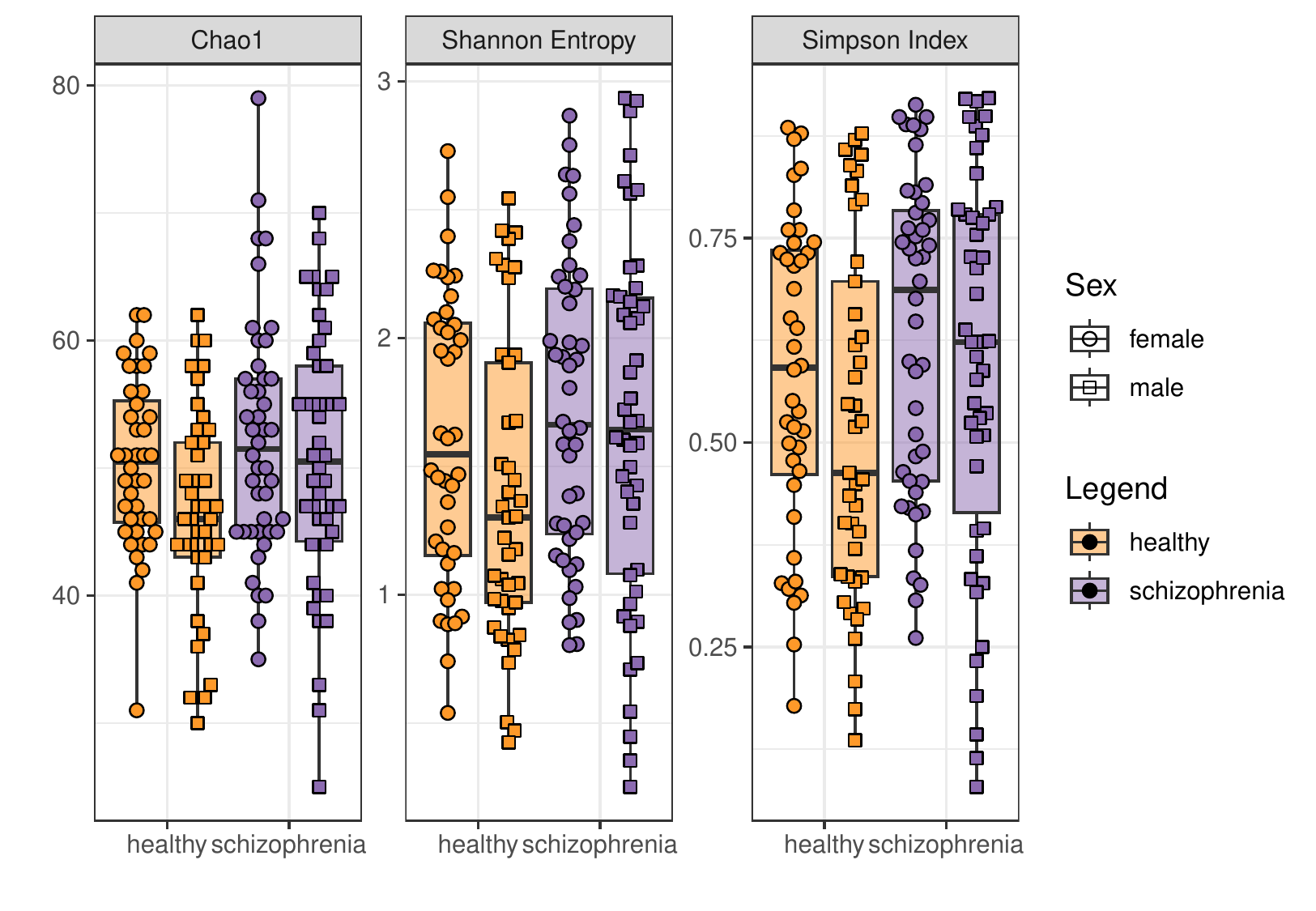}

When eyeballing these figures, we might be able to spot an effect of
both sex and group. Note how in all three alpha diversity metrics the
female healthy volunteers seem to have a higher diversity than the
males. Nevertheless, after taking this sex effect into account, there is
still seems to be an effect of group. Rather than relying on eyeballing
though, statistical testing will help us determine what to say about
these results.

\hypertarget{testing-for-differences-in-alpha-diversity}{%
\subsection{2.1 Testing for differences in Alpha
diversity}\label{testing-for-differences-in-alpha-diversity}}

Usually, one would use standard statistical tests such as t-tests and
ANOVAs to assess differences in alpha diversity metrics. However, since
we expect that smoking status and sex will also influence the
microbiome, we will here opt for a linear model approach to try and
account for the effect of sex and smoking status. We will generate
tables based on the fitting of these models. When interpreting these
tables, we are typically mainly interested in the \textbf{Estimate}
column, which is an estimate of the \(\beta\) (beta), and the
\textbf{Pr(\textgreater\textbar t\textbar)} column, which essentially
depicts the p-values. In this case, the beta of a group can be readily
interpreted as the difference between the means between the respective
groups. Since we'll be estimating a 95\% confidence interval as well,
We'll also get two columns with the 2.5\% and 97.5\% points of the
\(\beta\) estimate. The top row, containing the \textbf{(Intercept)}
gives an estimation of the overall mean of the data and can usually be
ignored altogether, don't get too excited if this has a low p-value.

\hypertarget{reporting-statistical-models}{%
\subsection{2.2 Reporting statistical
models}\label{reporting-statistical-models}}

According to the STROBE guidelines for observational studies, we should
report both the unadjusted model and the adjusted model, it is best
practice to do so. We will do so here in the alpha diversity section,
but for the sake of brevity we will skip this step for beta diversity
and differential abundance. In microbiome studies we typically assess a
very large number of features. Comprehensive statistical tables for
feature-wise tests will often find a home in the supplementary files.

\newpage

\hypertarget{code-chunk-testing-for-differences-in-alpha-diversity-chao1}{%
\subsubsection{Code chunk: Testing for differences in Alpha diversity:
Chao1}\label{code-chunk-testing-for-differences-in-alpha-diversity-chao1}}

\begin{Shaded}
\begin{Highlighting}[]
\CommentTok{\#Fit an unadjusted linear model}
\NormalTok{chao1\_unadj }\OtherTok{=} \FunctionTok{lm}\NormalTok{(Chao1 }\SpecialCharTok{\textasciitilde{}}\NormalTok{ Legend, alpha\_diversity)}

\CommentTok{\#Combine the summary of the model with the 95\% confidence interval of the estimates}
\NormalTok{res\_chao1\_unadj }\OtherTok{=} \FunctionTok{cbind}\NormalTok{(}\FunctionTok{coefficients}\NormalTok{(}\FunctionTok{summary}\NormalTok{(chao1\_unadj)), }\FunctionTok{confint}\NormalTok{(chao1\_unadj))}

\CommentTok{\#Plot the unadjusted results in a nice looking table}
\FunctionTok{kable}\NormalTok{(res\_chao1\_unadj, }\AttributeTok{digits =} \DecValTok{3}\NormalTok{)}
\end{Highlighting}
\end{Shaded}

\begin{longtable}[]{@{}
  >{\raggedright\arraybackslash}p{(\columnwidth - 12\tabcolsep) * \real{0.2469}}
  >{\raggedleft\arraybackslash}p{(\columnwidth - 12\tabcolsep) * \real{0.1111}}
  >{\raggedleft\arraybackslash}p{(\columnwidth - 12\tabcolsep) * \real{0.1358}}
  >{\raggedleft\arraybackslash}p{(\columnwidth - 12\tabcolsep) * \real{0.0988}}
  >{\raggedleft\arraybackslash}p{(\columnwidth - 12\tabcolsep) * \real{0.2346}}
  >{\raggedleft\arraybackslash}p{(\columnwidth - 12\tabcolsep) * \real{0.0864}}
  >{\raggedleft\arraybackslash}p{(\columnwidth - 12\tabcolsep) * \real{0.0864}}@{}}
\toprule()
\begin{minipage}[b]{\linewidth}\raggedright
\end{minipage} & \begin{minipage}[b]{\linewidth}\raggedleft
Estimate
\end{minipage} & \begin{minipage}[b]{\linewidth}\raggedleft
Std. Error
\end{minipage} & \begin{minipage}[b]{\linewidth}\raggedleft
t value
\end{minipage} & \begin{minipage}[b]{\linewidth}\raggedleft
Pr(\textgreater\textbar t\textbar)
\end{minipage} & \begin{minipage}[b]{\linewidth}\raggedleft
2.5 \%
\end{minipage} & \begin{minipage}[b]{\linewidth}\raggedleft
97.5 \%
\end{minipage} \\
\midrule()
\endhead
(Intercept) & 48.383 & 0.981 & 49.298 & 0.000 & 46.445 & 50.320 \\
Legendschizophrenia & 3.084 & 1.353 & 2.280 & 0.024 & 0.413 & 5.755 \\
\bottomrule()
\end{longtable}

\begin{Shaded}
\begin{Highlighting}[]
\CommentTok{\#Fit a linear model with sex and smoking status as covariates}
\NormalTok{chao1\_adj }\OtherTok{=} \FunctionTok{lm}\NormalTok{(Chao1 }\SpecialCharTok{\textasciitilde{}}\NormalTok{ Legend }\SpecialCharTok{+}\NormalTok{ Sex }\SpecialCharTok{+}\NormalTok{ Smoker, alpha\_diversity)}

\CommentTok{\#Combine the summary of the model with the 95\% confidence interval of the estimates}
\NormalTok{res\_chao1\_adj }\OtherTok{=} \FunctionTok{cbind}\NormalTok{(}\FunctionTok{coefficients}\NormalTok{(}\FunctionTok{summary}\NormalTok{(chao1\_adj)), }\FunctionTok{confint}\NormalTok{(chao1\_adj))}

\CommentTok{\#Plot the adjusted results in a nice looking table}
\FunctionTok{kable}\NormalTok{(res\_chao1\_adj, }\AttributeTok{digits =} \DecValTok{3}\NormalTok{)}
\end{Highlighting}
\end{Shaded}

\begin{longtable}[]{@{}
  >{\raggedright\arraybackslash}p{(\columnwidth - 12\tabcolsep) * \real{0.2469}}
  >{\raggedleft\arraybackslash}p{(\columnwidth - 12\tabcolsep) * \real{0.1111}}
  >{\raggedleft\arraybackslash}p{(\columnwidth - 12\tabcolsep) * \real{0.1358}}
  >{\raggedleft\arraybackslash}p{(\columnwidth - 12\tabcolsep) * \real{0.0988}}
  >{\raggedleft\arraybackslash}p{(\columnwidth - 12\tabcolsep) * \real{0.2346}}
  >{\raggedleft\arraybackslash}p{(\columnwidth - 12\tabcolsep) * \real{0.0864}}
  >{\raggedleft\arraybackslash}p{(\columnwidth - 12\tabcolsep) * \real{0.0864}}@{}}
\toprule()
\begin{minipage}[b]{\linewidth}\raggedright
\end{minipage} & \begin{minipage}[b]{\linewidth}\raggedleft
Estimate
\end{minipage} & \begin{minipage}[b]{\linewidth}\raggedleft
Std. Error
\end{minipage} & \begin{minipage}[b]{\linewidth}\raggedleft
t value
\end{minipage} & \begin{minipage}[b]{\linewidth}\raggedleft
Pr(\textgreater\textbar t\textbar)
\end{minipage} & \begin{minipage}[b]{\linewidth}\raggedleft
2.5 \%
\end{minipage} & \begin{minipage}[b]{\linewidth}\raggedleft
97.5 \%
\end{minipage} \\
\midrule()
\endhead
(Intercept) & 49.649 & 1.190 & 41.723 & 0.000 & 47.300 & 51.999 \\
Legendschizophrenia & 3.012 & 1.350 & 2.231 & 0.027 & 0.347 & 5.677 \\
Sexmale & -3.067 & 1.520 & -2.018 & 0.045 & -6.067 & -0.066 \\
Smokeryes & 1.157 & 1.686 & 0.686 & 0.494 & -2.172 & 4.487 \\
\bottomrule()
\end{longtable}

\hypertarget{code-chunk-testing-for-differences-in-alpha-diversity-shannon-entropy}{%
\subsubsection{Code chunk: Testing for differences in Alpha diversity:
Shannon
Entropy}\label{code-chunk-testing-for-differences-in-alpha-diversity-shannon-entropy}}

\begin{Shaded}
\begin{Highlighting}[]
\CommentTok{\#Fit an unadjusted linear model}
\NormalTok{shann\_unadj }\OtherTok{=} \FunctionTok{lm}\NormalTok{(}\StringTok{\textasciigrave{}}\AttributeTok{Shannon Entropy}\StringTok{\textasciigrave{}} \SpecialCharTok{\textasciitilde{}}\NormalTok{ Legend, alpha\_diversity)}

\CommentTok{\#Combine the summary of the model with the 95\% confidence interval of the estimates}
\NormalTok{res\_shann\_unadj }\OtherTok{=} \FunctionTok{cbind}\NormalTok{(}\FunctionTok{coefficients}\NormalTok{(}\FunctionTok{summary}\NormalTok{(shann\_unadj)), }\FunctionTok{confint}\NormalTok{(shann\_unadj))}

\CommentTok{\#Plot the unadjusted results in a nice looking table}
\FunctionTok{kable}\NormalTok{(res\_shann\_unadj, }\AttributeTok{digits =} \DecValTok{3}\NormalTok{)}
\end{Highlighting}
\end{Shaded}

\begin{longtable}[]{@{}
  >{\raggedright\arraybackslash}p{(\columnwidth - 12\tabcolsep) * \real{0.2500}}
  >{\raggedleft\arraybackslash}p{(\columnwidth - 12\tabcolsep) * \real{0.1125}}
  >{\raggedleft\arraybackslash}p{(\columnwidth - 12\tabcolsep) * \real{0.1375}}
  >{\raggedleft\arraybackslash}p{(\columnwidth - 12\tabcolsep) * \real{0.1000}}
  >{\raggedleft\arraybackslash}p{(\columnwidth - 12\tabcolsep) * \real{0.2375}}
  >{\raggedleft\arraybackslash}p{(\columnwidth - 12\tabcolsep) * \real{0.0750}}
  >{\raggedleft\arraybackslash}p{(\columnwidth - 12\tabcolsep) * \real{0.0875}}@{}}
\toprule()
\begin{minipage}[b]{\linewidth}\raggedright
\end{minipage} & \begin{minipage}[b]{\linewidth}\raggedleft
Estimate
\end{minipage} & \begin{minipage}[b]{\linewidth}\raggedleft
Std. Error
\end{minipage} & \begin{minipage}[b]{\linewidth}\raggedleft
t value
\end{minipage} & \begin{minipage}[b]{\linewidth}\raggedleft
Pr(\textgreater\textbar t\textbar)
\end{minipage} & \begin{minipage}[b]{\linewidth}\raggedleft
2.5 \%
\end{minipage} & \begin{minipage}[b]{\linewidth}\raggedleft
97.5 \%
\end{minipage} \\
\midrule()
\endhead
(Intercept) & 1.498 & 0.069 & 21.728 & 0.00 & 1.362 & 1.634 \\
Legendschizophrenia & 0.188 & 0.095 & 1.974 & 0.05 & 0.000 & 0.375 \\
\bottomrule()
\end{longtable}

\begin{Shaded}
\begin{Highlighting}[]
\CommentTok{\#Fit a linear model with sex and smoking status as covariates}
\NormalTok{shann\_adj }\OtherTok{=} \FunctionTok{lm}\NormalTok{(}\StringTok{\textasciigrave{}}\AttributeTok{Shannon Entropy}\StringTok{\textasciigrave{}} \SpecialCharTok{\textasciitilde{}}\NormalTok{ Legend }\SpecialCharTok{+}\NormalTok{ Sex }\SpecialCharTok{+}\NormalTok{ Smoker, alpha\_diversity)}

\CommentTok{\#Combine the summary of the model with the 95\% confidence interval of the estimates}
\NormalTok{res\_shann\_adj }\OtherTok{=} \FunctionTok{cbind}\NormalTok{(}\FunctionTok{coefficients}\NormalTok{(}\FunctionTok{summary}\NormalTok{(shann\_adj)), }\FunctionTok{confint}\NormalTok{(shann\_adj))}

\CommentTok{\#Plot the adjusted results in a nice looking table}
\FunctionTok{kable}\NormalTok{(res\_shann\_adj, }\AttributeTok{digits =} \DecValTok{3}\NormalTok{)}
\end{Highlighting}
\end{Shaded}

\begin{longtable}[]{@{}
  >{\raggedright\arraybackslash}p{(\columnwidth - 12\tabcolsep) * \real{0.2469}}
  >{\raggedleft\arraybackslash}p{(\columnwidth - 12\tabcolsep) * \real{0.1111}}
  >{\raggedleft\arraybackslash}p{(\columnwidth - 12\tabcolsep) * \real{0.1358}}
  >{\raggedleft\arraybackslash}p{(\columnwidth - 12\tabcolsep) * \real{0.0988}}
  >{\raggedleft\arraybackslash}p{(\columnwidth - 12\tabcolsep) * \real{0.2346}}
  >{\raggedleft\arraybackslash}p{(\columnwidth - 12\tabcolsep) * \real{0.0864}}
  >{\raggedleft\arraybackslash}p{(\columnwidth - 12\tabcolsep) * \real{0.0864}}@{}}
\toprule()
\begin{minipage}[b]{\linewidth}\raggedright
\end{minipage} & \begin{minipage}[b]{\linewidth}\raggedleft
Estimate
\end{minipage} & \begin{minipage}[b]{\linewidth}\raggedleft
Std. Error
\end{minipage} & \begin{minipage}[b]{\linewidth}\raggedleft
t value
\end{minipage} & \begin{minipage}[b]{\linewidth}\raggedleft
Pr(\textgreater\textbar t\textbar)
\end{minipage} & \begin{minipage}[b]{\linewidth}\raggedleft
2.5 \%
\end{minipage} & \begin{minipage}[b]{\linewidth}\raggedleft
97.5 \%
\end{minipage} \\
\midrule()
\endhead
(Intercept) & 1.567 & 0.084 & 18.633 & 0.000 & 1.401 & 1.734 \\
Legendschizophrenia & 0.189 & 0.095 & 1.986 & 0.049 & 0.001 & 0.378 \\
Sexmale & -0.128 & 0.107 & -1.195 & 0.234 & -0.341 & 0.084 \\
Smokeryes & -0.017 & 0.119 & -0.142 & 0.887 & -0.252 & 0.218 \\
\bottomrule()
\end{longtable}

\hypertarget{code-chunk-testing-for-differences-in-alpha-diversity-simpson-index}{%
\subsubsection{Code chunk: Testing for differences in Alpha diversity:
Simpson
Index}\label{code-chunk-testing-for-differences-in-alpha-diversity-simpson-index}}

\begin{Shaded}
\begin{Highlighting}[]
\CommentTok{\#Fit an unadjusted linear model}
\NormalTok{simps\_unadj }\OtherTok{=} \FunctionTok{lm}\NormalTok{(}\StringTok{\textasciigrave{}}\AttributeTok{Simpson Index}\StringTok{\textasciigrave{}} \SpecialCharTok{\textasciitilde{}}\NormalTok{ Legend, alpha\_diversity)}

\CommentTok{\#Combine the summary of the model with the 95\% confidence interval of the estimates}
\NormalTok{res\_simps\_unadj }\OtherTok{=} \FunctionTok{cbind}\NormalTok{(}\FunctionTok{coefficients}\NormalTok{(}\FunctionTok{summary}\NormalTok{(simps\_unadj)), }\FunctionTok{confint}\NormalTok{(simps\_unadj))}

\CommentTok{\#Plot the unadjusted results in a nice looking table}
\FunctionTok{kable}\NormalTok{(res\_simps\_unadj, }\AttributeTok{digits =} \DecValTok{3}\NormalTok{)}
\end{Highlighting}
\end{Shaded}

\begin{longtable}[]{@{}
  >{\raggedright\arraybackslash}p{(\columnwidth - 12\tabcolsep) * \real{0.2469}}
  >{\raggedleft\arraybackslash}p{(\columnwidth - 12\tabcolsep) * \real{0.1111}}
  >{\raggedleft\arraybackslash}p{(\columnwidth - 12\tabcolsep) * \real{0.1358}}
  >{\raggedleft\arraybackslash}p{(\columnwidth - 12\tabcolsep) * \real{0.0988}}
  >{\raggedleft\arraybackslash}p{(\columnwidth - 12\tabcolsep) * \real{0.2346}}
  >{\raggedleft\arraybackslash}p{(\columnwidth - 12\tabcolsep) * \real{0.0864}}
  >{\raggedleft\arraybackslash}p{(\columnwidth - 12\tabcolsep) * \real{0.0864}}@{}}
\toprule()
\begin{minipage}[b]{\linewidth}\raggedright
\end{minipage} & \begin{minipage}[b]{\linewidth}\raggedleft
Estimate
\end{minipage} & \begin{minipage}[b]{\linewidth}\raggedleft
Std. Error
\end{minipage} & \begin{minipage}[b]{\linewidth}\raggedleft
t value
\end{minipage} & \begin{minipage}[b]{\linewidth}\raggedleft
Pr(\textgreater\textbar t\textbar)
\end{minipage} & \begin{minipage}[b]{\linewidth}\raggedleft
2.5 \%
\end{minipage} & \begin{minipage}[b]{\linewidth}\raggedleft
97.5 \%
\end{minipage} \\
\midrule()
\endhead
(Intercept) & 0.550 & 0.024 & 23.247 & 0.000 & 0.504 & 0.597 \\
Legendschizophrenia & 0.063 & 0.033 & 1.924 & 0.056 & -0.002 & 0.127 \\
\bottomrule()
\end{longtable}

\begin{Shaded}
\begin{Highlighting}[]
\CommentTok{\#Fit a linear model with sex and smoking status as covariates}
\NormalTok{simps\_adj }\OtherTok{=} \FunctionTok{lm}\NormalTok{(}\StringTok{\textasciigrave{}}\AttributeTok{Simpson Index}\StringTok{\textasciigrave{}} \SpecialCharTok{\textasciitilde{}}\NormalTok{ Legend }\SpecialCharTok{+}\NormalTok{ Sex }\SpecialCharTok{+}\NormalTok{ Smoker, alpha\_diversity)}

\CommentTok{\#Combine the summary of the model with the 95\% confidence interval of the estimates}
\NormalTok{res\_simps\_adj }\OtherTok{=} \FunctionTok{cbind}\NormalTok{(}\FunctionTok{coefficients}\NormalTok{(}\FunctionTok{summary}\NormalTok{(simps\_adj)), }\FunctionTok{confint}\NormalTok{(simps\_adj))}

\CommentTok{\#Plot the adjusted results in a nice looking table}
\FunctionTok{kable}\NormalTok{(res\_simps\_adj, }\AttributeTok{digits =} \DecValTok{3}\NormalTok{)}
\end{Highlighting}
\end{Shaded}

\begin{longtable}[]{@{}
  >{\raggedright\arraybackslash}p{(\columnwidth - 12\tabcolsep) * \real{0.2469}}
  >{\raggedleft\arraybackslash}p{(\columnwidth - 12\tabcolsep) * \real{0.1111}}
  >{\raggedleft\arraybackslash}p{(\columnwidth - 12\tabcolsep) * \real{0.1358}}
  >{\raggedleft\arraybackslash}p{(\columnwidth - 12\tabcolsep) * \real{0.0988}}
  >{\raggedleft\arraybackslash}p{(\columnwidth - 12\tabcolsep) * \real{0.2346}}
  >{\raggedleft\arraybackslash}p{(\columnwidth - 12\tabcolsep) * \real{0.0864}}
  >{\raggedleft\arraybackslash}p{(\columnwidth - 12\tabcolsep) * \real{0.0864}}@{}}
\toprule()
\begin{minipage}[b]{\linewidth}\raggedright
\end{minipage} & \begin{minipage}[b]{\linewidth}\raggedleft
Estimate
\end{minipage} & \begin{minipage}[b]{\linewidth}\raggedleft
Std. Error
\end{minipage} & \begin{minipage}[b]{\linewidth}\raggedleft
t value
\end{minipage} & \begin{minipage}[b]{\linewidth}\raggedleft
Pr(\textgreater\textbar t\textbar)
\end{minipage} & \begin{minipage}[b]{\linewidth}\raggedleft
2.5 \%
\end{minipage} & \begin{minipage}[b]{\linewidth}\raggedleft
97.5 \%
\end{minipage} \\
\midrule()
\endhead
(Intercept) & 0.576 & 0.029 & 19.986 & 0.000 & 0.519 & 0.633 \\
Legendschizophrenia & 0.063 & 0.033 & 1.939 & 0.054 & -0.001 & 0.128 \\
Sexmale & -0.049 & 0.037 & -1.332 & 0.185 & -0.122 & 0.024 \\
Smokeryes & -0.005 & 0.041 & -0.127 & 0.899 & -0.086 & 0.076 \\
\bottomrule()
\end{longtable}

\begin{center}\rule{0.5\linewidth}{0.5pt}\end{center}

\newpage

\hypertarget{beta-diversity}{%
\section{3. Beta Diversity}\label{beta-diversity}}

Beta diversity can be thought of as the degree of difference between two
samples. Typically, Beta diversity is depicted using a 2d Principal
Component Analysis (PCA). We'll perform the procedure and visualize the
results in a few different ways.

\hypertarget{code-chunk-performing-a-principal-component-analysis-and-plotting-beta-diversity}{%
\subsubsection{Code chunk: Performing a Principal Component Analysis and
plotting Beta
diversity}\label{code-chunk-performing-a-principal-component-analysis-and-plotting-beta-diversity}}

\begin{Shaded}
\begin{Highlighting}[]
\CommentTok{\#Apply the base R principal component analysis function on our CLR{-}transformed data.}
\NormalTok{data.a.pca  }\OtherTok{\textless{}{-}} \FunctionTok{prcomp}\NormalTok{(}\FunctionTok{t}\NormalTok{(genus.exp))}

\CommentTok{\#Extract the amount of variance the first four components explain for plotting. }
\NormalTok{pc1 }\OtherTok{\textless{}{-}} \FunctionTok{round}\NormalTok{(data.a.pca}\SpecialCharTok{$}\NormalTok{sdev[}\DecValTok{1}\NormalTok{]}\SpecialCharTok{\^{}}\DecValTok{2}\SpecialCharTok{/}\FunctionTok{sum}\NormalTok{(data.a.pca}\SpecialCharTok{$}\NormalTok{sdev}\SpecialCharTok{\^{}}\DecValTok{2}\NormalTok{),}\DecValTok{4}\NormalTok{) }\SpecialCharTok{*} \DecValTok{100}
\NormalTok{pc2 }\OtherTok{\textless{}{-}} \FunctionTok{round}\NormalTok{(data.a.pca}\SpecialCharTok{$}\NormalTok{sdev[}\DecValTok{2}\NormalTok{]}\SpecialCharTok{\^{}}\DecValTok{2}\SpecialCharTok{/}\FunctionTok{sum}\NormalTok{(data.a.pca}\SpecialCharTok{$}\NormalTok{sdev}\SpecialCharTok{\^{}}\DecValTok{2}\NormalTok{),}\DecValTok{4}\NormalTok{) }\SpecialCharTok{*} \DecValTok{100}
\NormalTok{pc3 }\OtherTok{\textless{}{-}} \FunctionTok{round}\NormalTok{(data.a.pca}\SpecialCharTok{$}\NormalTok{sdev[}\DecValTok{3}\NormalTok{]}\SpecialCharTok{\^{}}\DecValTok{2}\SpecialCharTok{/}\FunctionTok{sum}\NormalTok{(data.a.pca}\SpecialCharTok{$}\NormalTok{sdev}\SpecialCharTok{\^{}}\DecValTok{2}\NormalTok{),}\DecValTok{4}\NormalTok{) }\SpecialCharTok{*} \DecValTok{100}
\NormalTok{pc4 }\OtherTok{\textless{}{-}} \FunctionTok{round}\NormalTok{(data.a.pca}\SpecialCharTok{$}\NormalTok{sdev[}\DecValTok{4}\NormalTok{]}\SpecialCharTok{\^{}}\DecValTok{2}\SpecialCharTok{/}\FunctionTok{sum}\NormalTok{(data.a.pca}\SpecialCharTok{$}\NormalTok{sdev}\SpecialCharTok{\^{}}\DecValTok{2}\NormalTok{),}\DecValTok{4}\NormalTok{) }\SpecialCharTok{*} \DecValTok{100}

\CommentTok{\#Extract the scores for every sample for the first four components for plotting. }
\NormalTok{pca  }\OtherTok{=} \FunctionTok{data.frame}\NormalTok{(}\AttributeTok{PC1 =}\NormalTok{ data.a.pca}\SpecialCharTok{$}\NormalTok{x[,}\DecValTok{1}\NormalTok{], }
                  \AttributeTok{PC2 =}\NormalTok{ data.a.pca}\SpecialCharTok{$}\NormalTok{x[,}\DecValTok{2}\NormalTok{], }
                  \AttributeTok{PC3 =}\NormalTok{ data.a.pca}\SpecialCharTok{$}\NormalTok{x[,}\DecValTok{3}\NormalTok{], }
                  \AttributeTok{PC4 =}\NormalTok{ data.a.pca}\SpecialCharTok{$}\NormalTok{x[,}\DecValTok{4}\NormalTok{])}

\CommentTok{\#Add relevant information from the metadata}
\NormalTok{pca}\SpecialCharTok{$}\NormalTok{ID                  }\OtherTok{=}\NormalTok{ metadata}\SpecialCharTok{$}\NormalTok{master\_ID}
\NormalTok{pca}\SpecialCharTok{$}\NormalTok{Legend              }\OtherTok{=}\NormalTok{ metadata}\SpecialCharTok{$}\NormalTok{Group}
\NormalTok{pca}\SpecialCharTok{$}\NormalTok{Sex                 }\OtherTok{=}\NormalTok{ metadata}\SpecialCharTok{$}\NormalTok{Sex}
\NormalTok{pca}\SpecialCharTok{$}\NormalTok{Smoker              }\OtherTok{=}\NormalTok{ metadata}\SpecialCharTok{$}\NormalTok{Smoker}

\CommentTok{\#First, the main plot. Plot the first two components of the PCA}
\NormalTok{mainbeta  }\OtherTok{\textless{}{-}} \FunctionTok{ggplot}\NormalTok{(pca, }\FunctionTok{aes}\NormalTok{(}\AttributeTok{x       =}\NormalTok{ PC1, }
                             \AttributeTok{y       =}\NormalTok{ PC2, }
                             \AttributeTok{fill    =}\NormalTok{ Legend,}
                             \AttributeTok{colour  =}\NormalTok{ Legend,}
                             \AttributeTok{shape   =}\NormalTok{ Sex, }
                             \AttributeTok{group   =}\NormalTok{ Legend)) }\SpecialCharTok{+}  
  
  \CommentTok{\#Create the points and ellipses}
  \FunctionTok{stat\_ellipse}\NormalTok{(}\AttributeTok{geom =} \StringTok{"polygon"}\NormalTok{, }\AttributeTok{alpha =} \DecValTok{1}\SpecialCharTok{/}\DecValTok{4}\NormalTok{) }\SpecialCharTok{+}
  \FunctionTok{geom\_point}\NormalTok{(}\AttributeTok{size=}\DecValTok{3}\NormalTok{, }\AttributeTok{col =} \StringTok{"black"}\NormalTok{) }\SpecialCharTok{+} 
  
  \CommentTok{\#Adjust appearance}
  \FunctionTok{scale\_fill\_manual}\NormalTok{(}\AttributeTok{values   =} \FunctionTok{c}\NormalTok{(}\StringTok{"healthy"} \OtherTok{=} \StringTok{"\#fe9929"}\NormalTok{, }\StringTok{"schizophrenia"} \OtherTok{=} \StringTok{"\#8c6bb1"}\NormalTok{)) }\SpecialCharTok{+} 
  \FunctionTok{scale\_colour\_manual}\NormalTok{(}\AttributeTok{values =} \FunctionTok{c}\NormalTok{(}\StringTok{"healthy"} \OtherTok{=} \StringTok{"\#fe9929"}\NormalTok{, }\StringTok{"schizophrenia"} \OtherTok{=} \StringTok{"\#8c6bb1"}\NormalTok{)) }\SpecialCharTok{+} 
  \FunctionTok{scale\_shape\_manual}\NormalTok{(}\AttributeTok{values  =} \FunctionTok{c}\NormalTok{(}\StringTok{"female"} \OtherTok{=} \DecValTok{21}\NormalTok{, }\StringTok{"male"}   \OtherTok{=} \DecValTok{22}\NormalTok{), }\AttributeTok{guide =} \StringTok{"none"}\NormalTok{) }\SpecialCharTok{+}  
  \FunctionTok{guides}\NormalTok{(}\AttributeTok{fill =} \FunctionTok{guide\_legend}\NormalTok{(}\AttributeTok{override.aes =} \FunctionTok{list}\NormalTok{(}\AttributeTok{shape =} \FunctionTok{c}\NormalTok{(}\DecValTok{21}\NormalTok{)))) }\SpecialCharTok{+}

  \CommentTok{\#Adjust labels}
  \FunctionTok{ggtitle}\NormalTok{(}\StringTok{"Main"}\NormalTok{) }\SpecialCharTok{+} 
  \FunctionTok{xlab}\NormalTok{(}\FunctionTok{paste}\NormalTok{(}\StringTok{"PC1: "}\NormalTok{, pc1,  }\StringTok{"\%"}\NormalTok{, }\AttributeTok{sep=}\StringTok{""}\NormalTok{)) }\SpecialCharTok{+} 
  \FunctionTok{ylab}\NormalTok{(}\FunctionTok{paste}\NormalTok{(}\StringTok{"PC2: "}\NormalTok{, pc2,  }\StringTok{"\%"}\NormalTok{, }\AttributeTok{sep=}\StringTok{""}\NormalTok{)) }\SpecialCharTok{+} 
  \FunctionTok{theme\_bw}\NormalTok{() }

\CommentTok{\#Second, a smaller version to investigate the effect of sex. }
\CommentTok{\#Plot the first two components of the PCA}
\NormalTok{sexbeta   }\OtherTok{\textless{}{-}} \FunctionTok{ggplot}\NormalTok{(pca, }\FunctionTok{aes}\NormalTok{(}\AttributeTok{x       =}\NormalTok{ PC1, }
                             \AttributeTok{y       =}\NormalTok{ PC2, }
                             \AttributeTok{fill    =}\NormalTok{ Sex,}
                             \AttributeTok{colour  =}\NormalTok{ Sex,}
                             \AttributeTok{shape   =}\NormalTok{ Sex, }
                             \AttributeTok{group   =}\NormalTok{ Sex)) }\SpecialCharTok{+}  
  
  \CommentTok{\#Create the points}
  \FunctionTok{stat\_ellipse}\NormalTok{(}\AttributeTok{geom =} \StringTok{"polygon"}\NormalTok{, }\AttributeTok{alpha =} \DecValTok{1}\SpecialCharTok{/}\DecValTok{4}\NormalTok{) }\SpecialCharTok{+}
  \FunctionTok{geom\_point}\NormalTok{(}\AttributeTok{size=}\DecValTok{2}\NormalTok{, }\AttributeTok{col =} \StringTok{"black"}\NormalTok{) }\SpecialCharTok{+} 
  
  \CommentTok{\#Adjust appearance}
  \FunctionTok{scale\_fill\_manual}\NormalTok{(}\AttributeTok{values   =} \FunctionTok{c}\NormalTok{(}\StringTok{"female"} \OtherTok{=} \StringTok{"\#fb9a99"}\NormalTok{, }\StringTok{"male"} \OtherTok{=} \StringTok{"\#a6cee3"}\NormalTok{)) }\SpecialCharTok{+} 
  \FunctionTok{scale\_colour\_manual}\NormalTok{(}\AttributeTok{values =} \FunctionTok{c}\NormalTok{(}\StringTok{"female"} \OtherTok{=} \StringTok{"\#fb9a99"}\NormalTok{, }\StringTok{"male"} \OtherTok{=} \StringTok{"\#a6cee3"}\NormalTok{)) }\SpecialCharTok{+}
  \FunctionTok{scale\_shape\_manual}\NormalTok{(}\AttributeTok{values  =} \FunctionTok{c}\NormalTok{(}\StringTok{"female"} \OtherTok{=} \DecValTok{21}\NormalTok{, }\StringTok{"male"}   \OtherTok{=} \DecValTok{22}\NormalTok{)) }\SpecialCharTok{+}  
  
  \CommentTok{\#Adjust labels}
  \FunctionTok{ggtitle}\NormalTok{(}\StringTok{"Sex"}\NormalTok{) }\SpecialCharTok{+} \FunctionTok{xlab}\NormalTok{(}\StringTok{""}\NormalTok{) }\SpecialCharTok{+} \FunctionTok{ylab}\NormalTok{(}\StringTok{""}\NormalTok{) }\SpecialCharTok{+} \FunctionTok{theme\_bw}\NormalTok{() }

\CommentTok{\#Third, a smaller version to investivate the effect of smoking.}
\CommentTok{\#Plot the first two components of the PCA}
\NormalTok{smokebeta }\OtherTok{\textless{}{-}} \FunctionTok{ggplot}\NormalTok{(pca, }\FunctionTok{aes}\NormalTok{(}\AttributeTok{x       =}\NormalTok{ PC1, }
                             \AttributeTok{y       =}\NormalTok{ PC2, }
                             \AttributeTok{fill    =}\NormalTok{ Smoker,}
                             \AttributeTok{colour  =}\NormalTok{ Smoker,}
                             \AttributeTok{shape   =}\NormalTok{ Sex, }
                             \AttributeTok{group   =}\NormalTok{ Smoker)) }\SpecialCharTok{+}  
  
  \CommentTok{\#Create the points}
  \FunctionTok{stat\_ellipse}\NormalTok{(}\AttributeTok{geom =} \StringTok{"polygon"}\NormalTok{, }\AttributeTok{alpha =} \DecValTok{1}\SpecialCharTok{/}\DecValTok{4}\NormalTok{) }\SpecialCharTok{+}
  \FunctionTok{geom\_point}\NormalTok{(}\AttributeTok{size=}\DecValTok{2}\NormalTok{, }\AttributeTok{col =} \StringTok{"black"}\NormalTok{) }\SpecialCharTok{+} 
  
  \CommentTok{\#Adjust appearance}
  \FunctionTok{scale\_fill\_manual}\NormalTok{(}\AttributeTok{values   =} \FunctionTok{c}\NormalTok{(}\StringTok{"yes"} \OtherTok{=} \StringTok{"\#ff7f00"}\NormalTok{, }\StringTok{"no"} \OtherTok{=} \StringTok{"\#33a02c"}\NormalTok{)) }\SpecialCharTok{+}
  \FunctionTok{scale\_colour\_manual}\NormalTok{(}\AttributeTok{values =} \FunctionTok{c}\NormalTok{(}\StringTok{"yes"} \OtherTok{=} \StringTok{"\#ff7f00"}\NormalTok{, }\StringTok{"no"} \OtherTok{=} \StringTok{"\#33a02c"}\NormalTok{)) }\SpecialCharTok{+}
  \FunctionTok{scale\_shape\_manual}\NormalTok{(}\AttributeTok{values  =} \FunctionTok{c}\NormalTok{(}\StringTok{"female"} \OtherTok{=} \DecValTok{21}\NormalTok{, }\StringTok{"male"}   \OtherTok{=} \DecValTok{22}\NormalTok{), }\AttributeTok{guide =} \StringTok{"none"}\NormalTok{) }\SpecialCharTok{+}   
  \FunctionTok{guides}\NormalTok{(}\AttributeTok{fill =} \FunctionTok{guide\_legend}\NormalTok{(}\AttributeTok{override.aes =} \FunctionTok{list}\NormalTok{(}\AttributeTok{shape =} \FunctionTok{c}\NormalTok{(}\DecValTok{21}\NormalTok{)))) }\SpecialCharTok{+}

  \CommentTok{\#Adjust labels}
  \FunctionTok{ggtitle}\NormalTok{(}\StringTok{"Smoker status"}\NormalTok{) }\SpecialCharTok{+} \FunctionTok{xlab}\NormalTok{(}\StringTok{""}\NormalTok{) }\SpecialCharTok{+} \FunctionTok{ylab}\NormalTok{(}\StringTok{""}\NormalTok{) }\SpecialCharTok{+} \FunctionTok{theme\_bw}\NormalTok{() }


\CommentTok{\#Use patchwork to compose the three plots}
\NormalTok{(mainbeta }\SpecialCharTok{/}\NormalTok{ (sexbeta }\SpecialCharTok{|}\NormalTok{ smokebeta)) }\SpecialCharTok{+} 
  \FunctionTok{plot\_layout}\NormalTok{(}\AttributeTok{guides =} \StringTok{"collect"}\NormalTok{, }\AttributeTok{heights =} \FunctionTok{c}\NormalTok{(}\DecValTok{3}\NormalTok{, }\DecValTok{1}\NormalTok{))}
\end{Highlighting}
\end{Shaded}

\includegraphics{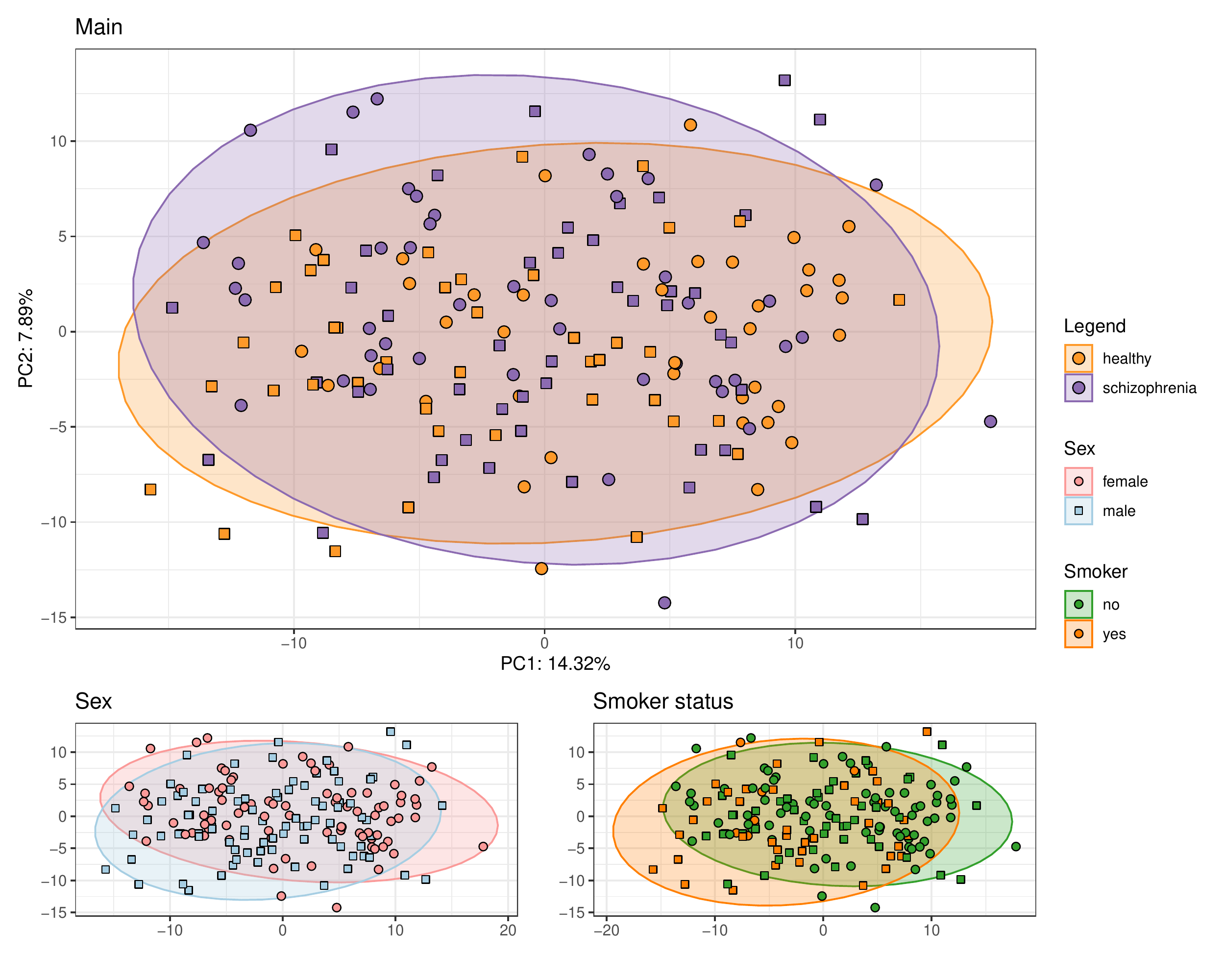}

Here we see the first two components of our Principal Component
Analysis. This type of figure is often used as a visual aid when
assessing Beta diversity. We've also made two additional smaller
versions of the same data, filled in with our two covariates; sex and
smoking status, to help get an idea of the influence of those factors on
our data.

Even though we are only looking at the first, largest, two components,
this type of figure will often be called a Beta diversity plot. In this
case, as we used CLR-transformed data as a basis, it would be based on
Aitchison distance. Interpreting a Beta diversity plot for a microbiome
study like this one can seem daunting, but will quickly become fairly
straightforward. In a nutshell, every sample is depicted as a single
point. If two points are close together, this means that the samples are
more similar to each other. We can see that group, smoking status and
sex seem to be important in explaining what's going on here.

These types of Beta diversity plots are also a useful way to detect
samples that are in some way \emph{off}. If a sample is on the far side
of the PCA, this may be reason to inspect it further. Based on the
amount of reads and alpha diversity of the sample, one may even decide
to exclude it from the analysis as it may not reliably reflect your
population of interest.

Another thing to be on the lookout for are the axis values, depicting
the percentage of variance explained. Components that either explain a
huge amount of variance or large differences between the amount of
variance explained between the first two components can be an indication
something drastic is going on, like an antibiotics effect. Typically, we
expect the sizes of the components to follow a power law. In this case,
the axes look totally reasonable. \newpage

\hypertarget{permanova}{%
\subsection{3.1 PERMANOVA}\label{permanova}}

We can use a PERMANOVA test to investigate whether the variance in the
data can be explained by the group and sex they come from. Typically,
we'd say that we use a PERMANOVA to see whether the groups are different
from each other.

It is always a good idea to consider and check the assumptions and
vulnerabilities of a statistical test. PERMANOVA, like it's univariate,
non-permutational cousin, ANOVA, make soft assumptions about the
dispersion per group (e.g.~variance, distance from a group centroid)
being equal. Like ANOVA, PERMANOVA is also reasonably robust to small
differences in variance between groups. In a simulation study, PERMANOVA
was found to be overly conservative in the case of the larger group (by
N) has a greater dispersion, whereas it is overly permissive in the case
the smaller group (by N) has a larger dispersion.

\hypertarget{code-chunk-preparing-for-permanova}{%
\subsubsection{Code chunk: Preparing for
PERMANOVA}\label{code-chunk-preparing-for-permanova}}

\begin{Shaded}
\begin{Highlighting}[]
\CommentTok{\#Display NAs as empty space in the table to improve appearance.}
\FunctionTok{options}\NormalTok{(}\AttributeTok{knitr.kable.NA =} \StringTok{""}\NormalTok{)}

\CommentTok{\#Compute euclidean distance over CLR{-}transformed values (i.e. Aitchison distance).}
\NormalTok{dis\_ait }\OtherTok{=} \FunctionTok{dist}\NormalTok{(}\FunctionTok{t}\NormalTok{(genus.exp), }\AttributeTok{method =} \StringTok{"euclidean"}\NormalTok{)}

\CommentTok{\#Use the betadisper function to assess whether the groups have a difference in variance}
\NormalTok{beta\_disp }\OtherTok{=} \FunctionTok{betadisper}\NormalTok{(dis\_ait, }\AttributeTok{group =}\NormalTok{ metadata}\SpecialCharTok{$}\NormalTok{Group)}

\CommentTok{\#Check average aitchison distance from the centroid }
\NormalTok{beta\_disp}\SpecialCharTok{$}\NormalTok{group.distances}
\end{Highlighting}
\end{Shaded}

\begin{verbatim}
##       healthy schizophrenia 
##      18.33498      19.31644
\end{verbatim}

\begin{Shaded}
\begin{Highlighting}[]
\CommentTok{\#Run an ANOVA on the difference in variance per group, plot the results in a table}
\FunctionTok{kable}\NormalTok{(}\FunctionTok{anova}\NormalTok{(beta\_disp), }\AttributeTok{digits =} \DecValTok{4}\NormalTok{)}
\end{Highlighting}
\end{Shaded}

\begin{longtable}[]{@{}lrrrrr@{}}
\toprule()
& Df & Sum Sq & Mean Sq & F value & Pr(\textgreater F) \\
\midrule()
\endhead
Groups & 1 & 41.0655 & 41.0655 & 4.0635 & 0.0454 \\
Residuals & 169 & 1707.8987 & 10.1059 & & \\
\bottomrule()
\end{longtable}

Here, we can see that while there is a significant difference in the
spreads per group, the difference is not very large: We see an average
distance to the centroid of 18.33 healthy controls vs 19 in
schizophrenia. Keeping this in mind, let's perform a PERMANOVA.

\newpage

\hypertarget{code-chunk-performing-a-permanova-test}{%
\subsubsection{Code chunk: Performing a PERMANOVA
test}\label{code-chunk-performing-a-permanova-test}}

\begin{Shaded}
\begin{Highlighting}[]
\CommentTok{\#Perform a PERMANOVA (PERmutational Multivariate ANalysis Of VAriance) test.}
\NormalTok{PERMANOVA\_res }\OtherTok{=} \FunctionTok{adonis2}\NormalTok{(dis\_ait }\SpecialCharTok{\textasciitilde{}}\NormalTok{ Group }\SpecialCharTok{+}\NormalTok{ Sex }\SpecialCharTok{+}\NormalTok{ Smoker, }
                        \AttributeTok{data =}\NormalTok{ metadata, }\AttributeTok{method =} \StringTok{"euclidean"}\NormalTok{, }\AttributeTok{permutations =} \DecValTok{1000}\NormalTok{)}

\CommentTok{\#Plot the PERMANOVA results in a nice looking table}
\FunctionTok{kable}\NormalTok{(PERMANOVA\_res, }\AttributeTok{digits =} \DecValTok{4}\NormalTok{ )}
\end{Highlighting}
\end{Shaded}

\begin{longtable}[]{@{}lrrrrr@{}}
\toprule()
& Df & SumOfSqs & R2 & F & Pr(\textgreater F) \\
\midrule()
\endhead
Group & 1 & 872.4806 & 0.0138 & 2.3900 & 0.002 \\
Sex & 1 & 721.8148 & 0.0114 & 1.9773 & 0.011 \\
Smoker & 1 & 792.3537 & 0.0125 & 2.1705 & 0.003 \\
Residual & 167 & 60963.1982 & 0.9623 & & \\
Total & 170 & 63349.8473 & 1.0000 & & \\
\bottomrule()
\end{longtable}

In general, the most interesting columns from a PERMANOVA table like
this one are \textbf{R2}, which shows the amount of variance that can be
explained by the factor in that row, and \textbf{Pr(\textgreater F)},
which can be thought of as a p-value. We can see that the group but also
the smoking status and sex factors explain enough variance that we deem
it unlikely to have happened by chance (p \textless{} 0.05). Thus, we
can say we found a group effect, a smoking effect and a sex effect. The
effect are really small though, both of them explain about 1\% of all
variance observed which isn't very much. This also tracks with our
figures, where we could see only mild differences for each factor.
Importantly, because the PERMANOVA is a permutation-based test, all test
results will likely vary slightly between runs. You could use the
\texttt{set.seed()} function to ensure consistency between runs, like we
did in chapter 0.

\begin{center}\rule{0.5\linewidth}{0.5pt}\end{center}

\newpage

\hypertarget{differential-abundance}{%
\section{4. Differential Abundance}\label{differential-abundance}}

Differential abundance testing is an integral part of microbiome
studies. here we check whether individual features, be they taxa or
functions, are present in different abundances between groups.

\hypertarget{genera}{%
\subsection{4.1 Genera}\label{genera}}

Differential abundance of taxa, in this case genera, are perhaps the
most common part of a microbiome study.

\hypertarget{code-chunk-testing-for-differentially-abundant-genera-and-plotting-the-results}{%
\subsubsection{Code chunk: Testing for differentially abundant genera
and plotting the
results}\label{code-chunk-testing-for-differentially-abundant-genera-and-plotting-the-results}}

\begin{Shaded}
\begin{Highlighting}[]
\CommentTok{\#This function fits the equivalent of lm(feature \textasciitilde{} Group + Sex + Smoker) for each feature.}
\CommentTok{\#It also performs an appropriate Benjamini{-}Hochberg correction on the p{-}values. }

\NormalTok{genus.glm }\OtherTok{=} \FunctionTok{fw\_glm}\NormalTok{(}\AttributeTok{x             =}\NormalTok{ genus.exp,}
                   \AttributeTok{f             =} \SpecialCharTok{\textasciitilde{}}\NormalTok{ Group }\SpecialCharTok{+}\NormalTok{ Sex }\SpecialCharTok{+}\NormalTok{ Smoker, }
                   \AttributeTok{metadata      =}\NormalTok{ metadata, }
                   \AttributeTok{adjust.method =} \StringTok{"BH"}\NormalTok{, }\AttributeTok{format =} \StringTok{"brief"}\NormalTok{)}
\end{Highlighting}
\end{Shaded}

\begin{verbatim}
## [1] "Using the following formula: x ~ Group + Sex + Smoker"
## [1] "Adjusting for FDR using Benjamini & Hochberg's procedure."
\end{verbatim}

\begin{Shaded}
\begin{Highlighting}[]
\CommentTok{\#(genus.glm, file = "genus.glm.csv") \#To save the results to a file. }
\end{Highlighting}
\end{Shaded}

\begin{center}\rule{0.5\linewidth}{0.5pt}\end{center}

Before we proceed with the demonstration, let's take a quick peek at the
output of this function:

\begin{Shaded}
\begin{Highlighting}[]
\FunctionTok{glimpse}\NormalTok{(genus.glm)}
\end{Highlighting}
\end{Shaded}

\begin{verbatim}
## Rows: 84
## Columns: 10
## $ feature                          <chr> "Acidaminococcaceae_Phascolarctobacte~
## $ `Groupschizophrenia Estimate`    <dbl> 0.38645474, 0.23365778, 1.72428829, 0~
## $ `Groupschizophrenia Pr(>|t|)`    <dbl> 3.290809e-01, 2.766298e-01, 4.349376e~
## $ `Sexmale Estimate`               <dbl> 0.485194237, 0.181234364, -0.32451485~
## $ `Sexmale Pr(>|t|)`               <dbl> 0.2766350, 0.4531579, 0.4837235, 0.88~
## $ `Smokeryes Estimate`             <dbl> -0.18867208, 0.65548364, 0.02917598, ~
## $ `Smokeryes Pr(>|t|)`             <dbl> 0.70254279, 0.01527056, 0.95471101, 0~
## $ `Groupschizophrenia Pr(>|t|).BH` <dbl> 0.537383182, 0.523552585, 0.003653476~
## $ `Sexmale Pr(>|t|).BH`            <dbl> 0.7524988, 0.8275057, 0.8417461, 0.96~
## $ `Smokeryes Pr(>|t|).BH`          <dbl> 0.8930102, 0.1483272, 0.9547110, 0.37~
\end{verbatim}

The output is a \texttt{data.frame} with the input features as rows and
the estimates of betas, p-values and adjusted p-values as columns. This
is all direct output from the \texttt{lm()} fits that were run under the
hood. If we hadn't specified \texttt{format\ =\ "brief"}, we'd also
receive the standard error of the estimate and the corresponding
t-statistic used to calculate a p-value. It is always a good idea to
investigate the output of any bioinformatics pipeline.

Now, let's proceed with the demonstration.

\begin{center}\rule{0.5\linewidth}{0.5pt}\end{center}

\newpage

It is best practice to investigate the distribution of p-values using a
histogram.

\begin{Shaded}
\begin{Highlighting}[]
\FunctionTok{hist}\NormalTok{(genus.glm}\SpecialCharTok{$}\StringTok{\textasciigrave{}}\AttributeTok{Groupschizophrenia Pr(\textgreater{}|t|)}\StringTok{\textasciigrave{}}\NormalTok{, }\AttributeTok{xlim =} \FunctionTok{c}\NormalTok{(}\DecValTok{0}\NormalTok{, }\DecValTok{1}\NormalTok{), }\AttributeTok{breaks =} \DecValTok{20}\NormalTok{)}
\end{Highlighting}
\end{Shaded}

\includegraphics{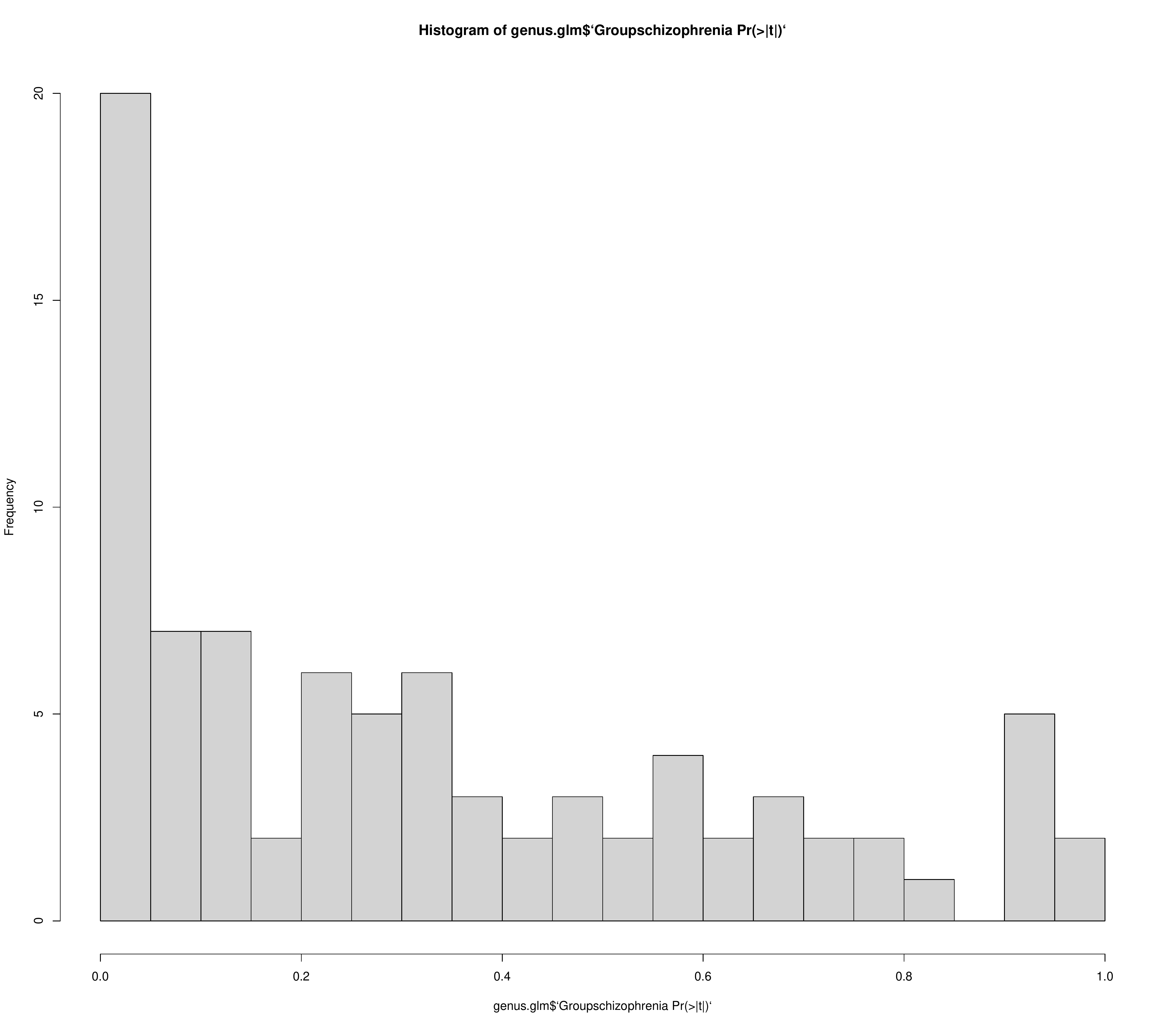}

Histograms of p-values are useful to assess whether there is something
fishy going on in your data. Broadly speaking, one would typically
expect some features to be altered based on a condition (null hypothesis
is false; H\textsubscript{1}) and some others to not be affected by the
condition (null hypothesis is true; H\textsubscript{0}). The p-value was
designed in such a way that in the case of a true H\textsubscript{0},
the p-values will be uniformly distributed from 0 - 1. Conversely, in
the case of H\textsubscript{1}, the p-values will typically aggregate
close to 0. Normally, we would expect a combination of these two
patterns in our histogram. So we would want to see a low density of
p-values form 0 - 1 with a nice peak around 0 indicating some
differences between our groups. This also means that if the p-value
histogram is ever overly `lumpy' at a value other than 0, this is an
indication that something fishy is going on and that you should try to
find out why this is the case. Sometimes, this can happen when using
one-tailed tests or when the individual features are heavily dependent
on each other.

In this case, the p-value distribution looks fine. We can see that there
is a peak on the left. There are many p-values under 0.05. There is a
group effect here.

\newpage

Check the distribution of Benjamini-Hochberg corrected p-values to get a
sense of the results. This is not strictly necessary, but it can be
helpful to get a sense of how your p-values were affected by the
post-hoc correction and how many figures to expect.

\begin{Shaded}
\begin{Highlighting}[]
\FunctionTok{hist}\NormalTok{(genus.glm}\SpecialCharTok{$}\StringTok{\textasciigrave{}}\AttributeTok{Groupschizophrenia Pr(\textgreater{}|t|).BH}\StringTok{\textasciigrave{}}\NormalTok{, }\AttributeTok{xlim =} \FunctionTok{c}\NormalTok{(}\DecValTok{0}\NormalTok{, }\DecValTok{1}\NormalTok{), }\AttributeTok{breaks =} \DecValTok{20}\NormalTok{)}
\end{Highlighting}
\end{Shaded}

\includegraphics{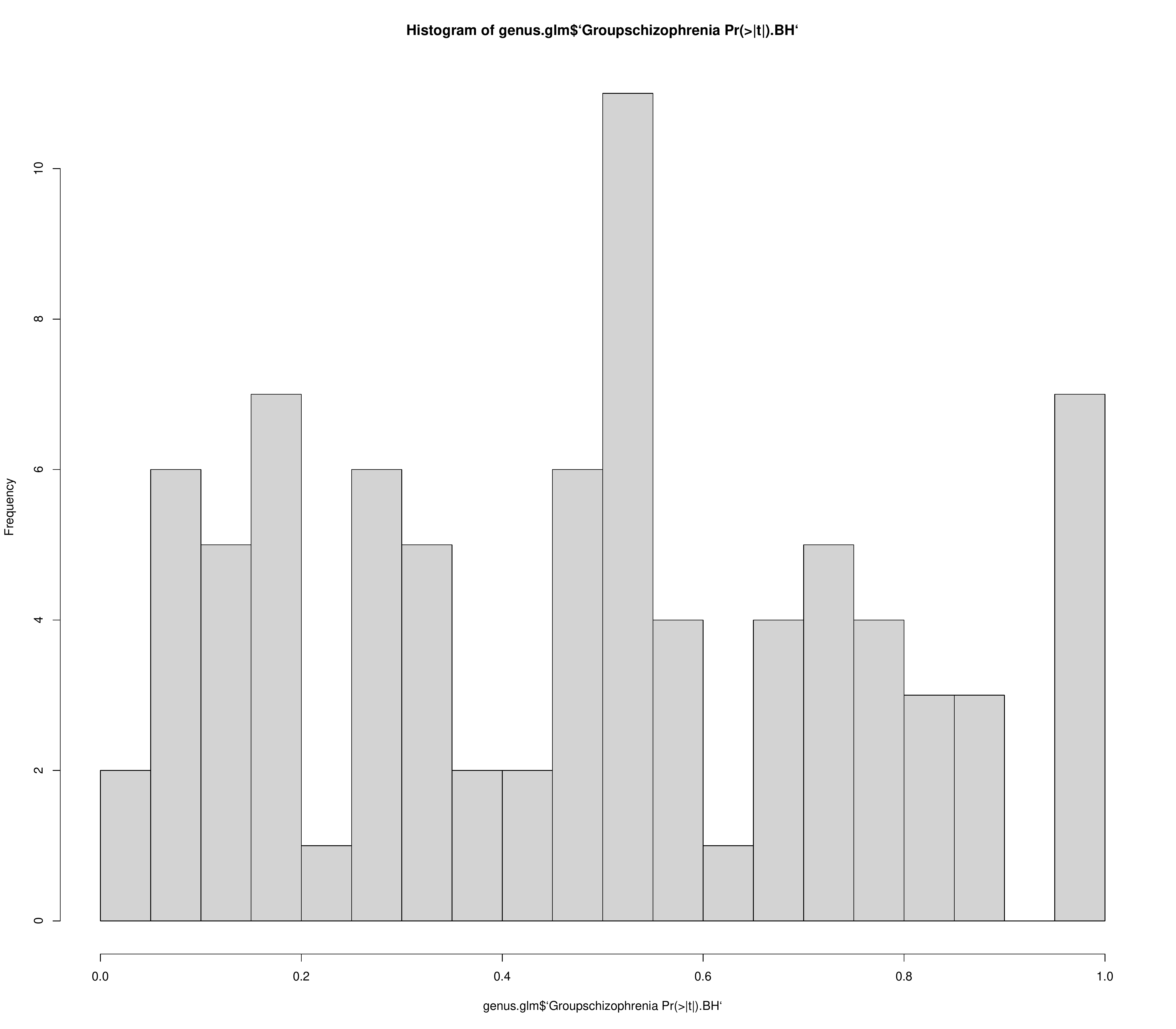}

Using a fairly standard cutoff of q \textless{} 0.1 we see a fair amount
of significant differences.

\newpage

\begin{Shaded}
\begin{Highlighting}[]
\CommentTok{\#Plot the features that show a group effect at q \textless{} 0.1}
\NormalTok{genBH }\OtherTok{\textless{}{-}}\NormalTok{ genus.exp[genus.glm[genus.glm}\SpecialCharTok{$}\StringTok{\textasciigrave{}}\AttributeTok{Groupschizophrenia Pr(\textgreater{}|t|).BH}\StringTok{\textasciigrave{}} \SpecialCharTok{\textless{}} \FloatTok{0.1}\NormalTok{,}\StringTok{"feature"}\NormalTok{],]}

\NormalTok{genBH }\SpecialCharTok{\%\textgreater{}\%}
  \FunctionTok{t}\NormalTok{() }\SpecialCharTok{\%\textgreater{}\%}
  \FunctionTok{as.data.frame}\NormalTok{() }\SpecialCharTok{\%\textgreater{}\%}
  \FunctionTok{add\_column}\NormalTok{(}\AttributeTok{Group =}\NormalTok{ metadata}\SpecialCharTok{$}\NormalTok{Group, }
             \AttributeTok{Sex   =}\NormalTok{ metadata}\SpecialCharTok{$}\NormalTok{Sex)  }\SpecialCharTok{\%\textgreater{}\%}
  \FunctionTok{pivot\_longer}\NormalTok{(}\SpecialCharTok{!}\FunctionTok{c}\NormalTok{(}\StringTok{"Group"}\NormalTok{, }\StringTok{"Sex"}\NormalTok{))  }\SpecialCharTok{\%\textgreater{}\%}
  \FunctionTok{mutate}\NormalTok{(}\AttributeTok{name =} \FunctionTok{str\_replace}\NormalTok{(name, }\StringTok{".*ales\_"}\NormalTok{, }\StringTok{""}\NormalTok{)) }\SpecialCharTok{\%\textgreater{}\%} 
  \FunctionTok{ggplot}\NormalTok{(}\FunctionTok{aes}\NormalTok{(}\AttributeTok{x     =}\NormalTok{ Group, }
             \AttributeTok{y     =}\NormalTok{ value, }
             \AttributeTok{fill  =}\NormalTok{ Group, }
             \AttributeTok{shape =}\NormalTok{ Sex, }
             \AttributeTok{group =}\NormalTok{ Group)) }\SpecialCharTok{+} 
  \FunctionTok{geom\_boxplot}\NormalTok{(}\AttributeTok{alpha =} \DecValTok{1}\SpecialCharTok{/}\DecValTok{2}\NormalTok{, }\AttributeTok{coef =} \DecValTok{100}\NormalTok{) }\SpecialCharTok{+}
  \FunctionTok{geom\_beeswarm}\NormalTok{(}\AttributeTok{size =} \DecValTok{3}\NormalTok{, }\AttributeTok{cex =} \DecValTok{3}\NormalTok{) }\SpecialCharTok{+} 
  
  \FunctionTok{facet\_wrap}\NormalTok{(}\SpecialCharTok{\textasciitilde{}}\NormalTok{name, }\AttributeTok{scales =} \StringTok{"free\_y"}\NormalTok{, }\AttributeTok{ncol =} \DecValTok{4}\NormalTok{) }\SpecialCharTok{+}
  \FunctionTok{scale\_fill\_manual}\NormalTok{(  }\AttributeTok{values =} \FunctionTok{c}\NormalTok{(}\StringTok{"healthy"}  \OtherTok{=} \StringTok{"\#fe9929"}\NormalTok{, }
                                 \StringTok{"schizophrenia"} \OtherTok{=} \StringTok{"\#8c6bb1"}\NormalTok{)) }\SpecialCharTok{+} 
  \FunctionTok{scale\_shape\_manual}\NormalTok{(}\AttributeTok{values =} \FunctionTok{c}\NormalTok{(}\StringTok{"female"} \OtherTok{=} \DecValTok{21}\NormalTok{, }
                                \StringTok{"male"} \OtherTok{=} \DecValTok{22}\NormalTok{)) }\SpecialCharTok{+}  
  \FunctionTok{ylab}\NormalTok{(}\StringTok{""}\NormalTok{) }\SpecialCharTok{+} \FunctionTok{xlab}\NormalTok{(}\StringTok{""}\NormalTok{) }\SpecialCharTok{+} \FunctionTok{theme\_bw}\NormalTok{() }\SpecialCharTok{+} \FunctionTok{theme}\NormalTok{(}\AttributeTok{text =} \FunctionTok{element\_text}\NormalTok{(}\AttributeTok{size =} \DecValTok{12}\NormalTok{))}
\end{Highlighting}
\end{Shaded}

\includegraphics{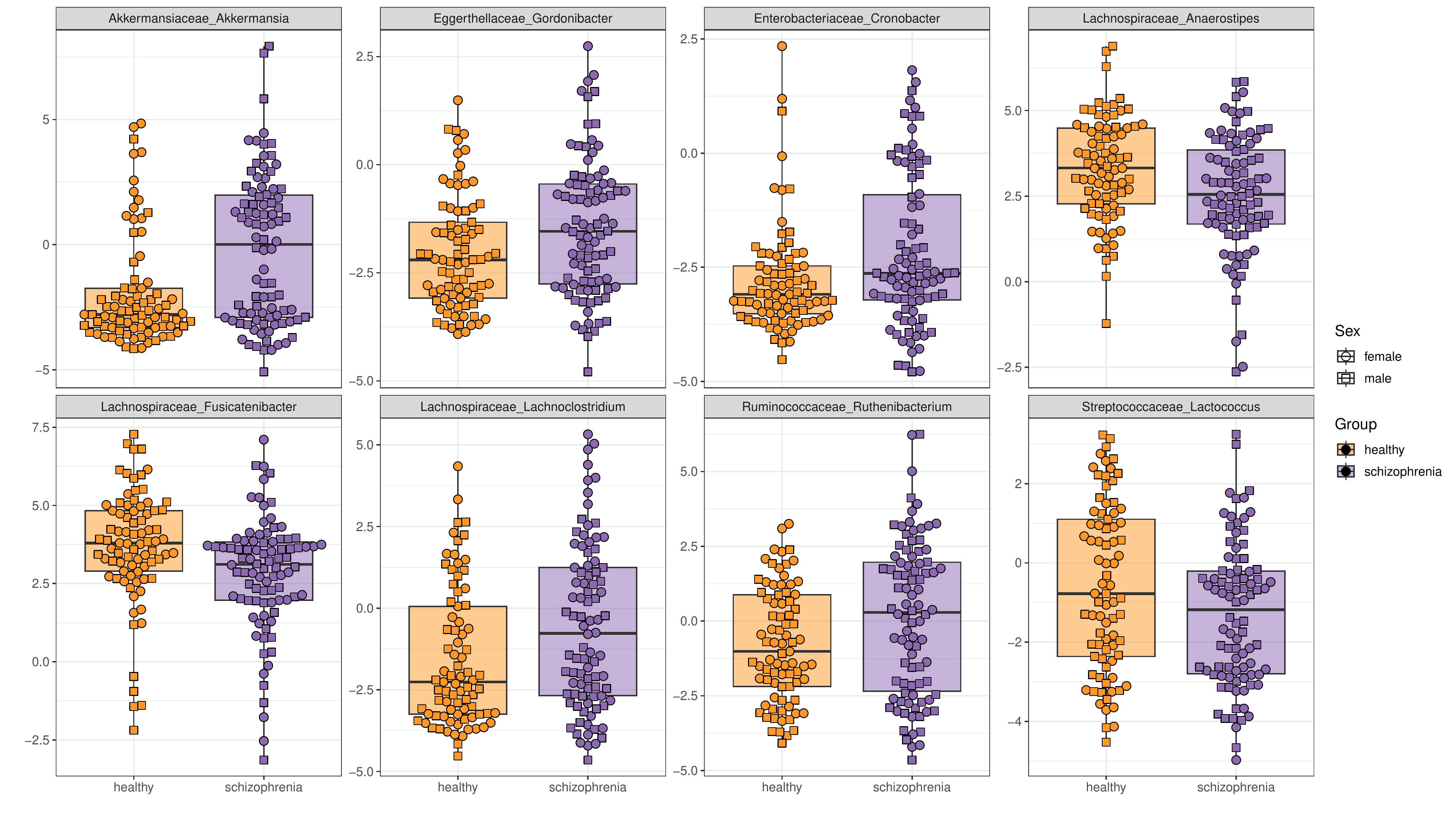}

\begin{Shaded}
\begin{Highlighting}[]
\CommentTok{\#write.csv(genus.glm, "genus.glm.csv") \#To save the results to a file. }
\end{Highlighting}
\end{Shaded}

It seems eight genera are significantly differential abundant between
our healthy and schizophrenia groups given the q \textless{} 0.1
threshold. Note that the y-axis depicts CLR-transformed abundance.
\newpage

\hypertarget{functional-modules}{%
\subsection{4.2 Functional Modules}\label{functional-modules}}

Functional modules provide many advantages over taxa when it comes to
differential abundance analysis. Perhaps prime among them, they are
easier to interpret as they cover concrete molecular processes rather
than the abundance of a microbe that may not be characterized.

\hypertarget{code-chunk-load-the-gut-brain-modules-and-prepare-them-for-analysis}{%
\subsubsection{Code chunk: Load the Gut Brain Modules and prepare them
for
analysis}\label{code-chunk-load-the-gut-brain-modules-and-prepare-them-for-analysis}}

\begin{Shaded}
\begin{Highlighting}[]
\CommentTok{\#Ensure reproducibility within this document}
\FunctionTok{set.seed}\NormalTok{(}\DecValTok{1}\NormalTok{)}

\CommentTok{\#Load GBMS like we did with the genus{-}level counts and metadata above. }
\NormalTok{GBMs }\OtherTok{\textless{}{-}}\NormalTok{ GBMs}

\CommentTok{\#Make sure our count data is all numbers}
\NormalTok{GBMs   }\OtherTok{\textless{}{-}} \FunctionTok{apply}\NormalTok{(GBMs,}\FunctionTok{c}\NormalTok{(}\DecValTok{1}\NormalTok{,}\DecValTok{2}\NormalTok{),}\ControlFlowTok{function}\NormalTok{(x) }\FunctionTok{as.numeric}\NormalTok{(}\FunctionTok{as.character}\NormalTok{(x)))}

\CommentTok{\#Remove features with prevalence \textless{} 10\% in two steps:}
\CommentTok{\#First, determine how often every feature is absent in a sample}
\NormalTok{n\_zeroes\_GBMs }\OtherTok{\textless{}{-}} \FunctionTok{rowSums}\NormalTok{(GBMs }\SpecialCharTok{==} \DecValTok{0}\NormalTok{)}

\CommentTok{\#Then, remove features that are absent in more than your threshold (90\% in this case).}
\NormalTok{GBMs    }\OtherTok{\textless{}{-}}\NormalTok{ GBMs[n\_zeroes\_GBMs }\SpecialCharTok{\textless{}=} \FunctionTok{round}\NormalTok{(}\FunctionTok{ncol}\NormalTok{(GBMs) }\SpecialCharTok{*} \FloatTok{0.90}\NormalTok{),]  }

\CommentTok{\#Perform a CLR transformation}
\NormalTok{GBMs.exp }\OtherTok{\textless{}{-}} \FunctionTok{clr\_c}\NormalTok{(GBMs)}

\CommentTok{\#Reorder the CLR{-}transformed feature table to match the metadata}
\NormalTok{GBMs.exp }\OtherTok{=}\NormalTok{ GBMs.exp[,metadata}\SpecialCharTok{$}\NormalTok{master\_ID]}

\CommentTok{\#This function fits the equivalent of lm(feature \textasciitilde{} Group + Sex + Smoker) for each feature.}
\CommentTok{\#It also performs an appropriate Benjamini{-}Hochberg correction on the p{-}values. }
\NormalTok{GBMs.glm }\OtherTok{=}  \FunctionTok{fw\_glm}\NormalTok{(}\AttributeTok{x             =}\NormalTok{ GBMs.exp,}
                   \AttributeTok{f             =} \SpecialCharTok{\textasciitilde{}}\NormalTok{ Group }\SpecialCharTok{+}\NormalTok{ Sex }\SpecialCharTok{+}\NormalTok{ Smoker, }
                   \AttributeTok{metadata      =}\NormalTok{ metadata, }
                   \AttributeTok{adjust.method =} \StringTok{"BH"}\NormalTok{)}
\end{Highlighting}
\end{Shaded}

\begin{verbatim}
## [1] "Using the following formula: x ~ Group + Sex + Smoker"
## [1] "Adjusting for FDR using Benjamini & Hochberg's procedure."
\end{verbatim}

\newpage

It is best practice to investigate the distribution of p-values using a
histogram.

\begin{Shaded}
\begin{Highlighting}[]
\FunctionTok{hist}\NormalTok{(GBMs.glm}\SpecialCharTok{$}\StringTok{\textasciigrave{}}\AttributeTok{Groupschizophrenia Pr(\textgreater{}|t|)}\StringTok{\textasciigrave{}}\NormalTok{, }\AttributeTok{xlim =} \FunctionTok{c}\NormalTok{(}\DecValTok{0}\NormalTok{, }\DecValTok{1}\NormalTok{), }\AttributeTok{breaks =} \DecValTok{20}\NormalTok{)}
\end{Highlighting}
\end{Shaded}

\includegraphics{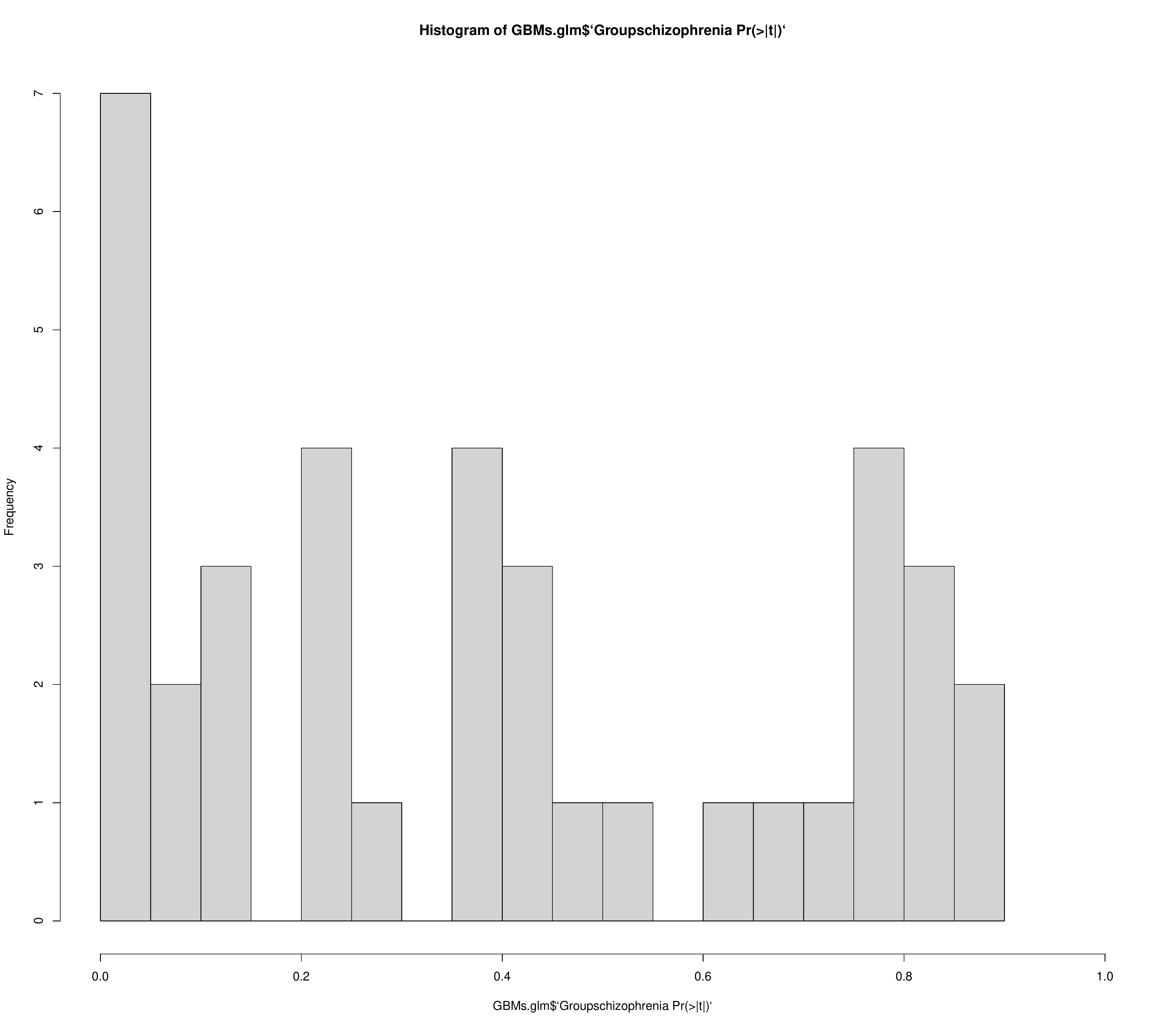}

Histograms of p-values are useful to assess whether there is something
fishy going on in your data. Broadly speaking, one would typically
expect some features to be altered based on a condition (null hypothesis
is false; H\textsubscript{1}) and some others to not be affected by the
condition (null hypothesis is true; H\textsubscript{0}). The p-value was
designed in such a way that in the case of a true H\textsubscript{0},
the p-values will be uniformly distributed from 0 - 1. Conversely, in
the case of H\textsubscript{1}, the p-values will typically aggregate
close to 0. Normally, we would expect a combination of these two
patterns in our histogram. So we would want to see a low density of
p-values form 0 - 1 with a nice peak around 0 indicating some
differences between our groups. This also means that if the p-value
histogram is ever overly `lumpy' at a value other than 0, this is an
indication that something fishy is going on and that you should try to
find out why this is the case. Sometimes, this can happen when using
one-tailed tests or when the individual features are heavily dependent
on each other.

In this case, the p-value distribution looks fine. We can see that there
is a peak on the left. There are many p-values under 0.05. There is a
group effect here.

\newpage

Check the distribution of Benjamini-Hochberg corrected p-values to get a
sense of the results. This is not strictly necessary, but it can be
helpful to get a sense of how your p-values were affected by the
post-hoc correction and how many figures to expect.

\begin{Shaded}
\begin{Highlighting}[]
\FunctionTok{hist}\NormalTok{(GBMs.glm}\SpecialCharTok{$}\StringTok{\textasciigrave{}}\AttributeTok{Groupschizophrenia Pr(\textgreater{}|t|).BH}\StringTok{\textasciigrave{}}\NormalTok{, }\AttributeTok{xlim =} \FunctionTok{c}\NormalTok{(}\DecValTok{0}\NormalTok{, }\DecValTok{1}\NormalTok{), }\AttributeTok{breaks =} \DecValTok{20}\NormalTok{)}
\end{Highlighting}
\end{Shaded}

\includegraphics{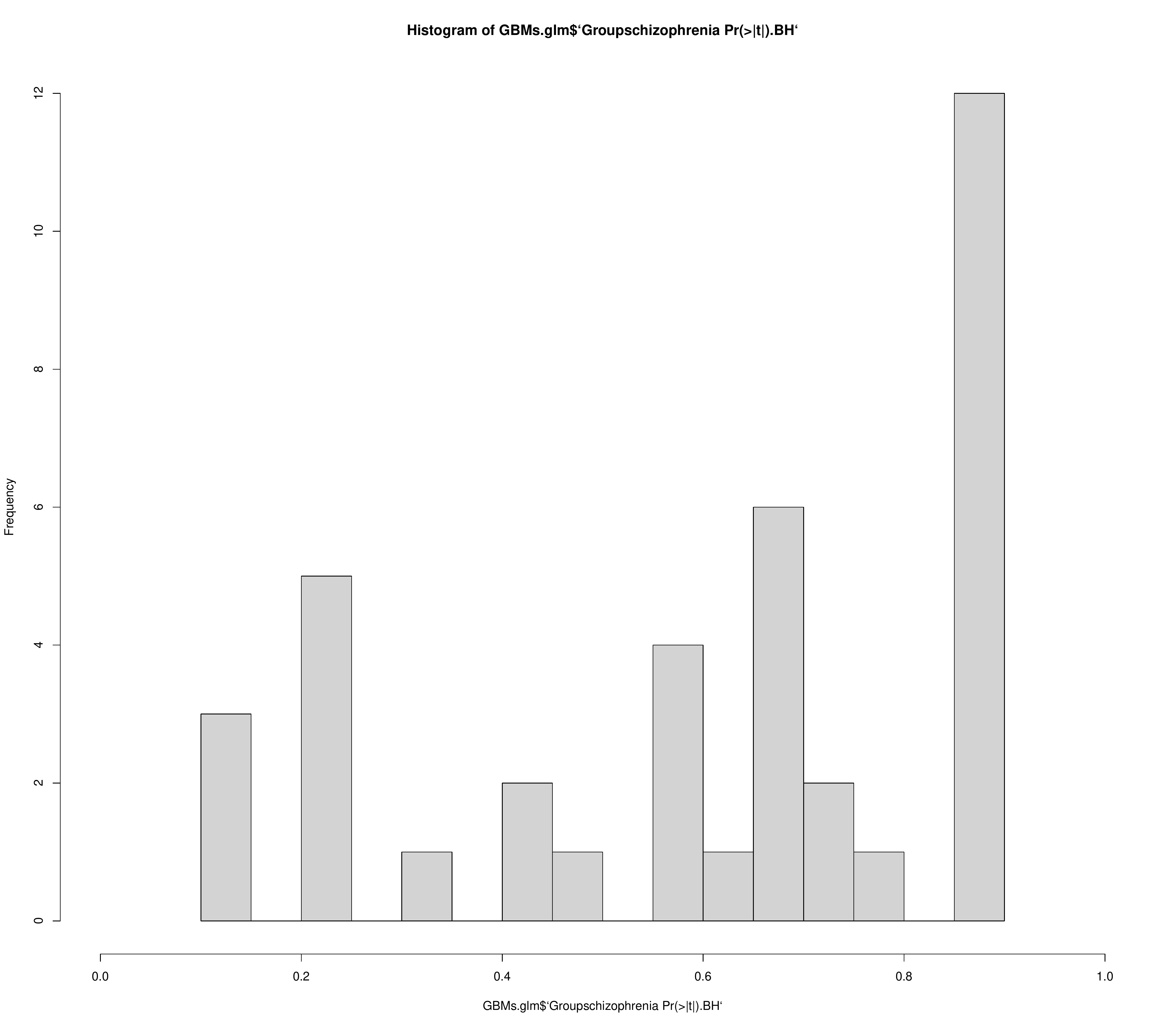}
Using a fairly standard cutoff of q \textless{} 0.2 we see a few hits.

\newpage

\hypertarget{code-chunk-plot-the-differentially-abundant-gut-brain-modules}{%
\subsubsection{Code chunk: Plot the differentially abundant Gut Brain
modules}\label{code-chunk-plot-the-differentially-abundant-gut-brain-modules}}

\begin{Shaded}
\begin{Highlighting}[]
\CommentTok{\#Plot the features that show a group effect at q \textless{} 0.2}
\NormalTok{GBM\_BH }\OtherTok{\textless{}{-}}\NormalTok{ GBMs.exp[GBMs.glm[GBMs.glm}\SpecialCharTok{$}\StringTok{\textasciigrave{}}\AttributeTok{Groupschizophrenia Pr(\textgreater{}|t|).BH}\StringTok{\textasciigrave{}} \SpecialCharTok{\textless{}} \FloatTok{0.2}\NormalTok{,}\StringTok{"feature"}\NormalTok{],] }

\NormalTok{GBM\_BH }\SpecialCharTok{\%\textgreater{}\%}
  \FunctionTok{t}\NormalTok{() }\SpecialCharTok{\%\textgreater{}\%}
  \FunctionTok{as.data.frame}\NormalTok{() }\SpecialCharTok{\%\textgreater{}\%}
  \FunctionTok{add\_column}\NormalTok{(}\AttributeTok{Group =}\NormalTok{ metadata}\SpecialCharTok{$}\NormalTok{Group, }
             \AttributeTok{Sex   =}\NormalTok{ metadata}\SpecialCharTok{$}\NormalTok{Sex)  }\SpecialCharTok{\%\textgreater{}\%}
  \FunctionTok{pivot\_longer}\NormalTok{(}\SpecialCharTok{!}\FunctionTok{c}\NormalTok{(}\StringTok{"Group"}\NormalTok{, }\StringTok{"Sex"}\NormalTok{))  }\SpecialCharTok{\%\textgreater{}\%}
  \FunctionTok{mutate}\NormalTok{(}\AttributeTok{name =} \FunctionTok{str\_replace}\NormalTok{(name, }\StringTok{".*ales\_"}\NormalTok{, }\StringTok{""}\NormalTok{)) }\SpecialCharTok{\%\textgreater{}\%} 
  \FunctionTok{ggplot}\NormalTok{(}\FunctionTok{aes}\NormalTok{(}\AttributeTok{x     =}\NormalTok{ Group, }
             \AttributeTok{y     =}\NormalTok{ value, }
             \AttributeTok{fill  =}\NormalTok{ Group, }
             \AttributeTok{shape =}\NormalTok{ Sex, }
             \AttributeTok{group =}\NormalTok{ Group)) }\SpecialCharTok{+} 
  \FunctionTok{geom\_boxplot}\NormalTok{(}\AttributeTok{alpha =} \DecValTok{1}\SpecialCharTok{/}\DecValTok{2}\NormalTok{, }\AttributeTok{coef =} \DecValTok{100}\NormalTok{) }\SpecialCharTok{+}
  \FunctionTok{geom\_beeswarm}\NormalTok{(}\AttributeTok{size =} \DecValTok{3}\NormalTok{, }\AttributeTok{cex =} \DecValTok{3}\NormalTok{) }\SpecialCharTok{+} 
  
  \FunctionTok{facet\_wrap}\NormalTok{(}\SpecialCharTok{\textasciitilde{}}\NormalTok{name, }\AttributeTok{scales =} \StringTok{"free\_y"}\NormalTok{, }\AttributeTok{ncol =} \DecValTok{3}\NormalTok{) }\SpecialCharTok{+}
  \FunctionTok{scale\_fill\_manual}\NormalTok{(  }\AttributeTok{values =} \FunctionTok{c}\NormalTok{(}\StringTok{"healthy"}  \OtherTok{=} \StringTok{"\#fe9929"}\NormalTok{, }
                                 \StringTok{"schizophrenia"} \OtherTok{=} \StringTok{"\#8c6bb1"}\NormalTok{)) }\SpecialCharTok{+} 
  \FunctionTok{scale\_shape\_manual}\NormalTok{(}\AttributeTok{values =} \FunctionTok{c}\NormalTok{(}\StringTok{"female"} \OtherTok{=} \DecValTok{21}\NormalTok{, }
                                \StringTok{"male"} \OtherTok{=} \DecValTok{22}\NormalTok{)) }\SpecialCharTok{+}  
  \FunctionTok{ylab}\NormalTok{(}\StringTok{""}\NormalTok{) }\SpecialCharTok{+} \FunctionTok{xlab}\NormalTok{(}\StringTok{""}\NormalTok{) }\SpecialCharTok{+} \FunctionTok{theme\_bw}\NormalTok{() }\SpecialCharTok{+} \FunctionTok{theme}\NormalTok{(}\AttributeTok{text =} \FunctionTok{element\_text}\NormalTok{(}\AttributeTok{size =} \DecValTok{12}\NormalTok{))}
\end{Highlighting}
\end{Shaded}

\includegraphics{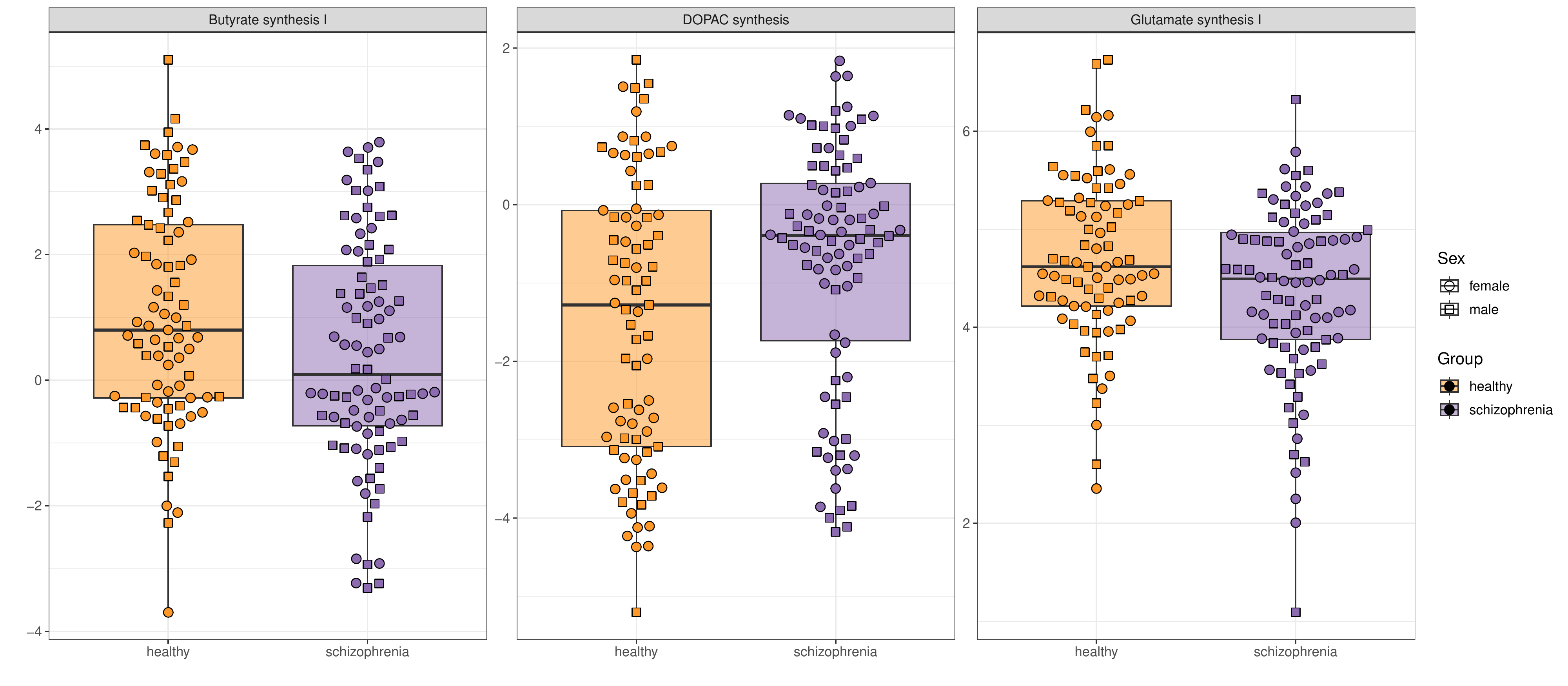}

\begin{Shaded}
\begin{Highlighting}[]
\CommentTok{\#write.csv(GBMs.glm, "GBMs.glm.csv") \#To save the results to a file. }
\end{Highlighting}
\end{Shaded}

\newpage

\hypertarget{code-chunk-load-the-gut-metabolic-modules-and-prepare-them-for-analysis}{%
\subsubsection{Code chunk: Load the Gut Metabolic Modules and prepare
them for
analysis}\label{code-chunk-load-the-gut-metabolic-modules-and-prepare-them-for-analysis}}

\begin{Shaded}
\begin{Highlighting}[]
\CommentTok{\#Ensure reproducibility within this document}
\FunctionTok{set.seed}\NormalTok{(}\DecValTok{1}\NormalTok{)}

\CommentTok{\#Load GBMS like we did with the genus{-}level counts and metadata above. }
\NormalTok{GMMs }\OtherTok{\textless{}{-}}\NormalTok{ GMMs}

\CommentTok{\#Make sure our count data is all numbers}
\NormalTok{GMMs   }\OtherTok{\textless{}{-}} \FunctionTok{apply}\NormalTok{(GMMs,}\FunctionTok{c}\NormalTok{(}\DecValTok{1}\NormalTok{,}\DecValTok{2}\NormalTok{),}\ControlFlowTok{function}\NormalTok{(x) }\FunctionTok{as.numeric}\NormalTok{(}\FunctionTok{as.character}\NormalTok{(x)))}


\CommentTok{\#Remove features with prevalence \textless{} 10\% in two steps:}
\CommentTok{\#First, determine how often every feature is absent in a sample}
\NormalTok{n\_zeroes\_GMMs }\OtherTok{\textless{}{-}} \FunctionTok{rowSums}\NormalTok{(GMMs }\SpecialCharTok{==} \DecValTok{0}\NormalTok{)}

\CommentTok{\#Then, remove features that are absent in more than your threshold (90\% in this case).}
\NormalTok{GMMs    }\OtherTok{\textless{}{-}}\NormalTok{ GMMs[n\_zeroes\_GMMs }\SpecialCharTok{\textless{}=} \FunctionTok{round}\NormalTok{(}\FunctionTok{ncol}\NormalTok{(GMMs) }\SpecialCharTok{*} \FloatTok{0.90}\NormalTok{),]   }

\CommentTok{\#Perform a CLR transformation}
\NormalTok{GMMs.exp }\OtherTok{\textless{}{-}} \FunctionTok{clr\_c}\NormalTok{(GMMs)}

\CommentTok{\#Reorder the CLR{-}transformed feature table to match the metadata}
\NormalTok{GMMs.exp }\OtherTok{=}\NormalTok{ GMMs.exp[,metadata}\SpecialCharTok{$}\NormalTok{master\_ID]}

\CommentTok{\#This function fits the equivalent of lm(feature \textasciitilde{} Group + Sex + Smoker) for each feature.}
\CommentTok{\#It also performs an appropriate Benjamini{-}Hochberg correction on the p{-}values. }
\NormalTok{GMMs.glm }\OtherTok{=}  \FunctionTok{fw\_glm}\NormalTok{(}\AttributeTok{x             =}\NormalTok{ GMMs.exp,}
                   \AttributeTok{f             =} \SpecialCharTok{\textasciitilde{}}\NormalTok{ Group }\SpecialCharTok{+}\NormalTok{ Sex }\SpecialCharTok{+}\NormalTok{ Smoker, }
                   \AttributeTok{metadata      =}\NormalTok{ metadata, }
                   \AttributeTok{adjust.method =} \StringTok{"BH"}\NormalTok{)}
\end{Highlighting}
\end{Shaded}

\begin{verbatim}
## [1] "Using the following formula: x ~ Group + Sex + Smoker"
## [1] "Adjusting for FDR using Benjamini & Hochberg's procedure."
\end{verbatim}

\newpage

It is best practice to investigate the distribution of p-values using a
histogram.

\begin{Shaded}
\begin{Highlighting}[]
\FunctionTok{hist}\NormalTok{(GMMs.glm}\SpecialCharTok{$}\StringTok{\textasciigrave{}}\AttributeTok{Groupschizophrenia Pr(\textgreater{}|t|)}\StringTok{\textasciigrave{}}\NormalTok{, }\AttributeTok{xlim =} \FunctionTok{c}\NormalTok{(}\DecValTok{0}\NormalTok{, }\DecValTok{1}\NormalTok{), }\AttributeTok{breaks =} \DecValTok{20}\NormalTok{)}
\end{Highlighting}
\end{Shaded}

\includegraphics{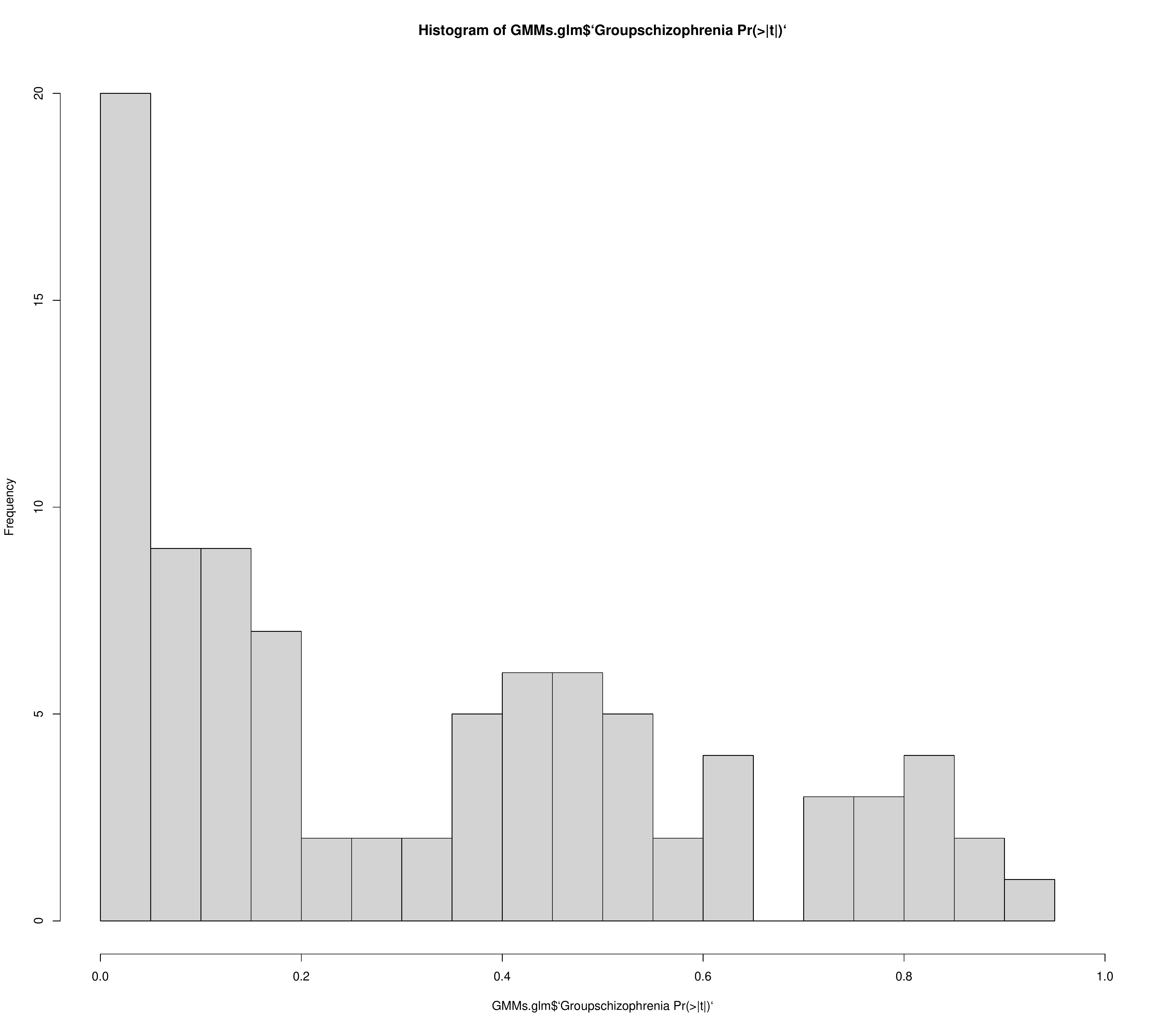}

Histograms of p-values are useful to assess whether there is something
fishy going on in your data. Broadly speaking, one would typically
expect some features to be altered based on a condition (null hypothesis
is false; H\textsubscript{1}) and some others to not be affected by the
condition (null hypothesis is true; H\textsubscript{0}). The p-value was
designed in such a way that in the case of a true H\textsubscript{0},
the p-values will be uniformly distributed from 0 - 1. Conversely, in
the case of H\textsubscript{1}, the p-values will typically aggregate
close to 0. Normally, we would expect a combination of these two
patterns in our histogram. So we would want to see a low density of
p-values form 0 - 1 with a nice peak around 0 indicating some
differences between our groups. This also means that if the p-value
histogram is ever overly `lumpy' at a value other than 0, this is an
indication that something fishy is going on and that you should try to
find out why this is the case. Sometimes, this can happen when using
one-tailed tests or when the individual features are heavily dependent
on each other.

In this case, the p-value distribution looks fine. We can see that there
is a peak on the left. There are many p-values under 0.05. There is a
group effect here.

\newpage

Check the distribution of Benjamini-Hochberg corrected p-values to get a
sense of the results. This is not strictly necessary, but it can be
helpful to get a sense of how your p-values were affected by the
post-hoc correction and how many figures to expect.

\begin{Shaded}
\begin{Highlighting}[]
\FunctionTok{hist}\NormalTok{(GMMs.glm}\SpecialCharTok{$}\StringTok{\textasciigrave{}}\AttributeTok{Groupschizophrenia Pr(\textgreater{}|t|).BH}\StringTok{\textasciigrave{}}\NormalTok{, }\AttributeTok{xlim =} \FunctionTok{c}\NormalTok{(}\DecValTok{0}\NormalTok{, }\DecValTok{1}\NormalTok{), }\AttributeTok{breaks =} \DecValTok{20}\NormalTok{)}
\end{Highlighting}
\end{Shaded}

\includegraphics{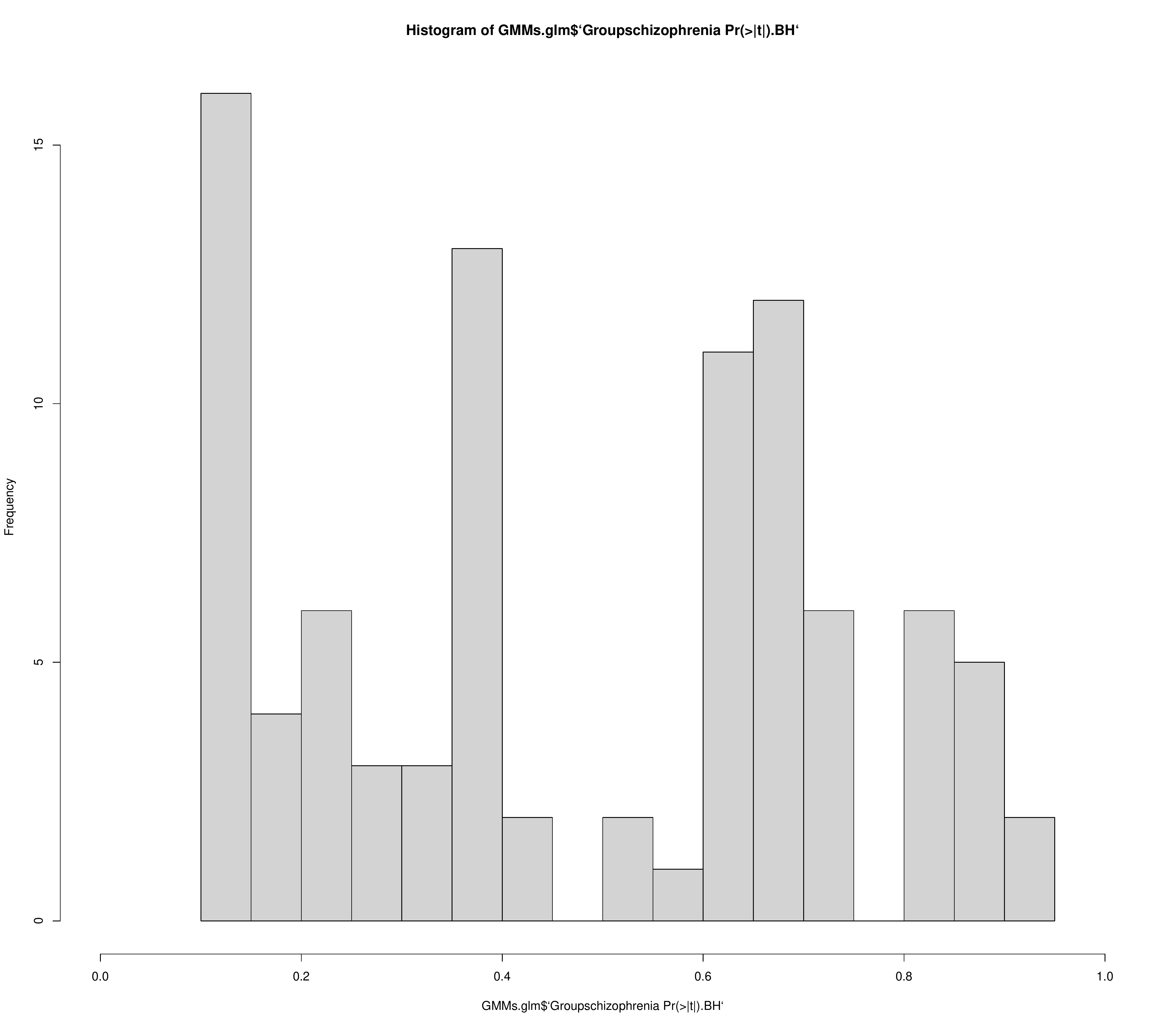}

Using a fairly standard cutoff of q \textless{} 0.2 we see a fair amount
of significant differences.

\newpage

\hypertarget{code-chunk-plot-the-differentially-abundant-gut-metabolic-module}{%
\subsubsection{Code chunk: Plot the differentially abundant Gut
Metabolic
module}\label{code-chunk-plot-the-differentially-abundant-gut-metabolic-module}}

\begin{Shaded}
\begin{Highlighting}[]
\CommentTok{\#Plot the features that show a group effect at q \textless{} 0.2}
\NormalTok{GMM\_BH }\OtherTok{\textless{}{-}}\NormalTok{ GMMs.exp[GMMs.glm[GMMs.glm}\SpecialCharTok{$}\StringTok{\textasciigrave{}}\AttributeTok{Groupschizophrenia Pr(\textgreater{}|t|).BH}\StringTok{\textasciigrave{}} \SpecialCharTok{\textless{}} \FloatTok{0.2}\NormalTok{,}\StringTok{"feature"}\NormalTok{],] }

\NormalTok{GMM\_BH }\SpecialCharTok{\%\textgreater{}\%}
  \FunctionTok{t}\NormalTok{() }\SpecialCharTok{\%\textgreater{}\%}
  \FunctionTok{as.data.frame}\NormalTok{() }\SpecialCharTok{\%\textgreater{}\%}
  \FunctionTok{add\_column}\NormalTok{(}\AttributeTok{Group =}\NormalTok{ metadata}\SpecialCharTok{$}\NormalTok{Group, }
             \AttributeTok{Sex   =}\NormalTok{ metadata}\SpecialCharTok{$}\NormalTok{Sex)  }\SpecialCharTok{\%\textgreater{}\%}
  \FunctionTok{pivot\_longer}\NormalTok{(}\SpecialCharTok{!}\FunctionTok{c}\NormalTok{(}\StringTok{"Group"}\NormalTok{, }\StringTok{"Sex"}\NormalTok{))  }\SpecialCharTok{\%\textgreater{}\%}
  \FunctionTok{mutate}\NormalTok{(}\AttributeTok{name =} \FunctionTok{str\_replace}\NormalTok{(name, }\StringTok{".*ales\_"}\NormalTok{, }\StringTok{""}\NormalTok{)) }\SpecialCharTok{\%\textgreater{}\%} 
  \FunctionTok{ggplot}\NormalTok{(}\FunctionTok{aes}\NormalTok{(}\AttributeTok{x     =}\NormalTok{ Group, }
             \AttributeTok{y     =}\NormalTok{ value, }
             \AttributeTok{fill  =}\NormalTok{ Group, }
             \AttributeTok{shape =}\NormalTok{ Sex, }
             \AttributeTok{group =}\NormalTok{ Group)) }\SpecialCharTok{+} 
  \FunctionTok{geom\_boxplot}\NormalTok{(}\AttributeTok{alpha =} \DecValTok{1}\SpecialCharTok{/}\DecValTok{2}\NormalTok{, }\AttributeTok{coef =} \DecValTok{100}\NormalTok{) }\SpecialCharTok{+}
  \FunctionTok{geom\_beeswarm}\NormalTok{(}\AttributeTok{size =} \DecValTok{3}\NormalTok{, }\AttributeTok{cex =} \DecValTok{3}\NormalTok{) }\SpecialCharTok{+} 
  
  \FunctionTok{facet\_wrap}\NormalTok{(}\SpecialCharTok{\textasciitilde{}}\NormalTok{name, }\AttributeTok{scales =} \StringTok{"free\_y"}\NormalTok{, }\AttributeTok{ncol =} \DecValTok{4}\NormalTok{) }\SpecialCharTok{+}
  \FunctionTok{scale\_fill\_manual}\NormalTok{(  }\AttributeTok{values =} \FunctionTok{c}\NormalTok{(}\StringTok{"healthy"}  \OtherTok{=} \StringTok{"\#fe9929"}\NormalTok{, }
                                 \StringTok{"schizophrenia"} \OtherTok{=} \StringTok{"\#8c6bb1"}\NormalTok{)) }\SpecialCharTok{+} 
  \FunctionTok{scale\_shape\_manual}\NormalTok{(}\AttributeTok{values =} \FunctionTok{c}\NormalTok{(}\StringTok{"female"} \OtherTok{=} \DecValTok{21}\NormalTok{, }
                                \StringTok{"male"} \OtherTok{=} \DecValTok{22}\NormalTok{)) }\SpecialCharTok{+}  
  \FunctionTok{ylab}\NormalTok{(}\StringTok{""}\NormalTok{) }\SpecialCharTok{+} \FunctionTok{xlab}\NormalTok{(}\StringTok{""}\NormalTok{) }\SpecialCharTok{+} \FunctionTok{theme\_bw}\NormalTok{() }\SpecialCharTok{+} \FunctionTok{theme}\NormalTok{(}\AttributeTok{text =} \FunctionTok{element\_text}\NormalTok{(}\AttributeTok{size =} \DecValTok{12}\NormalTok{))}
\end{Highlighting}
\end{Shaded}

\includegraphics{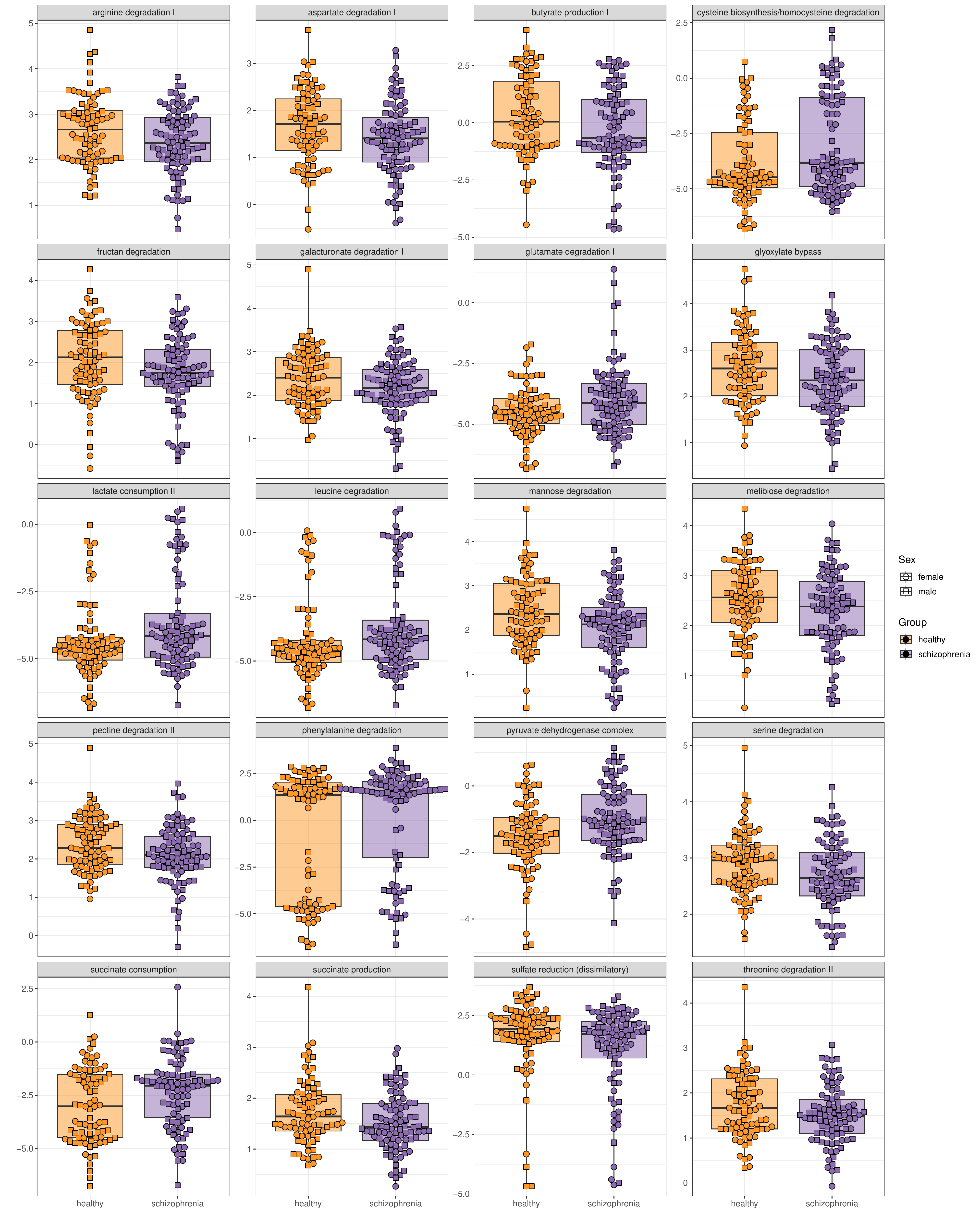}

\begin{Shaded}
\begin{Highlighting}[]
\CommentTok{\#write.csv(GMMs.glm, "GMMs.glm.csv") \#To save the results to a file. }
\end{Highlighting}
\end{Shaded}

\begin{center}\rule{0.5\linewidth}{0.5pt}\end{center}

\newpage

\hypertarget{discussion}{%
\section{5. Discussion}\label{discussion}}

Here, we have presented what a fairly standard microbiome analysis might
look like. The main points we would take from the analysis would be that
there is indeed a difference in the microbiome between our cohort of
patients with schizophrenia and healthy volunteers, in terms of
composition (beta diversity), diversity (alpha diversity) as well as in
differential feature abundance, both on the taxonomical level as well as
the functional level. We could then go on and comment on what specific
functions in the microbiome may explain the differences in our cohort.
For instance, differences in the metabolism of DOPAC and glutamate,
precursor molecules for the important neurotransmitters dopamine and
GABA, could be pointed out and compared to literature. In our
limitations section, we would stress that the effect sizes we found were
quite small and that we found an effect of smoking and of sex as well.

Of course, this document is just a template. Depending on the
experimental setup, findings and experimental questions, you may want to
choose a differing approach. Given the highly complex nature of
microbiome data, one should ideally avoid blindly applying models and
pipelines without understanding what they are doing. D.R. Cox is
famously ascribed the statement: \emph{``Most real life statistical
problems have one or more nonstandard features. There are no routine
statistical questions; only questionable statistical routines.''} We
find this holds true for the microbiome as well.

Clear communication, both in terms of describing and explaining our
methods as well as in terms of figure presentation, are essential for
the health of the field. Indeed, failing to do so can lead to confusion
among our peers. We hope that both aspiring and veteran
bioinformaticians will find our guide helpful. We have tried to model
this piece after what we would have loved to have access to ourselves
when we first set out to study the microbiome.

\begin{center}\rule{0.5\linewidth}{0.5pt}\end{center}

\newpage

\hypertarget{session-info}{%
\section{Session Info}\label{session-info}}

\begin{Shaded}
\begin{Highlighting}[]
\NormalTok{sessioninfo}\SpecialCharTok{::}\FunctionTok{session\_info}\NormalTok{()}
\end{Highlighting}
\end{Shaded}

\begin{verbatim}
## - Session info ---------------------------------------------------------------
##  setting  value
##  version  R version 4.2.2 Patched (2022-11-10 r83330)
##  os       Ubuntu 18.04.6 LTS
##  system   x86_64, linux-gnu
##  ui       X11
##  language en_IE:en
##  collate  en_IE.UTF-8
##  ctype    en_IE.UTF-8
##  tz       Europe/Dublin
##  date     2023-07-25
##  pandoc   2.19.2 @ /usr/lib/rstudio/resources/app/bin/quarto/bin/tools/ (via rmarkdown)
## 
## - Packages -------------------------------------------------------------------
##  package       * version  date (UTC) lib source
##  abind           1.4-5    2016-07-21 [1] CRAN (R 4.2.0)
##  assertthat      0.2.1    2019-03-21 [1] CRAN (R 4.2.0)
##  backports       1.4.1    2021-12-13 [1] CRAN (R 4.2.0)
##  beeswarm        0.4.0    2021-06-01 [1] CRAN (R 4.2.0)
##  broom           1.0.2    2022-12-15 [1] CRAN (R 4.2.1)
##  car             3.0-13   2022-05-02 [1] CRAN (R 4.2.0)
##  carData         3.0-5    2022-01-06 [1] CRAN (R 4.2.0)
##  cellranger      1.1.0    2016-07-27 [1] CRAN (R 4.2.0)
##  cli             3.6.0    2023-01-09 [1] CRAN (R 4.2.1)
##  cluster         2.1.4    2022-08-22 [4] CRAN (R 4.2.1)
##  colorspace      2.0-3    2022-02-21 [1] CRAN (R 4.2.0)
##  crayon          1.5.2    2022-09-29 [1] CRAN (R 4.2.1)
##  DBI             1.1.3    2022-06-18 [1] CRAN (R 4.2.0)
##  dbplyr          2.3.0    2023-01-16 [1] CRAN (R 4.2.1)
##  digest          0.6.31   2022-12-11 [1] CRAN (R 4.2.1)
##  dplyr         * 1.0.10   2022-09-01 [1] CRAN (R 4.2.1)
##  ellipsis        0.3.2    2021-04-29 [1] CRAN (R 4.2.0)
##  evaluate        0.20     2023-01-17 [1] CRAN (R 4.2.1)
##  fansi           1.0.3    2022-03-24 [1] CRAN (R 4.2.0)
##  farver          2.1.1    2022-07-06 [1] CRAN (R 4.2.1)
##  fastmap         1.1.0    2021-01-25 [1] CRAN (R 4.2.0)
##  forcats       * 0.5.2    2022-08-19 [1] CRAN (R 4.2.1)
##  fs              1.5.2    2021-12-08 [1] CRAN (R 4.2.0)
##  gargle          1.2.1    2022-09-08 [1] CRAN (R 4.2.1)
##  generics        0.1.3    2022-07-05 [1] CRAN (R 4.2.1)
##  ggbeeswarm    * 0.7.1    2022-12-16 [1] CRAN (R 4.2.1)
##  ggforce       * 0.4.1    2022-10-04 [1] CRAN (R 4.2.1)
##  ggplot2       * 3.4.0    2022-11-04 [1] CRAN (R 4.2.1)
##  glue            1.6.2    2022-02-24 [1] CRAN (R 4.2.0)
##  googledrive     2.0.0    2021-07-08 [1] CRAN (R 4.2.0)
##  googlesheets4   1.0.1    2022-08-13 [1] CRAN (R 4.2.1)
##  gtable          0.3.1    2022-09-01 [1] CRAN (R 4.2.1)
##  haven           2.5.1    2022-08-22 [1] CRAN (R 4.2.1)
##  highr           0.10     2022-12-22 [1] CRAN (R 4.2.1)
##  hms             1.1.2    2022-08-19 [1] CRAN (R 4.2.1)
##  htmltools       0.5.4    2022-12-07 [1] CRAN (R 4.2.1)
##  httr            1.4.4    2022-08-17 [1] CRAN (R 4.2.1)
##  iNEXT         * 3.0.0    2022-08-29 [1] CRAN (R 4.2.1)
##  jsonlite        1.8.4    2022-12-06 [1] CRAN (R 4.2.1)
##  knitr         * 1.41     2022-11-18 [1] CRAN (R 4.2.1)
##  labeling        0.4.2    2020-10-20 [1] CRAN (R 4.2.0)
##  lattice       * 0.20-45  2021-09-22 [4] CRAN (R 4.2.0)
##  lifecycle       1.0.3    2022-10-07 [1] CRAN (R 4.2.1)
##  lubridate       1.9.0    2022-11-06 [1] CRAN (R 4.2.1)
##  magrittr        2.0.3    2022-03-30 [1] CRAN (R 4.2.0)
##  MASS            7.3-58.2 2023-01-23 [4] CRAN (R 4.2.2)
##  Matrix          1.5-3    2022-11-11 [1] CRAN (R 4.2.1)
##  metafolio     * 0.1.1    2022-04-11 [1] CRAN (R 4.2.0)
##  mgcv            1.8-41   2022-10-21 [4] CRAN (R 4.2.1)
##  modelr          0.1.10   2022-11-11 [1] CRAN (R 4.2.1)
##  munsell         0.5.0    2018-06-12 [1] CRAN (R 4.2.0)
##  nlme            3.1-162  2023-01-31 [4] CRAN (R 4.2.2)
##  patchwork     * 1.1.2    2022-08-19 [1] CRAN (R 4.2.1)
##  permute       * 0.9-7    2022-01-27 [1] CRAN (R 4.2.0)
##  pillar          1.8.1    2022-08-19 [1] CRAN (R 4.2.1)
##  pkgconfig       2.0.3    2019-09-22 [1] CRAN (R 4.2.0)
##  plyr            1.8.8    2022-11-11 [1] CRAN (R 4.2.1)
##  polyclip        1.10-4   2022-10-20 [1] CRAN (R 4.2.1)
##  purrr         * 1.0.1    2023-01-10 [1] CRAN (R 4.2.1)
##  R6              2.5.1    2021-08-19 [1] CRAN (R 4.2.0)
##  Rcpp            1.0.9    2022-07-08 [1] CRAN (R 4.2.1)
##  readr         * 2.1.3    2022-10-01 [1] CRAN (R 4.2.1)
##  readxl          1.4.1    2022-08-17 [1] CRAN (R 4.2.1)
##  reprex          2.0.2    2022-08-17 [1] CRAN (R 4.2.1)
##  reshape2        1.4.4    2020-04-09 [1] CRAN (R 4.2.0)
##  rlang           1.0.6    2022-09-24 [1] CRAN (R 4.2.1)
##  rmarkdown       2.20     2023-01-19 [1] CRAN (R 4.2.1)
##  rstudioapi      0.14     2022-08-22 [1] CRAN (R 4.2.1)
##  rvest           1.0.3    2022-08-19 [1] CRAN (R 4.2.1)
##  scales          1.2.1    2022-08-20 [1] CRAN (R 4.2.1)
##  sessioninfo     1.2.2    2021-12-06 [1] CRAN (R 4.2.0)
##  stringi         1.7.12   2023-01-11 [1] CRAN (R 4.2.1)
##  stringr       * 1.5.0    2022-12-02 [1] CRAN (R 4.2.1)
##  tibble        * 3.1.8    2022-07-22 [1] CRAN (R 4.2.1)
##  tidyr         * 1.2.1    2022-09-08 [1] CRAN (R 4.2.1)
##  tidyselect      1.2.0    2022-10-10 [1] CRAN (R 4.2.1)
##  tidyverse     * 1.3.2    2022-07-18 [1] CRAN (R 4.2.1)
##  timechange      0.2.0    2023-01-11 [1] CRAN (R 4.2.1)
##  Tjazi         * 0.1.0.0  2023-04-26 [1] Github (thomazbastiaanssen/Tjazi@91f5c82)
##  tweenr          2.0.2    2022-09-06 [1] CRAN (R 4.2.1)
##  tzdb            0.3.0    2022-03-28 [1] CRAN (R 4.2.0)
##  utf8            1.2.2    2021-07-24 [1] CRAN (R 4.2.0)
##  vctrs           0.5.1    2022-11-16 [1] CRAN (R 4.2.1)
##  vegan         * 2.6-4    2022-10-11 [1] CRAN (R 4.2.1)
##  vipor           0.4.5    2017-03-22 [1] CRAN (R 4.2.0)
##  waldo         * 0.4.0    2022-03-16 [1] CRAN (R 4.2.0)
##  withr           2.5.0    2022-03-03 [1] CRAN (R 4.2.0)
##  xfun            0.36     2022-12-21 [1] CRAN (R 4.2.1)
##  xml2            1.3.3    2021-11-30 [1] CRAN (R 4.2.0)
##  yaml            2.3.6    2022-10-18 [1] CRAN (R 4.2.1)
## 
##  [1] /home/thomaz/R/x86_64-pc-linux-gnu-library/4.2
##  [2] /usr/local/lib/R/site-library
##  [3] /usr/lib/R/site-library
##  [4] /usr/lib/R/library
## 
## ------------------------------------------------------------------------------
\end{verbatim}